\documentclass[10pt,aps,prd,twocolumn,superscriptaddress,nofootinbib,floatfix,notitlepage]{revtex4-2}

\usepackage{epsfig}
\usepackage{amsfonts}
\usepackage{bbm,bm}
\usepackage{comment}
\usepackage{amsmath}
\usepackage{cancel}
\usepackage{verbatim}
\usepackage[normalem]{ulem}
\usepackage[hidelinks,urlbordercolor={1 1 1}]{hyperref}
\usepackage[utf8]{inputenc}
\usepackage{graphicx}
\usepackage{xcolor}
\usepackage{orcidlink}
\allowdisplaybreaks

\let\vec\mathbf %
\newcommand\MSbar{$\overline{\rm MS}$} %

\newcommand{\nn}{\nonumber \\}
\newcommand{\rmi}[1]{{\mbox{\scriptsize #1}}}
\newcommand{\rmii}[1]{{\mbox{\tiny\rm{#1}}}}

\newcommand{\nf}{n_{\rm f}}
\newcommand{\Nc}{N_{\rm c}}

\newcommand{\CF}{C_\rmii{F}}

\newcommand{\mD}{m_\rmii{D}}

\newcommand{\mA}{m_\rmii{$A$}}

\newcommand{\Mh}{M_h}
\newcommand{\Ms}{M_s}
\newcommand{\MZ}{M_\rmii{$Z$}}
\newcommand{\MW}{M_\rmii{$W$}}
\newcommand{\Mt}{M_t}

\newcommand{\gs}{g_\rmi{s}}

\newcommand{\Lamd}{\bar{\mu}_{3\rmii{D}}}
\newcommand{\LamD}{\bar{\mu}}

\newcommand{\Mpl}{M_\rmii{Pl}}

\newcommand{\Tc}{T_{\rm c}}
\newcommand{\Tn}{T_{\rm n}}
\newcommand{\Tp}{T_p}

\newcommand{\vw}{v_{w}}
\newcommand{\cs}{c_{s}}
\newcommand{\fp}{f_{\rm p}}

\newcommand{\Tint}[1]{{\hbox{$\sum$}\!\!\!\!\!\!\!\int\,}_{\!\!\!\!\raise-0.9ex\hbox{$\scriptstyle{#1}$}}}
\newcommand{\Tinti}[1]{{{\Sigma}\!\!\!\!\raise0.3ex\hbox{$\int$}_\rmii{${#1}$}}}
\newcommand{\Tintip}[1]{{{\Sigma'}\!\!\!\!\!\raise0.3ex\hbox{$\int$}_\rmii{${#1}$}}}

\newcommand{\bT}{\overline{T}}

\newcommand{\Veff}{V_{\rmi{eff}}}

\renewcommand{\vec}[1]{{\bf #1}}
\newcommand{\gammaE}{\gamma}
\newcommand{\T}{\rmii{$T$}}

\newcommand{\re}{\mathop{\mbox{Re}}}
\newcommand{\im}{\mathop{\mbox{Im}}}

\newcommand{\article}{{\em article}}

\begin{document}
\newcommand{\UPP}{\affiliation{
Department of Physics and Astronomy, Uppsala University,
Box 516, SE-751 20 Uppsala,
Sweden}}

\newcommand{\HEL}{\affiliation{%
Department of Physics and Helsinki Institute of Physics,
P.O.~Box 64, FI-00014 University of Helsinki,
Finland}}

\newcommand{\NOR}{\affiliation{Nordita,
KTH Royal Institute of Technology and Stockholm University,
Roslagstullsbacken 23,
SE-106 91 Stockholm,
Sweden}}

\newcommand{\UW}{\affiliation{Faculty of Physics, University of Warsaw ul.\ Pasteura 5, 02-093 Warsaw, Poland}}

\newcommand{\GU}{\affiliation{Institute for Theoretical Physics, Goethe Universit{\"a}t Frankfurt, 60438 Frankfurt, Germany}}

\newcommand{\KTHa}{\affiliation{KTH Royal Institute of Technology, Department of Physics, SE-10691 Stockholm, Sweden}}

\newcommand{\KTHb}{\affiliation{The Oskar Klein Centre for Cosmoparticle Physics, AlbaNova University Centre, SE-10691 Stockholm, Sweden}}

\title{%
  Impact of theoretical uncertainties on model parameter reconstruction
  \\ 
  from GW signals sourced by
  cosmological phase transitions
}

\preprint{}

\author{Marek Lewicki\,\orcidlink{0000-0002-8378-0107}}
\email{marek.lewicki@fuw.edu.pl}
\UW

\author{Marco Merchand\,\orcidlink{0000-0002-0154-3520}}
\email{marcomm@kth.se}
\KTHa
\KTHb

\author{Laura Sagunski\,\orcidlink{0000-0002-3506-3306}}
\email{sagunski@itp.uni-frankfurt.de}
\GU

\author{Philipp Schicho\,\orcidlink{0000-0001-5869-7611}}
\email{schicho@itp.uni-frankfurt.de}
\GU

\author{Daniel Schmitt\,\orcidlink{0000-0003-3369-2253}}
\email{dschmitt@itp.uni-frankfurt.de}
\GU

\begin{abstract}
\noindent
Different computational techniques for cosmological phase transition parameters can impact
the Gravitational Wave (GW) spectra predicted in a given particle physics model.
To scrutinize the importance of this effect,
we perform
large-scale parameter scans of
the dynamical real-singlet extended Standard Model using
three perturbative approximations for the effective potential:
the \MSbar{} and
on-shell schemes at leading order, and
three-dimensional thermal effective theory (3D~EFT) at next-to-leading order.
While predictions of GW amplitudes are typically unreliable in the absence of higher-order corrections, we show that the reconstructed model parameter spaces are %
robust up to a few percent in uncertainty.
While 3D~EFT is accurate from one loop order,
theoretical uncertainties of reconstructed model parameters,
using four-dimensional standard techniques,
remain dominant over the experimental ones even for signals merely strong enough to claim a detection by LISA.
\end{abstract}

\maketitle

\section{Introduction}
Recent strong indication for a stochastic gravitational wave (GW) background by
pulsar timing array collaborations~\cite{NANOGrav:2023gor, NANOGrav:2023hde, EPTA:2023sfo, EPTA:2023fyk,Zic:2023gta, Reardon:2023gzh, Xu:2023wog}
is a milestone in deciphering the history of the early universe.
One of the prominent early universe scenarios that produce a stochastic GW background %
and remains a viable source behind the recent observations is a first-order phase transition (PT)~\cite{NANOGrav:2023hvm, Figueroa:2023zhu, Ellis:2023oxs}.
The main motivation for such a signal, however, is associated with (electroweak) symmetry breaking which proceeds by a first-order PT in numerous extensions beyond the Standard Model (SM) and would lead to the production of a GW background in the mHz regime accessible to future experiments such as the upcoming LISA~\cite{Caprini:2015zlo, Caprini:2019egz, LISACosmologyWorkingGroup:2022jok}.

Reliable descriptions of first-order PTs including the
nucleating bubbles,
hydro- and thermodynamics require non-perturbative analyses~\cite{Farakos:1994xh, Kajantie:1995kf, Moore:2000jw, Gould:2022ran}.
Such computations are expensive and
scans of multi-dimensional parameter spaces untenable,
which motivates using perturbation theory when possible.
Concerning thermodynamics, recent progress involving
high-temperature effective field theory (EFT)~\cite{Ginsparg:1980ef,Appelquist:1981vg}
allows us to fully describe PTs that exhibit a scale hierarchy~\cite{Gould:2023ovu}.
While this description is by now state of the art,
practical applications still rely on four-dimensional (4D), incomplete
thermally resummed effective potentials~\cite{Delaunay:2007wb,Patel:2011th,Athron:2023xlk,Caprini:2024hue}.
In this \article{}, we quantify
the theoretical uncertainties remaining in perturbative
computations of the PT equilibrium thermodynamics.

Inverting a measured GW spectrum to determine
the underlying quantum theory Lagrangian is known as
the GW {\em inverse problem}~\cite{Grojean:2006bp, Caprini:2015zlo, Kang:2017mkl, Hashino:2016xoj, Bian:2019bsn, Caprini:2019egz, Friedrich:2022cak}.
Its theoretical challenge during the GW signal reconstruction is to
discern the most reliable approach to PT thermodynamics
since predictions from different levels of computational diligence
can be ambiguous in the reconstructed parameter space~\cite{Croon:2020cgk, Gould:2023jbz}.
In fact, some approaches can even predict inconsistent parameter spaces~\cite{Athron:2022jyi}.
Ensuring theoretical robustness affects all phenomenological analyses that
investigate the impact of new states
beyond the SM (BSM) during the electroweak PT
on GW signals through
a perturbative effective potential.
For such BSM theories to predict a GW signal detectable by upcoming experiments,
their new states need to be dynamical in the infrared (IR)
to modify the low-energy behavior of the SM~\cite{Gould:2019qek, Ellis:2020awk}.

This \article{} addresses the problem of such ambiguous predictions by performing
a large-scale scan of a minimal BSM scalar extension
with more than one light scalar in the IR;
this scan is the first to use the 3D~EFT of~\cite{Schicho:2021gca,Niemi:2021qvp}.
In a proof-of-principle analysis,
we employ different perturbative approaches to the effective potential,
to demonstrate that the theoretical uncertainties %
still dominate over the experimental ones
by contrasting reconstructed parameter spaces from the different approaches.
We show that in three-dimensional (3D) thermal EFT,
the perturbative series converges quickly and
already at next-to-leading order (NLO),
predictions are unambiguous and robust.
Common but perturbatively inconsistent 4D approaches at
incomplete leading order (LO)~\cite{Gould:2021oba} yield errors in
the reconstructed parameters of $\mathcal{O}(1\%)$.
However, such errors would still dominate over
the experimental uncertainties on the parameters if the signal was detected with
a signal-to-noise ratio (SNR) of ${\rm SNR}\simeq 10$ by LISA.

Section~\ref{sec:resummation} reviews the resummation methods used in our analysis.
Sec.~\ref{sec:fopt} introduces the thermal parameters that are
used in sec.~\ref{sec:GW} to determine GW signals.
After comparing the impact of different resummation methods
on model parameter predictions
and estimating sources of theoretical uncertainty,
we conclude in sec.~\ref{sec:conclusion}.

\section{Comparing resummation methods}
\label{sec:resummation}

By minimally extending the SM with one neutral scalar
(xSM)~\cite{Ashoorioon:2009nf,Espinosa:2011ax,Profumo:2014opa,Brauner:2016fla,Gould:2019qek,Schicho:2021gca,Niemi:2021qvp,Gould:2021oba,Sagunski:2023ynd},
we discuss different levels of diligence in computing the corresponding thermal potentials
and their influence on phenomenological predictions.

We study
the $Z_2$-symmetric potential for the xSM
\begin{align}
\label{eq:V0:tree}
  V_0(\Phi, S) &=
    \mu_{h}^2 \Phi^{\dagger} \Phi
  + \lambda (\Phi^{\dagger} \Phi)^2
  \nn &
  + \frac{1}{2}\mu_{s}^2 S^2
  + \frac{\lambda_{s}}{4} S^4
  + \frac{\lambda_{hs}}{2} S^2 \Phi^{\dagger} \Phi
  \;,
\end{align}
where
the tree-level scalar mass matrix is diagonal and
$\mu_{h}^2 < 0$ at zero temperature.
The field
$\Phi = \bigl(G^+ , \frac{1}{\sqrt{2}}(v + h + i G^0)\bigr)^\rmii{T}$
is the ${\rm SU}(2)_\rmii{L}$ SM Higgs doublet with
vacuum expectation value (VEV)
$v_0 \simeq 246$~GeV,
$S = (x + s)$
is the gauge singlet,
odd under a parity $Z_2$-symmetry, and
$G^{\pm}$, $G$ are the
would-be Goldstones fields.
The tree-level potential
\begin{align}
\label{eq:Veff-01}
  V_{0}(\bm{\phi}) &=
    \frac12 \mu^2_{h} v^2
  + \frac14 \lambda v^4
    \nn &
  + \frac12 \mu^2_{s} x^2
  + \frac14 \lambda_{s} x^4
  + \frac14 \lambda_{hs} v^2 x^2
  \,,
\end{align}
with
$\bm{\phi} = (v,x)$
depends on the homogeneous background fields $v$ and $x$.
The
${\rm SU}(2)_\rmii{L}$ and
${\rm U}(1)_\rmii{Y}$ gauge field,
would-be Goldstone, and
physical scalar eigenstate masses read
\begin{align}
\label{eq:mass:h}
  m_{h}^{2} &=
    \mu_{h}^{2}
  + 3\lambda v^2
  + \frac12 \lambda_{hs} x^2
  \,,&
    m_{\rmii{$G$}}^2 &=
    m_{h}^{2}
  - 2\lambda v^2
  \,,\\
\label{eq:mass:s}
  m_{s}^{2} &=
    \mu_{s}^{2}
  + 3\lambda_{s} x^2
  + \frac12 \lambda_{hs} v^2
    \,,&
  m_{\rmii{$W$}}^2 &=
    \frac14 g^2 v^2
  \,,\\
  \label{eq:mass:Z}
  m_{\rmii{$Z$}}^2 &=
    \frac14 \left(g^2 + g'^2 \right) v^2
  \,,
\end{align}
where
$g$ is
the ${\rm SU}(2)_\rmii{L}$ and
$g'$
the ${\rm U}(1)_\rmii{Y}$
gauge coupling.
All approaches employ Landau gauge ($\xi = 0$)
motivated by the arguments in~\cite{Niemi:2021qvp}.
In this gauge,
ghosts and scalars decouple as well as
kinetic mixing between
scalar and vector fields is removed~\cite{Martin:2018emo}.
The xSM gives rise to
two-step PTs from
symmetric to singlet to vacuum Higgs minimum {\em viz.}
\begin{equation}
\label{eq:PTsteps}
  (0,0) \stackrel{{\tt step1}}{\longrightarrow}
  (0,x) \stackrel{{\tt step2}}{\longrightarrow}
  (v_0,0)
  \,.
\end{equation}
This \article{} focuses on {\tt step2} as the source of GWs.

For our GW signal analysis,
we list the different approaches to construct
the effective potential, $\Veff(\bm{\phi},T)$, at finite temperature.
This potential is prone to collective IR sensitivities~\cite{Linde:1980ts}
that render perturbation theory slowly convergent and
can be treated by various forms of resummation.
This \article{} compares different forms of resummation
bearing in mind, that a significant mismatch among them is related
to the failure of perturbation theory
when
misidentifying corrections as higher order
in the underlying strict EFT power counting~\cite{Gould:2023ovu,Lofgren:2023sep}.
As an example, at finite temperature
the complete LO requires two-loop thermal mass effects~\cite{Gould:2021oba}.

Such collective IR effects
modify the vacuum masses by thermal corrections.
At one-loop level, the effective masses are
\begin{align}
\label{eq:mass:higgs}
  \mu_{h,3}^2 & =
        \mu_{h}^{2}
      + T^{2} \Bigr[
        \frac{3g^2}{16}
        + \frac{g'^2}{16}
        + \frac{\lambda}{2}
        + \frac{y_t^2}{4}
        + \frac{\lambda_{hs}}{24}
      \Bigr]
  \,,
  \\
  \mu_{s,3}^{2} &=
      \mu_{s}^{2}
    + T^2 \Bigr[
          \frac{1}{4} \lambda_s
        + \frac{1}{6} \lambda_{hs}
        \Bigr]
  \,,
  \\
\label{eq:mass:mDsu2}
  \mD^{2} &=
    \frac{g^2 T^2}{3} \Bigr[
          \frac{5}{2}
        + \frac{\Nc + 1}{4}\nf
        \Bigr]
  \,,
\end{align}
where
$\nf = 3$ is the number of fermion families,
$\Nc = 3$ the number of colors,
and
$y_t$ the SM top Yukawa coupling.
Subscripts indicate
that thermal corrections are an inherently soft 3D effect,
and
$\mD$ is the ${\rm SU}(2)_\rmii{L}$ Debye mass --
the thermal mass of longitudinal gauge bosons.
Two-loop corrections are obtained via~\cite{Ekstedt:2022bff,Niemi:2021qvp,Schicho:2021gca}.
In our analysis,
we adopt the power counting of~\cite{Niemi:2021qvp},
\begin{align}
  \mu_h^{ }, \mu_s^{ } &\sim g T
  \,,&
  \lambda,\lambda_s^{ },\lambda_{hs}^{ }&\sim g^2
  \,, &
  y_t^{ }, g', \gs^{ } &\sim g
  \,,
\end{align}
where $\gs$ is the QCD coupling.

\subsection{Four dimensions: \MSbar{} scheme}
\label{sec:MSbar}

This scheme uses dimensional regularization with $\LamD$ being
a \MSbar{} renormalization-group (RG) scale.
After renormalization, physical quantities should be $\LamD$-independent
order by order.
In the imaginary time formalism and
at one-loop level,
the effective potential, $V_1$, consists of corrections from the
vacuum
and
thermal medium~\cite{Patel:2011th},
\begin{align}
\label{eq:CW_potential}
  V_{\rmi{1}}^{ }(\bm{\phi},T) &= \sum_i n_{i}^{ }\, J_4^{ }(m_i^{2}(\bm{\phi},T),T)
  \,,\\
  J_4(m^2,T) &= \frac{1}{2} \Tint{P} \ln \big(P^2 + m^2 \bigr)
  \nn 
  &=
    J_{\rmii{CW}}(m^2)
  + J_{\T,\rmii{B(F)}}(m^2)
    \,.
\end{align}
Here,
$J_{\rmii{CW}}$
is the Coleman-Weinberg contribution~\cite{Coleman:1973jx,Jackiw:1974cv}
and
$J_{\T,\rmii{B(F)}}$
the thermal bosonic (fermionic) contribution.
The sum-integral
$\Tinti{P} = T \sum_{p_n} \int_\vec{p}$
contains a
sum over $p_n$ Matsubara modes,
$\int_{\vec{p}} = \mu^{2\epsilon} \int \frac{{\rm d}^d p}{(2\pi)^d}$,
and
$d=3 - 2\epsilon$.
The numbers of degrees of freedom, $n_i$, are
$d$-dependent and
positive (negative) for bosons (fermions).
Since the thermal integrals $J_{\T,\rmii{B(F)}}$ are finite in the ultraviolet (UV),
they are evaluated directly.

Relevant parameters of the theory
are RG-evolved to
the 4D RG-scale $\LamD = v_0$ at one-loop level.
In this scheme, vacuum renormalization
is achieved by the renormalization condition
\begin{align}
\label{eq:ren:cond:tree}
  \partial_{v}^{ }V_0^{ } &= 0
  \,,&
  \partial_{v}^{2}V_0^{ } &= \Mh^2
  \,,
\end{align}
where
capital masses indicate pole masses,
e.g.\
$\Mh$ is the Higgs pole mass.
For the $Z_2$-symmetric xSM
$m_h = \Mh$ and
$m_s = \Ms$.
At the electroweak (EW) scale
with a minimum at the field values
$\bm{\phi}=(v_0,0)$, the zero-temperature Higgs mass is reproduced at
$\LamD=v_0$.
Higher corrections are included by introducing an additional set of counterterms
that tune model parameters such that the tree-level vacuum structure
of eq.~\eqref{eq:ren:cond:tree}
of the potential is preserved
when $V_0 \to \Veff$.
See appendix~\ref{sec:vacren:4d}
for details.
This scheme is conventionally dubbed \MSbar{} scheme~\cite{Delaunay:2007wb} but not
to be confused with the original \MSbar{} scheme where
counterterms only remove UV divergences;
cf.~\ref{sec:3DEFT}.

Without RG improvement,
the PT parameters depend on $\LamD$
through higher-order effects.
In practice, this dependence appears to be small for the fixed
$\LamD = v_0$
but becomes relevant when considering a thermal scale
$\LamD \sim T$;
see sec.~\ref{sec:GW} for details.
The remaining scale dependence is related to
the implicit running of the model parameters in
the thermal, effective, parameters, e.g.\
eqs.~\eqref{eq:mass:higgs}--\eqref{eq:mass:mDsu2}.
The corresponding $\LamD$-dependent logarithms are counterfeit by
RG improvement contained in
the two-loop thermal masses of the 3D EFT~\cite{Gould:2021oba};
cf.\ sec.~\ref{sec:3DEFT}.

In all our comparisons,
we report the model parameters at the input scale,
the $Z$-boson mass, $\MZ$, such that in all plots below
e.g.\
$\lambda_{hs}=\lambda_{hs}(\LamD=\MZ)$.

\subsection{Four dimensions: On-shell scheme}
\label{sec:OS}

The on-shell scheme~\cite{Anderson:1991zb,Curtin:2014jma}
computes
the one-loop vacuum correction
of the effective potential~\eqref{eq:CW_potential} by installing
a cutoff regularization for the radial integration with a UV cutoff, $\Lambda$.
The potential
then directly depends on $\Lambda$ via~\cite{Quiros:1999jp}
\begin{align}
\label{eq:cutoff}
  J_{\rmii{CW}}(m^2,\Lambda) &= \frac{1}{2} \int_{P}^{\Lambda} \ln \big(P^2 + m^2 \bigr)
    \,,
\end{align}
where
$\int_P =
\int \frac{{\rm d}^4 P}{(2\pi)^4}$ and
UV divergences are regulated via counterterms.
The tree-level VEV and
mass eigenstates from eq.~\eqref{eq:ren:cond:tree} are
preserved at one-loop level
by imposing for scalar masses to not change with respect to their tree-level values.
This is achieved by relating
$\Lambda^2 \to m_{0i}^2$,
where $m_{0i}$ is the mass of particle $i$ at the EW vacuum.
As typically done in this simple approximation,
we ignore the running of all coupling constants.

In both the on-shell and \MSbar~scheme (cf.\ sec.~\ref{sec:MSbar}),
we implement IR resummation by
the truncated full dressing approach~\cite{Parwani:1991gq,Curtin:2016urg}.
This procedure replaces the vacuum by
one-loop thermally corrected masses
in the tree-level field-dependent masses as
$m_i^2 \to m_{i,3}^2 = m_i^2\bigr|_{\mu_{i}\to \mu_{i,3}}$.

\subsection{Dimensional reduction and 3D~EFT expansion}
\label{sec:3DEFT}

The finite-temperature scale hierarchy
separates
{\em hard},
{\em soft},
and
{\em ultrasoft} scales,
{\em viz.}
\begin{align}
\label{eq:hierarchy}
  \pi T
  \gg
  g^{ } T
  \gg
  g^2 T
  \,,
\end{align}
and
the dynamics of the PTs is driven by IR effects.
By recasting theories in
the EFT picture of dimensional reduction~\cite{Ginsparg:1980ef,Appelquist:1981vg,Kajantie:1995dw},
it becomes evident that for modes that are deeper in the IR,
the perturbative series converges increasingly slowly~\cite{Linde:1980ts}.
This is the case for transitioning Lorentz scalars, such as
the Higgs or the light singlet scalar of the xSM.
The validity of perturbation theory can then be extended
by integrating out heavy degrees of freedom which
eliminates hierarchy-induced large logarithms via RG equations.
The final theory is the 3D bosonic IR sector of the parent theory.

This theory, the {\em softer} 3D~EFT,
is structurally identical to the conventional ultrasoft EFT~\cite{Kajantie:1995dw}
but is valid
for transitions between the soft and ultrasoft scale,
$gT \gg |p| \gg g^2 T$,
where $\mD \gg m_{\rmii{$A$}} \sim m_\phi$.
Here,
$\mA$ are the spatial gauge boson and
$\mD$ the Debye masses.
After dimensional reduction,
we utilize
the effective potential up to two-loop order with
NLO matching of the 3D~EFT.
Here,
LO (NLO) matching refers to the matching of
one-loop (two-loop) thermal masses and
tree-level (one-loop) couplings.
This EFT construction is
obtained
by in-house {\tt FORM}~\cite{Ruijl:2017dtg} software,
by {\tt DRalgo}~\cite{Ekstedt:2022bff,Fonseca:2020vke}, and
by previous calculations of the EFT~\cite{Schicho:2021gca,Niemi:2021qvp}.%
\footnote{%
  The updated~\cite{Niemi:2021qvp},
  installs the correct scaling
  $x_3 \sim x_3' \sim \sqrt{T}$.
}
All barred and subscripted quantities are understood in the 3D~EFT
(e.g., $\mu_h \to \bar{\mu}_{h,3}$)
with
$\bar{\bm{\phi}} \sim \bm{\phi} T^{-\frac{1}{2}}$.

Up to two-loop order, the 3D effective potential is
\begin{align}
\label{eq:Veff:3d}
V_{\rmi{eff},{3}} &=
  V_{0,3} + V_{1,3} + V_{2,3}
  \,,
  \\
V_{1,3}^{ } &=
  \sum_i n_{i}^{ }\, J_3^{ }\bigl(\bar{m}_i^{2}(\bm{\phi},T)\bigr)
  \,,
\end{align}
where
$d = 3 - 2\epsilon$,
$V_{0,3}(\bm{\phi},T)$ is the three-dimensional version of
the tree-level potential~\eqref{eq:Veff-01}~\cite{Niemi:2021qvp}, and
the degrees of freedom, $n_i$, are $d$-dependent.
The corresponding mass eigenvalues $\bar{m}_i$ of
the dynamical fields $i\in \{W,Z,G,h,s\}$ in the 3D~EFT
take the same structure as for
eqs.~\eqref{eq:mass:h}--\eqref{eq:mass:Z}.
At one-loop level,
in $V_{1,3}$,
the integrals are
UV-finite and three-dimensional
\begin{align}
  J_3(m^2) = \frac{1}{2} \int_{\vec{p}}  \ln (p^2 + m^2)
  \stackrel{d=3-2\epsilon}{=}
  -\frac{1}{12\pi}[m^2]^\frac{3}{2}
\,.
\end{align}
The two-loop contributions, $V_{2,3}$, to the potential~\eqref{eq:Veff:3d},
as well as the two-loop 3D~EFT matching relations
are directly adopted
from~\cite{Niemi:2021qvp} and
can also be taken from {\tt DRalgo}~\cite{Ekstedt:2022bff}.
The parameters of
this final 3D~EFT
are evolved to the 3D RG scale,
$\Lamd$,
which is set to
$\Lamd = T$ in our analysis below.

Due to the $Z_2$-symmetry of the potential~\eqref{eq:V0:tree} in the xSM,
a tree-level barrier appears to be absent even at the softer scale
in general.
If this potential gives rise to a first-order transition,
then the perturbative expansion is expected to
converge slower
since the expansion around
the minimum receives radiative corrections already at LO.
As a result, the effective potential exhibits known pathologies such as
imaginary parts~\cite{Weinberg:1987vp},
IR divergences related to Goldstone modes~\cite{Laine:1994zq,Croon:2020cgk,Gould:2023ovu},
as well as gauge dependence~\cite{Wainwright:2012zn}.%
\footnote{
  For the potential,
  we verified that gauge dependence is a minor effect compared
  to incomplete resummation.
}
In perturbation theory,
such pathologies can be consistently treated in
a strict EFT expansion~\cite{Fukuda:1975di,Ekstedt:2022zro,Gould:2023ovu}.

In a practical approach to the problem,
we focus on a direct minimization at the softer scale~\cite{Croon:2020cgk},%
\footnote{
  Taking the absolute magnitude of the squared masses or
  omitting the imaginary part of the potential differs numerically.
  In practice,
  we discard the imaginary part
  if numerically $\im\Veff < \re\Veff$ at the minima~\cite{Delaunay:2007wb}.
}
and on two-step transitions
of the form eq.~\eqref{eq:PTsteps};
see~\cite{Niemi:2020hto,Niemi:2018asa} for non-perturbative analyses.
For light singlet scalar masses
and since perturbatively the barrier vanishes in singlet-direction, 
we assume for {\tt step1}~in eq.~\eqref{eq:PTsteps} to
be of second order~\cite{Niemi:2020hto}.
Then the second transition step,
{\tt step2} in eq.~\eqref{eq:PTsteps}, again features a tree-level barrier
through a spontaneously broken $Z_2$-symmetry~\cite{Gould:2021oba,Niemi:2021qvp}
where $\langle S \rangle \neq 0$.
Consequently,
the transition is rendered strong with
the advantage of avoiding difficulties related to
radiatively induced transitions at the softer scale~\cite{Gould:2021oba}.

Since the 3D~EFT is constructed in the high-temperature expansion $m/T \ll 1$,
we inspect only temperature regions where the high-temperature assumption holds
and effects of the scalar masses are small compared to the temperature.
Especially,
by retaining explicit terms of
$J_3(m_{X_0}^{2})$ in a soft-scale EFT
for (adjoint) temporal scalars $X_0 \in \{A_0,B_0,C_0\}$,
we confirm that $\mD^2 \gg h_3 v^2$ effects are small for
our analysis at the softer-scale EFT.
Here, $h_3$ is
the coupling between Lorentz and
temporal scalars
$X_0$~\cite{Kajantie:1995dw,Gould:2023ovu,Kierkla:2023von}.
Given the maximal ratio
$\bm{\phi}/T \sim \mathcal{O}(1)$ we encountered,
and after both expanding and explicitly keeping $J_3$-functions,
we identify soft effects to modify the predicted parameter space
of sec.~\ref{sec:GW} by $\mathcal{O}(0.1\%)$.
This analysis is reported in a second xSM scan
in appendix~\ref{sec:soft:softer}.

For the 3D~EFT,
we relate \MSbar- with physical parameters
using
the one-loop improved vacuum renormalization of appendix~\ref{sec:vacren:3d}
and~\cite{Niemi:2021qvp}.
This procedure has the advantage, that
the minimization condition~\eqref{eq:ren:cond:tree} is only needed at tree-level and
momentum corrections are consistently included.
Corrections to the tree-level relations
are then contained in
vacuum pole mass renormalization at higher orders.

\section{First-Order Phase Transitions}
\label{sec:fopt}

Cosmological first-order PTs occur via
nucleating true-vacuum bubbles which expand in a space filled with false vacuum.
At finite temperature,
the probability per unit time and volume of
jumping from a metastable to a state of lower energy,
has the semi-classical
approximation~\cite{Langer:1969bc,Coleman:1977py,Linde:1980tt,Linde:1981zj,Affleck:1980ac}
\begin{equation}
\label{eq:tunneling_rate}
  \Gamma(T) =
    A_\rmi{dyn}
    \times
    A_\rmi{stat}
    \,
    e^{-S_3/T}
    \,,
\end{equation}
where the prefactor factorizes into
$A_\rmi{dyn}$,
a dynamical and
$A_\rmi{stat}$,
a statistical
contribution at all orders~\cite{Ekstedt:2022tqk}.
Focusing only on equilibrium thermodynamic contributions,
we use the standard approximation~\cite{Linde:1980tt,Linde:1981zj}
\begin{align}
\label{eq:prefac:approx}
  A_\rmi{stat} &\approx T^3 \Bigl(\frac{S_3}{2\pi T}\Bigr)^{\frac{3}{2}}
  \,,&
  A_\rmi{dyn} &\approx T
  \,,
\end{align}
which suppresses effects of
inhomogeneous contributions to
the functional determinant of the fluctuation operator
of $A_\rmi{stat}$.
Such effects are
formally of higher order
but can dominate
the exponent~\cite{Croon:2020cgk,Ekstedt:2021kyx,Gould:2021ccf} in extreme cases.
We justify the choice in eq.~\eqref{eq:prefac:approx}
since the absence of both
higher-order effects and
a self-consistent bounce solution~\cite{Ai:2023yce}
are systematic errors to
all schemes of sec.~\ref{sec:resummation}.
The Euclidean action, $S_3$, of the critical bubble or bounce
for $O(3)$-symmetric thermal systems,
is then computed at
$\mathcal{O}(\hbar^1)$ for the one-loop and
$\mathcal{O}(\hbar^2)$ for the two-loop softer EFT,
bearing in mind that this systematically discards
real-time physics encoded in $A_\rmi{dyn}$ and
field-dependent gradient terms in the effective action;
see
\cite{Ekstedt:2023sqc}
for out-of-equilibrium effects
and~\cite{Gould:2021ccf,Lofgren:2021ogg,Hirvonen:2021zej}
for approaches using nucleation EFT.

From the rate~\eqref{eq:tunneling_rate}, we compute
the temperature scales of the transition.
At the nucleation temperature, $\Tn$,
the tunneling rate per horizon volume becomes relevant%
\footnote{
  To convert temporal to temperature integration
  $a(T)= 1/T$,
  ${\rm d}t = -a(T)/H(T){\rm d}T$, and
  $a(t)$ is the FLRW metric scale factor.
}
\begin{equation}
    \label{cond1}
    \int_{t_{\rm c}}^{t_{\rm n}}\!{\rm d}t \frac{\Gamma(t)}{H(t)^3}=
    \int_{\Tn}^{\Tc}\!{\rm d}T \frac{\Gamma(T)}{H(t)^4 T} = 1
    \,.
\end{equation}
The integration is bounded by the critical temperature, $\Tc$,
where the two-phase minima become degenerate.
Conversely,
$t_\rmi{c}$ and
$t_\rmi{n}$ are the times related to
$\Tc$ and $\Tn$, respectively.
The Hubble rate is defined as
\begin{align}
  H^2(T) &=\frac{\rho_\rmi{tot}}{3\Mpl^2}
  \,,&
  \rho_\rmi{tot} &=
      \rho_r
    + \Delta \Veff(T)
  \, ,
\end{align}
with
$\Mpl = 2.4 \times 10^{18}$~GeV the reduced Planck mass,
$\rho_r = \frac{\pi^2}{30} g_{*}(T) T^4$
the radiation energy density of relativistic species, and
$g_*(T)$ the number of radiative degrees of freedom~\cite{Saikawa:2018rcs}.
Differences between
the meta-stable $(+)$ and stable $(-)$ phase are henceforth denoted as
$\Delta X = X^{(+)} - X^{(-)}$
for e.g.\
$X = \Veff$.

To estimate the time of bubble collisions,
we employ the percolation temperature $\Tp$.
It is defined by requiring
the probability, that a point in space remains in
the false vacuum, to be
$P(\Tp)=e^{-I(\Tp)} \simeq 71 \%$~\cite{Ellis:2018mja}:
\begin{equation}
  I(\Tp) = \frac{4 \pi}{3} \int_{\Tp}^{\Tc}\!{\rm d}T'
    \frac{\Gamma(T')}{H(T')}
    \frac{r(T,T')^3}{T'^4} = 0.34
     \,.
\end{equation}
The comoving radius of a bubble nucleated at time $t'$ and
propagated until $t$ with velocity $\vw$
is
$r(t,t') = \int_{t'}^{t}\!  {\rm d}\tilde{t} \frac{\vw(\tilde{t})}{a(\tilde{t})}$,
and
we require a shrinking volume in the false vacuum, {\em viz.}
$\frac{{\rm d}I(T)}{{\rm d}\ln T} <-3$.
After the PT, the false vacuum energy is re-transferred to the thermal bath
that is reheated to
$T_* = \Tp[1+\alpha(\Tp)]^{1/4}$~\cite{Ellis:2018mja}.
The strength of a cosmological PT~\cite{Hindmarsh:2017gnf,Caprini:2019egz}
and its inverse time duration are approximated as
\begin{align}
\label{eq:alpha:beta}
  \alpha &\equiv \frac{1}{\rho_r}\Bigl(
    \Delta \Veff
  - \frac{1}{4}\frac{{\rm d}\Delta \Veff}{{\rm d}\ln T} \Bigr)
    \Big{|}_{T=\Tp}
  \,,\\
\label{eq:betaH}
    \frac{\beta}{H} &\equiv \frac{{\rm d}}{{\rm d}\ln T}
    \Bigl(\frac{S_3}{T} \Bigr)
    \Big{|}_{T=\Tp}
    \,.
\end{align}
The trace of the energy-momentum tensor for $\alpha$
is taken in the relativistic plasma limit and
in practice receives further corrections if
the speed of sound differs from $\cs^2=1/3$~\cite{Giese:2020rtr,Tenkanen:2022tly}.
When using the 3D approach,
in both eqs.~\eqref{eq:alpha:beta} and \eqref{eq:betaH}
one can naturally use the 3D potential,
{\em viz.}\
$V_{\rmi{eff},4} \simeq T\, V_{\rmi{eff},3}$.

The thermodynamic parameters required to compute the GW spectra are
$T_*$, $\alpha(\Tp)$, $\beta(\Tp)$,
the speed of sound~$\cs$ and
the bubble wall velocity~$\vw$.
The latter is particularly challenging to compute as it requires
an out-of-equilibrium computation~\cite{%
  Konstandin:2014zta,Kozaczuk:2015owa,Friedlander:2020tnq,Moore:1995ua,Moore:1995si,
  DeCurtis:2022hlx,Dorsch:2021nje,Bodeker:2017cim,Liu:1992tn,
  Cline:2021iff,Dine:1992wr,Laurent:2020gpg,Laurent:2022jrs,Ai:2023see,Krajewski:2024gma}.
It has been shown~\cite{Cline:2021iff,Lewicki:2021pgr}
that whenever the properties of the wall can be extracted with
the semi-classical approximation~\cite{Moore:1995ua,Moore:1995si},
the associated GW signals are too weak to be observed by the future experiments.
For very strong transitions with $\alpha\approx 0.1$,
the semi-classical approximation fails and
the bubbles are expected to run away with ultra-relativistic velocities.
Since we are interested in strong transitions that
allow for a reasonable reconstruction,
we henceforth assume $\vw = 1$;
see~\cite{Laurent:2022jrs} for a more careful evaluation of $\vw$ in the xSM.

To obtain the thermodynamic parameters, we use a modified version of
{\tt CosmoTransitions}~\cite{Wainwright:2011kj}
that takes as an input the effective potential $\Veff(\bm{\phi},T)$
for the different schemes.%
\footnote{
  Our
  {\tt python} software
  \href{https://github.com/DMGW-Goethe/DRansitions}{{\tt DRansitions}}~\cite{DRansitions}
  implements
  a generic potential in the softer 3D~EFT
  of the xSM
  using {\tt DRalgo}~\cite{Ekstedt:2022bff,Fonseca:2020vke}.
}

\section{Impact on the GW signals}
\label{sec:GW}

By focussing on two-step transitions~\eqref{eq:PTsteps},
from symmetric to singlet to vacuum Higgs minimum,
we inspect GWs sourced by sound waves in the plasma~\cite{Caprini:2015zlo, Caprini:2019egz}.
Transitions in the xSM are never significantly supercooled~\cite{Ellis:2018mja} and
we can refrain from including
GWs sourced by relativistic fluid motion or bubble collisions~\cite{Ellis:2019oqb, Lewicki:2019gmv, Lewicki:2020jiv, Lewicki:2020azd, Lewicki:2022pdb}.
For them to be relevant much stronger transitions are required.
We neglect possible contributions from turbulence~\cite{Caprini:2019egz} since
despite significant progress to understand that source~\cite{RoperPol:2019wvy, Kahniashvili:2020jgm, RoperPol:2021xnd},
its overall amplitude remains uncertain.
The spectrum produced by sound waves in the plasma also evolved in recent years~\cite{Hindmarsh:2016lnk, Hindmarsh:2019phv, Jinno:2020eqg, Gowling:2021gcy, RoperPol:2023dzg, Sharma:2023mao},
predicting a modified spectral shape depending on the wall velocity.
In the xSM, sound wave modifications are less important since
mostly transitions with $\vw\approx 1$ predict observable signals~\cite{Cline:2021iff, Lewicki:2021pgr, Ellis:2022lft}.
Thus, we can use the results of lattice simulations for
the shape of the signal~\cite{Hindmarsh:2013xza, Hindmarsh:2015qta, Hindmarsh:2017gnf}.

The sound wave spectrum,
its shape, and
peak frequency can be expressed as
\begin{align}
\Omega_\rmii{GW}(f)h^2 &= 4.13\times 10^{-7} \, \left(R_* H_*\right)
    \biggl(1- \frac{1}{\sqrt{1+2\tau_\rmi{sw}H_*}} \biggr)
    \nn &\hphantom{{}= 4.13}
    \times
    \Bigl(\frac{\kappa_\rmi{sw} \,\alpha }{1+\alpha }\Bigr)^2
    \biggl(\frac{100}{g_*}\biggr)^\frac{1}{3}
    S_\rmi{sw}(f)
  \,,\\[2mm]
S_\rmi{sw}(f) &=
    \biggl(\frac{f}{\fp}\biggr)^3
    \biggl[\frac{4}{7}+\frac{3}{7} \Bigl(\frac{f}{\fp}\Bigr)^2\biggr]^{-\frac{7}{2}}
  \,,\\[2mm]
\fp =
f_\rmi{sw} &=
    2.6\times 10^{-5}~{\rm Hz}\,
    \bigl(R_* H_*\bigr)^{-1}
    \nn &\hphantom{{}=2.6}
    \times
    \Bigl(\frac{T_*}{100~{\rm GeV}}\Bigr)
    \Bigl(\frac{g_*}{100}\Bigr)^\frac{1}{6}
  \,,
\end{align}
with efficiency factor $\kappa_\rmi{sw}$.
The average bubble radius normalized to the horizon size reads~\cite{Caprini:2019egz}
\begin{equation}
H_*R_* \approx (8\pi)^\frac13 \, \max\{\vw,\cs\}
  \Bigl(\frac{\beta}{H}\Bigr)^{-1}
  \,,
\end{equation}
where for strong transitions
$\vw > \cs$.
The sound wave period normalized to the Hubble rate
is approximated as~\cite{Hindmarsh:2017gnf,Ellis:2018mja,Ellis:2019oqb,Ellis:2020awk,Guo:2020grp}
\begin{align}
  \tau_\rmi{sw}H_* &=\frac{H_* R_*}{\bar{U}_f}
  \,,&
  \bar{U}_f &\approx \sqrt{\frac34 \frac{\alpha}{1+\alpha} \kappa_\rmi{sw}}
  \,,
\end{align}
with the
root-mean-square of the fluid velocity $\bar{U}_f$.

\begin{figure}[t]
\centering
\includegraphics[width=0.49\textwidth]{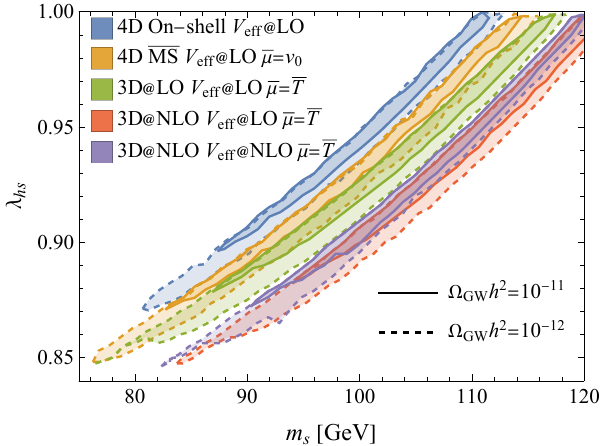}
\caption{%
  Regions in the xSM parameter space predicting GW abundances above
  $\Omega_\rmii{GW}h^2=10^{-11}$ (solid) and
  $\Omega_\rmii{GW}h^2=10^{-12}$ (dashed)
  for two-step transitions where only the second step sources
  a strong first-order PT.
  Results are displayed as functions of the physical broken phase mass $m_s$ and
  portal coupling $\lambda_{hs}$
  at fixed scalar self coupling $\lambda_s =1$.
  The input scale $\LamD = \MZ$.
  The effective potential was computed in the
  on-shell~(blue), and
  \MSbar{}~(orange) schemes at one-loop, and
  the 3D~EFT at
  LO using one-loop potential (green) and
  NLO using both one- (red) and two-loop (purple) effective potentials
  with 4D RG-scales
  $\LamD = v_0$ for \MSbar{}- and
  $\LamD =\! \bT = 4\pi e^{-\gammaE} T$ for 3D approaches.
  In regions above each respective area,
  the EW minimum and the initial one become degenerate and percolation fails.
  For even larger $\lambda_{hs}$,
  the EW minimum is not global at zero temperature.
  Such models are excluded.
}
\label{fig:abundances}
\end{figure}
To predict the experimental accessibility of the xSM,
we scan a large section of its parameter space
by varying%
\footnote{
  In the simplest $Z_2$-symmetric incarnation of the xSM,
  direct detection experiments exclude most of the parameter space due to
  an overabundance of the scalar dark matter (DM)~\cite{Beniwal:2017eik} while
  the transition parameters might be modified due to the presence of domain walls~\cite{Blasi:2022woz,Blasi:2023rqi,Agrawal:2023cgp,Li:2023yzq}.
  To evade these issues, we assume minimal modifications:
  new particles in the dark sector destabilize
  the scalar DM~\cite{Beniwal:2018hyi,Bian:2018mkl} and
  a small $Z_2$-breaking term erases
  the domain walls~\cite{Azatov:2022tii}.
  Neither of these would impact our main results.
}
$\lambda_{hs}\in [0.1,1.2]$,
$\frac{m_h}{2} < m_s < 130$~GeV, and
$\lambda_s\in \{0.1, 1.0\}$.
We restrict
the physical masses $m_h$ and $m_s$ from eq.~\eqref{eq:mass:s}
for both scalars to be dynamical in the IR.
Figure~\ref{fig:abundances} shows the predicted GW abundance as
a function of the scalar mass and Higgs-scalar coupling for fixed scalar self-coupling $\lambda_s=1$.%
\footnote{
  Varying $\lambda_s$ shifts the observable parameter space,
  since for different $\lambda_s$,
  the hyperplane in the $(m_s,\lambda_{hs},\lambda_{s,\rmii{fixed}})$-space changes
  (cf.~appendix~\ref{sec:lambdas}).
  The main results remain unchanged.
}
The different regions are obtained by computing
the xSM effective potential in
the \MSbar~scheme,
the on-shell~scheme and
the 3D~EFT.

\begin{table}[t]
\centering
\begin{tabular}{|c||c|c|c}
    \hline
    EFT matching
    & 1-loop $\Veff$
    & 2-loop $\Veff$
    \\ \hline\hline
    \hphantom{N}LO
    & 3D@LO \hfill $\Veff$@LO
    & --
    \\ \hline
    NLO
    & 3D@NLO \hfill $\Veff$@LO
    & 3D@NLO $\Veff$@NLO
    \\ \hline
\end{tabular}
\caption{%
    Different levels of diligence in the 3D~EFT approach
    for different
    orders of EFT matching using high-temperature dimensional reduction and
    loop orders of the effective potential $\Veff$.
    The \MSbar-approach is identical to (3D@LO $\Veff$@LO)
    at high temperatures.
    }
\label{tab:3dEFT}
\end{table}
For the 3D~EFT approach,
the EFT is constructed
via dimensional reduction at
LO (3D@LO) and
NLO (3D@NLO) while
the 3D effective potential is computed at
one- ($\Veff$@LO) and
two-loop ($\Veff$@NLO) levels;
see tab~\ref{tab:3dEFT} for nomenclature.
The minimal approach using
(3D@NLO~$\Veff$@LO) was proposed in~\cite{Schicho:2022wty}.
The RG-scale is fixed to
$\LamD = v_0$ for the \MSbar~scheme and
$\LamD = \bT$ for the 3D~EFT
where $\bT =4\pi e^{-\gammaE} T$ and
$\gammaE$ is the Euler-Mascheroni constant.
The 3D EFT RG scale is set to
$\Lamd = T$.

The (3D@NLO) methods are the most stable and their predictions change little when
using NLO corrections to the effective potential. Since the \MSbar~scheme and
the (3D@LO) results are identical in the high-temperature limit,
their results are naturally closest with the difference rooted in
using
the conventional scale
$\LamD=v_0$ in the \MSbar~scheme instead of
a scale more natural to a thermal transition
$\LamD=\bar{T}$.
Finally,
the relatively large differences in
signal amplitudes,
$\Omega_\rmii{GW}h^2$,
between the various methods
convert
to relatively small differences of
the corresponding model parameters.
Predictions of $(m_s,\lambda_{hs})$
amount to uncertainties of at most $10\%$
in extreme cases.
However, the parameter space predicting
a strong two-step transition is often narrow and
shifts of a few percent can become qualitatively important.
They can even
change the nature of the transition entirely.

To inspect the impact of
resummation methods of the potential on reconstructing sources behind
future observed signals with LISA~\cite{Smith:2019wny,Pieroni:2020rob,Flauger:2020qyi,Gowling:2021gcy},
we perform
Fisher matrix analyses~\cite{Fisher:1922saa}.
The matrix elements for the two spectral parameters,
the peak amplitude and frequency,
$\theta_{i} \in \{\Omega_{\rm sw}, f_{\rm sw}\}$,
read
\begin{equation}
\Gamma_{ij} = \mathcal{T} \int
    \frac{{\rm d} f}{\Omega_{\rm tot}(f)^2}
    \frac{\partial \Omega_{\rm tot}(f)}{\partial\theta_i}
    \frac{\partial \Omega_{\rm tot}(f)}{\partial\theta_j}
  \,,
\end{equation}
where the mission operation time $\mathcal{T} = 4\,$yr
and
the variance of parameters $\sigma_{i}^{2}=\Gamma^{-1}_{ii}$.
By including only
the instrumental noise (see e.g.~\cite{Smith:2019wny})
\begin{equation}
\label{eq:GWspectrum}
  \Omega_\rmi{tot}(f) =
      \Omega_\rmii{GW}(f)
    + \Omega_\rmi{instr}(f)
    \,,
\end{equation}
we neglect
astrophysical noise sources of
binary white dwarfs~\cite{Cornish:2017vip,Robson:2018ifk} and
the black hole population currently probed by LIGO-Virgo-KAGRA~\cite{Lewicki:2021kmu}.
Including the latter would effectively reduce the sensitivity of the experiment such that
a stronger spectrum would be necessary to reproduce benchmarks with the same relative uncertainty on
the parameters of the spectrum. Our results would otherwise remain unchanged.
The Fisher matrix approach
we follow, is a simplification and a more accurate reconstruction would require
a Markov Chain Monte Carlo fit.
However, the methods will agree provided the errors on the reconstruction are smaller
than about $10\%$ which is exactly the accuracy we find in our benchmark.
For details on the state-of-the-art reconstruction of phase transition parameters with LISA see~\cite{Caprini:2024hue}, which also provides a comparison of
the two methods and a discussion of the exact impact of the inclusion of foregrounds.%
\footnote{
  The reconstruction also proves robust under
  a more general noise model for LISA~\cite{Hartwig:2023pft}.
}

\begin{figure}[t]
\centering
\includegraphics[width=0.49\textwidth]{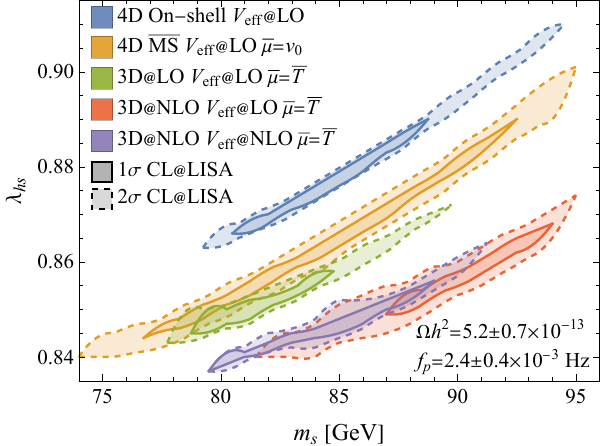}
\caption{%
  Parameter reconstruction with LISA of
  the benchmark point
  spectrum defined by
  abundance
  $\Omega_\rmii{GW}h^2=5.2 \times 10^{-13}$
  and peak frequency
  $\fp=2.4 \times 10^{-3}$~Hz.
  Using Fisher analysis, we
  estimate the error associated with the reconstruction and
  indicate uncertainties with central values at
  $1\sigma$ ($68\%$, solid) and
  $2\sigma$ ($95\%$, dashed) confidence levels (CL).
  As in fig.~\ref{fig:abundances},
  the colored regions indicate the method of resummation.
  With ${\rm SNR}=10$,
  the chosen benchmark corresponds to a barely detectable signal.
 }
\label{fig:fits}
\end{figure}
As a representative benchmark point, we chose a spectrum with
the abundance
$\Omega_\rmii{GW}h^2 = 5.2\pm 0.7 \times 10^{-13}$ and
peak frequency
$\fp=2.4\pm 0.4 \times 10^{-3}$~Hz.
The signal-to-noise ratio observed by LISA would be ${\rm SNR}=10$ and
render it one of the weakest signals where one can claim a detection.
Figure~\ref{fig:fits} shows the reconstructed parameter space
based on the various methods used for computing the transition parameters.
While
(3D@NLO) predictions converge at $2\sigma$ confidence level (CL),
the differences between
4D and
(3D@NLO) predictions do not overlap at $2\sigma$~CL.
Such a discrepancy indicates that
theoretical uncertainties in the computation of the potential are at least of
the same order as those stemming from the reconstruction of the GW spectrum with LISA.
For all stronger and more easily observable spectra, the error from the experimental reconstruction would be smaller indicating that theoretical errors in thermodynamic computations of the potential would be the main source of uncertainty in all observable spectra.
Hence, state-of-the-art methods
(cf.\ e.g.~\cite{Croon:2020cgk,Ekstedt:2022bff,Gould:2023ovu,Ekstedt:2023sqc,Ekstedt:2024etx})
need to be used to improve our determination of
the parameter space of the underlying model any further
in the future.

\begin{figure*}[t]
\centering
\includegraphics[width=0.45\textwidth]{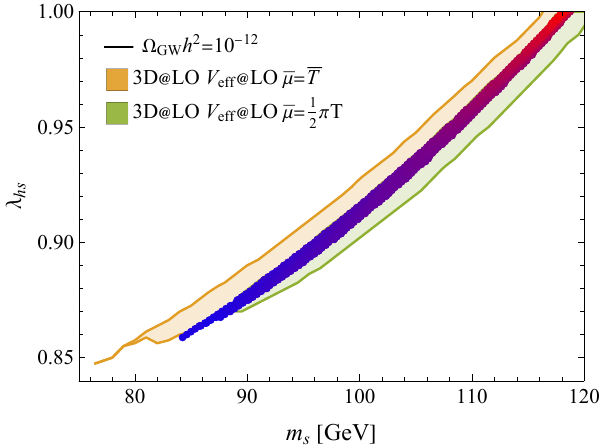}%
\includegraphics[width=0.45\textwidth]{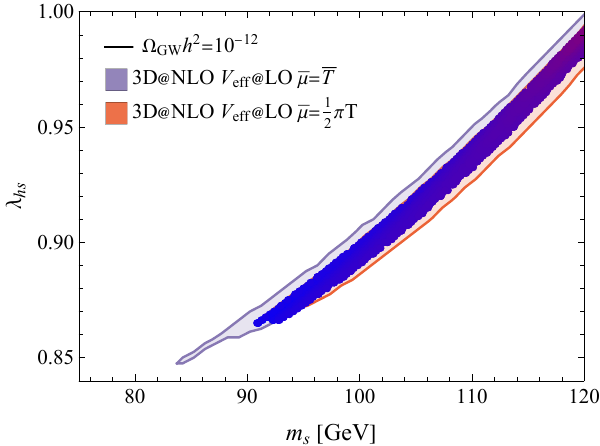}%
\includegraphics[width=0.10\textwidth]{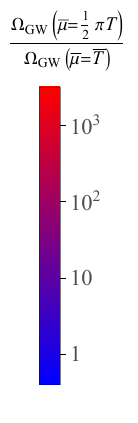}
\caption{%
  Regions in the xSM parameter space predicting GW abundances
  $\Omega_\rmii{GW}>10^{-12}$ computed at
  LO (left) and
  NLO (right)
  in dimensional reduction.
  To quantify the scale dependence of each method in both panels,
  we show results at the 4D RG-scales
  $\LamD = \{\frac{1}{2}\pi T, \bT\}$.
  The ratio~\eqref{eq:GW:ratio} of the resulting GW amplitudes at these scales is
  indicated by the heatmap with ratios up to
  $\Delta\Omega_\rmii{GW} \sim \mathcal{O}(10^3)$.
  }
\label{fig:RGscale:comp}
\end{figure*}

Another factor in carefully estimating theoretical uncertainty is
renormalization scale dependence.
Since perturbative approximations of the effective potential generally
depend on the employed 4D RG-scale,
its variation can impact the predicted parameter space of
GW signals detectable by LISA~\cite{Croon:2020cgk,Gould:2021oba}.
Such RG-scale variation
serves a proxy for quantifying the importance of absent higher-order corrections.
At finite temperature,
the potential is scale-dependent at
$\mathcal{O}(g^4 T^2)$ through the implicit running of
the thermal parameters
eqs.~\eqref{eq:mass:higgs}--\eqref{eq:mass:mDsu2}~\cite{Gould:2021oba}.

Both 4D methods,
the on-shell~scheme and
the \MSbar~scheme at $\LamD \sim v_0$,
effectively fix the scale and
therefore lack an estimate of their theoretical uncertainty.
Their nearly scale-invariant reconstructed parameter spaces
should not be mistaken for theoretical robustness.
First, the on-shell~scheme does not exhibit an explicit RG-scale dependence
since divergent integrals are regulated by a UV cutoff.
Since the cutoff is fixed at the pole masses of the respective particles,
phenomenological predictions are manifestly $\LamD$-independent.
This does not indicate the inclusion of higher-order corrections that
are theoretically relevant.
At the same time dimensional regularization has the advantage of
being manifestly Lorentz invariant which is absent in cutoff regularization.
Second,
the \MSbar~scheme at $\LamD \sim v_0$,
uses a non-thermal RG scale which renders it almost insensitive
to RG-scale variation.
We explicitly verified, that
the remaining $\LamD$-variation via
$\LamD \in \{1/2,1,2\}\times v_0$
results in a $\mathcal{O}(0.1\%)$ shift of the parameter space.

To estimate the relevance of higher-order contributions,
however,
we focus on the 3D EFT since it also contains the \MSbar~scheme
via (3D@LO $\Veff$@LO), and
impose two
different values of the 4D RG-scale
\begin{align}
\label{eq:mu:3d}
  \text{3D~EFT:}
  &&
  \LamD &= \{1/2,4e^{-\gammaE}\} \times \pi T
  \,,
\end{align}
where $\bT=4\pi e^{-\gammaE} T$
is the weighted sum of a nonzero bosonic Matsubara frequency
that arise within logarithms at one-loop order.
By varying also the 3D~RG scale, $\Lamd$,
we observed a minuscule effect on the thermodynamic parameters
well dominated by the 4D RG scale.
See also~\cite{Gould:2021oba} for a similar discussion.

Using the 3D~EFT approach, we restrict ourselves to
a comparison between
(3D@LO $\Veff$@LO) and
(3D@NLO $\Veff$@LO), i.e.\
the one-loop effective potential
with LO~(NLO) dimensional reduction.
The reconstructed parameter space predicting the GW abundance
$\Omega_\rmii{GW}>10^{-12}$ is shown in
fig.~\ref{fig:RGscale:comp}.
For the parts of the parameter space where these regions coincide,
we show the ratio between the predicted GW amplitudes
\begin{align}
\label{eq:GW:ratio}
\Delta\Omega_\rmii{GW} =
  \frac{
    \Omega_\rmii{GW}(\LamD = \frac{1}{2}\pi T)}{
    \Omega_\rmii{GW}(\LamD= \bT)
  }
\,.
\end{align}
The observable parameter space for the
(3D@LO $\Veff$@LO) potential in the 3D~EFT,
is shifted by $\mathcal{O}(1\%)$
depending on the singlet mass $m_s$
and
exhibits deviations up to
$\Delta\Omega_\rmii{GW} \sim \mathcal{O}(10^3)$
as previously predicted~\cite{Croon:2020cgk, Gould:2021oba}.
This is identical for the \MSbar~scheme at high temperatures.
The observed scale-dependence also increases logarithmically
with $m_s$ through the implicit running of
the thermal parameters.
Predictions from the
(3D@NLO $\Veff$@LO) potential,
remain mostly insensitive under the RG-scale variation.
They are more robust and
the $\LamD$-variation of eq.~\eqref{eq:mu:3d} changes the amplitude by at most
$\Delta\Omega_\rmii{GW} \lesssim \mathcal{O}(10^2)$.
As expected,
utilizing two-loop thermal masses,
shifts the $\LamD$-dependence
to higher orders~\cite{Gould:2021oba}
due to parametric scale cancellation.
This is true already at the one-loop level for $\Veff$~\cite{Schicho:2022wty}.

Notably, large variations of the predicted signal correspond
to shifts at the $\mathcal{O}(1\%)$ level in
the reconstructed values of the parameters.
We summarize our findings as follows:
\begin{itemize}
  \item[(i)]
    in the \MSbar~scheme
    (3D@LO~$\Veff$@LO),
    the parameter space is displaced by $\mathcal{O}(1\%)$
    when varying the RG-scale $\LamD$,
  \item[(ii)]
    the on-shell~scheme does not involve running of couplings such that parameter spaces are fixed and
    implicitly affected by missing higher-order corrections,
  \item[(iii)]
    in the 3D~EFT,
    (3D@NLO~$\Veff$@LO),
    the parameter space is displaced by $\mathcal{O}(0.1\%)$ when varying
    four-dimensional RG-scale $\LamD$.
\end{itemize}

Theoretical uncertainty is also intrinsically linked with
residual gauge dependence~\cite{Croon:2020cgk}.
In all approaches of sec.~\ref{sec:resummation},
gauge dependence would only affect the effective potential since
the matching relations up to (3D@NLO) are gauge-invariant.
Studying the dependence of our results on different gauges~\cite{Martin:2018emo}
other than Landau gauge
is deferred to future work;
cf.~\cite{Athron:2022jyi} for such a study for the thermodynamics in 4D approaches.
Focusing on $\xi = 0$ is justifiable, since gauge dependence affects all stages of our computation
in a similar manner and can therefore be treated as a systematic error
of our analysis.
Additionally,
we expect that strong transitions in the xSM
are dominated by the residual RG-scale dependence~\cite{Chiang:2018gsn}.

\section{Conclusions}
\label{sec:conclusion}

This \article{} compares methods of thermal resummation between
the state-of-the-art 3D~EFT and the 4D daisy resummed potentials in cosmological PTs realized in
the simplest SM extension featuring a neutral scalar, the xSM.
We confirm that the amplitude of the predicted GW signal can change between
the methods by many orders of magnitude.
Conversely, we report that this theoretical uncertainty corresponds to a small shift of
$\mathcal{O}(1\%)$
for the model parameters needed to realize the signal.
Despite the relatively small shifts, we find that for any signal visible to LISA,
these theoretical uncertainties would exceed the experimental ones assuming a
${\rm SNR}\simeq 10$ threshold corresponding to an observation.
For stronger signals, where experimental uncertainties would be significantly reduced,
a further theoretical effort is inevitable to obtain more information on
the underlying model Lagrangian.

The perturbative expansion of state-of-the-art high-precision 3D~EFT approaches
quickly converges with the loop order.
We show that once two-loop thermal resummation (3D@NLO) is included,
the predicted parameter spaces become robust.
Higher orders in Matsubara zero-mode loops are compatible with leading orders --
thus forming the most promising route towards robust predictions.
For radiatively-induced transitions, higher orders of the 3D effective potential
become relevant again~\cite{Niemi:2021qvp}.

The differences between the perturbative approaches can be traced back to
missing higher-order corrections to the effective potential
in a strict EFT expansion~\cite{Gould:2023ovu}.
The most severe deficiency is therefore imprinted on the 4D~on-shell and \MSbar{} approaches,
which lack higher-order contributions in thermal resummation.
A similar analysis concerning residual gauge dependence is kept for future work.

However, all approaches suffer from IR sensitivities from the soft scale
if the true transition is radiatively induced at the softer scale.
Then further degrees of freedom such as spatial gauge fields need to be integrated out
to ensure a tree-level barrier at
the softer-scale EFT.
This leads to a much more pronounced uncertainty in
classically scale-invariant models where the symmetry is broken radiatively~\cite{Kierkla:2023von}.
Since we focus on transitions that are barrier-inducing at the softer scale,
the impact of thermal resummation can be analyzed without further resummation
in the 3D~EFT~\cite{Gould:2023ovu}.
Otherwise, there would be a systematic error present from soft scale resummation
in the proof-of-principle analysis of this \article{}.
While recent advancements in the bubble nucleation rate~\cite{Gould:2021ccf,Ekstedt:2022tqk} warrant
to investigate the impact of these improvements,
multi-field tunneling at the nucleation scale EFT~\cite{Ekstedt:2023sqc} and
higher-order corrections to the nucleation rate~\cite{Ekstedt:2021kyx}
remain theoretical and practical challenges.

\begin{acknowledgments}
We acknowledge enlightening discussions with
Andreas Ekstedt,
Oliver Gould,
Maciej Kierkla,
Lauri Niemi,
Bogumi\l{}a \'Swie\.zewska,
Tuomas~V.~I.\ Tenkanen,
and
Jorinde van de Vis.
This work was supported by the Polish National Science Center grant 2018/31/D/ST2/02048.
ML was also supported by the Polish National Agency for Academic Exchange within Polish Returns Programme under agreement PPN/PPO/2020/1/00013/U/00001.
MM acknowledges support
from Norwegian Financial Mechanism for years 2014--2021,
grant no.\ DEC-2019/34/H/ST2/00707 and
from the Swedish Research Council (Vetenskapsr{\aa}det) through
contract no.\ 2017-03934.
LS, PS, and DS acknowledge support by
the Deutsche Forschungsgemeinschaft (DFG, German Research Foundation) through
the CRC-TR 211 `Strong-interaction matter under extreme conditions' --
project no.\ 315477589 -- TRR 211.
PS and DS acknowledge the hospitality of the University of Warsaw during the final stages of this work.
\end{acknowledgments}

\appendix
\allowdisplaybreaks

\section*{Appendix}
\label{sec:supp}

This section discusses the technical details for
vacuum renormalization and
further determining the robustness of the thermal effective potential for
the various approaches discussed in sec.~\ref{sec:resummation}.

\section{Renormalization group evolution}
Following \cite{Gonderinger:2009jp,Brauner:2016fla,Niemi:2021qvp} after proper rescaling of our parameters and with
$t=\log{\LamD^2}$, the one-loop RG equations are given by
\begin{align}
\label{eq:beta:g1}
\partial_t^{ }
g_1^2 & = \frac{g_1^4}{(4\pi)^2} \Bigl(
    \frac{1}{6} + \frac{20}{9}\nf
    \Bigr)
    \,,\\
\partial_t^{ }
g_2^2 & = \frac{g_2^4}{(4\pi)^2} \Bigl(
    - \frac{43}{6} + \frac{4}{3}\nf
    \Bigr)
  \,, \\
\partial_t^{ }
g_3^2 & = \frac{g_3^4}{(4\pi)^2} \Bigl(
    - \frac{11\Nc}{3}
    + \frac{4\nf}{3}
    \Bigr)
  \,, \\
\partial_t^{ }
y_t^2 & = \frac{y_t^2}{(4\pi)^2} \Bigl(
    \frac{2\Nc + 3}{2} y_t^2
    - 6\CF g_3^2
    - \frac{9}{4}g_2^2
    - \frac{17}{12}g_1^2
    \Bigr)
  \,, \\
\partial_t^{ }
\lambda & = \frac{1}{(4\pi)^2} \Bigl(
      12 \lambda^2
    + \frac{\lambda_{hs}^2}{4}
    - \lambda \frac{3}{2}(
        3 g_2^2
      + g_1^2)
    \nn &\hphantom{=\frac{1}{(4\pi)^2} \Bigl(}
    + \frac{3}{16}g_1^4
    + \frac{3}{8}g_2^2 g_1^2
    + \frac{9}{16}g_2^4
    \nn &\hphantom{=\frac{1}{(4\pi)^2} \Bigl(}
    - \Nc y_t^4
    + 2\Nc y_t^2\lambda
    \Bigr)
    \,, \\
\partial_t^{ }
\lambda_{hs} & = \frac{\lambda_{hs}}{(4\pi)^2} \Bigl(
      2 \lambda_{hs}^{ }
    + 3 \lambda_{s}^{ }
    - \frac{3}{4}(
        3 g_2^2
      + g_1^2)
    \nn &\hphantom{=\frac{1}{(4\pi)^2} \Bigl(}
    + \Nc y_t^2
    + 6 \lambda
    \Bigr)
    \,,  \\
\partial_t^{ }
\lambda_s & = \frac{1}{(4\pi)^2} \Bigl(
      \lambda_{hs}^{2}
    + 9\lambda_{s}^{2}
    \Bigr)
    \,, \\
\partial_t^{ }
\mu_{h}^2 & = \frac{1}{(4\pi)^2} \Bigl(
    \mu_{h}^2\Bigl(
          6 \lambda
        + \Nc y_t^2
        - \frac{9}{4} g_2^2
        - \frac{3}{4} g_1^2
      \Bigr)
    \nn &\hphantom{=\frac{1}{(4\pi)^2} \Bigl(}
    + \frac{1}{2}\lambda_{hs} \mu_{s}^{2}
    \Bigr)
    \,, \\
\label{eq:beta:ms}
\partial_t^{ }
\mu_{s}^2 & = \frac{1}{(4\pi)^2} \Bigl(
      2 \mu_{h}^2 \lambda_{hs}
    + 3 \mu_{s}^2 \lambda_{s}^{ }
    \Bigr)
    \,, \\
\partial_t^{ }
v & = \frac{1}{(4\pi)^2}\frac{1}{2} \Bigl(
      \Nc y_t^2
    - \frac{9}{4} g_2^2
    - \frac{3}{4} g_1^2
    \Bigr)
  \,, \\[2mm]
\partial_t^{ }
x & = 0
  \,,
\end{align}
where $g_i$ for $i=1,\dots,3$ are
the
${\rm U}(1)_\rmii{Y}$,
${\rm SU}(2)_\rmii{L}$,
${\rm SU}(3)_{\rm c}$ gauge couplings,
respectively.
Here,
$y_t$ is the top Yukawa coupling,
$\mu_h^2 < 0$,
$\nf = 3$ the number of fermion families,
$\Nc = 3$ the number of colors,
and
$\gamma_v = \partial_t v$ the Higgs and
$\gamma_x = \partial_t x$ singlet anomalous dimension.

In all our analyses,
the input scale is
$\LamD = \MZ$ and
as initial conditions,
we impose
\begin{align}
g_1(\MZ) &= 0.344
  \,, &
g_2(\MZ) &= 0.64
  \,, \nn
g_3(\MZ) &= 1.22
  \,, &
y_t(\MZ) &=1
  \,,
\end{align}
using
the physically observed masses,
the pole masses,
at their numerical values~\cite{Workman:2022ynf}
\begin{align}
  \label{eq:pole:mass}
  (\Mt,&\MW,\MZ, \Mh )
  \nn &=
  (172.69, 80.377, 91.1876, 125.25)~{\rm GeV}
  \,.
\end{align}

\section{\\Relating \MSbar{} parameters to physical observables}
\label{sec:ms:to:phys}

During our analysis,
we employed two different procedures
for vacuum renormalization:

\subsection{Vacuum renormalization through counterterms}
\label{sec:vacren:4d}

For the
vacuum renormalization
in the 4D~\MSbar~scheme,
the initial conditions for the parameters
$\lambda$,
$\mu_h^2$, and
$\mu_s^2$
are obtained by requiring
the tree-level potential, $V_0$,
to have a minimum at the electroweak vacuum $\bm{\phi}=(v_0,0)$ and
to yield the correct mass eigenvalues.
These requirements,
give rise to the conditions
\begin{align}
\label{eq:MSbar:RGcondition:V'}
    \frac{\partial V_0}{\partial {\phi_i}} \Bigr|_{\scriptsize
      \begin{aligned}
        \bm{\phi} &= (v_0 \xi(v_0),0)\\[-2mm]
        \LamD &= v_0
      \end{aligned}
      }
      &= 0
    \,,\\
\label{eq:MSbar:RGcondition:V''}
    \frac{\partial^2 V_0}{\partial {\phi_i^2}} \Bigr|_{\scriptsize
      \begin{aligned}
        \bm{\phi} &= (v_0 \xi(v_0),0)\\[-2mm]
        \LamD &= v_0
      \end{aligned}
      } &=
    M_i^2
    \,,
\end{align}
where
$M_i$ are the pole masses
with
$\phi_i=v,x$,
and
the off-diagonal Hessian derivative vanishes trivially when the $Z_2$-symmetry remains unbroken.
The minimization conditions thus read
\begin{align}
\label{eq:MSbar:rel:tree:mh}
   \mu_h^2(v_0) &= -\frac{1}{2}\frac{\Mh^2}{\xi(v_0)^2}
   \,,  \\
\label{eq:MSbar:rel:tree:ms}
   \mu_s^2(v_0) &= \Ms^2 -\frac{1}{2} \lambda_{hs}(v_0^{ }) v_0^{2}\xi(v_0^{ })^2
   \,, \\
\label{eq:MSbar:rel:tree:lambda}
   \lambda(v_0) &= \frac{1}{2}\frac{\Mh^2}{v_0^{2} \xi(v_0^{ })^4}
   \,,
\end{align}
where
\begin{equation}
  \xi(\LamD) = \exp{\biggl[ \int_{\log{(\LamD^2)}}^{\log{(\MZ^2)}}\!\!{\rm d}t\, \gamma_v(t) \biggr]}
    \,,
\end{equation}
encodes the solution of the Higgs field RG equation,
$v(\LamD) = \xi(\LamD)v$.
Since we fix
$\Mh=125.25$~GeV and
$v_0\simeq246$~GeV, the only free parameters are
the singlet mass $m_s$,
its quartic self-interaction $\lambda_{s}$ and
the Higgs portal coupling $\lambda_{hs}$ which we fix at the input scale $\MZ$.

At one-loop level,
this procedure ensures that
the minimization conditions of
eqs.~\eqref{eq:MSbar:RGcondition:V'} and \eqref{eq:MSbar:RGcondition:V''} are fulfilled
for $V_0 \to \Veff$.
By introducing counterterms~\cite{Delaunay:2007wb},
the model parameters
$\mu_{h}^{2}$,
$\mu_{s}^{2}$, and
$\lambda$ are tuned accordingly and
yield one-loop improvements to
the relations~\eqref{eq:MSbar:rel:tree:mh}--\eqref{eq:MSbar:rel:tree:lambda}.
The counterterm potential has the following form:
\begin{equation}
 V_\rmii{CT} =
      \frac{\delta \mu_h^2}{2}v^2
    + \frac{\delta \mu_s^2}{2}x^2
    + \frac{\delta \lambda}{4}v^4
  \,,
\end{equation}
where
\begin{align}
  \delta \mu_h^2 &=
  - \Bigl(
      \frac{3}{2}\frac{1}{v_0 \xi(v_0)}
      \partial_{v}^{ }
    - \frac{1}{2}
      \partial_{v}^{2}
    \Bigr)
    V_\rmii{CW}^{ }
    \Bigr|_{\scriptsize
      \begin{aligned}
        \bm{\phi} &= (v_0 \xi(v_0),0)\\[-2mm]
        \LamD &= v_0
      \end{aligned}
      }
  \,,\nn
  \delta \lambda &=
  \frac{1}{2v_0^{3} \xi(v_0^{ })^3}
  \Bigl(
      \partial_{v}^{ }
    - v_0^{ }\xi(v_0^{ })
      \partial_{v}^{2}
    \Bigr)
    V_\rmii{CW}^{ }
    \Bigr|_{\scriptsize
      \begin{aligned}
        \bm{\phi} &= (v_0 \xi(v_0),0)\\[-2mm]
        \LamD &= v_0
      \end{aligned}
      }
  \,,\nn
  \delta \mu_s^2 &=
  - \partial_{x}^{2} V_\rmii{CW}^{ }
  \Bigr|_{\scriptsize
      \begin{aligned}
        \bm{\phi} &= (v_0 \xi(v_0),0)\\[-2mm]
        \LamD &= v_0
      \end{aligned}
      }
    \,,
\end{align}
are obtained by the renormalization conditions and
$V_\rmii{CW}^{ }(\bm{\phi}) = \sum_{i} n_i^{ } J_\rmii{CW}^{ }(m_i^2(\bm{\phi}))$ is the vacuum contribution in eq.~\eqref{eq:CW_potential}
or the Coleman-Weinberg potential.

\begin{figure}[t]
\centering
\includegraphics[width=.49\textwidth]{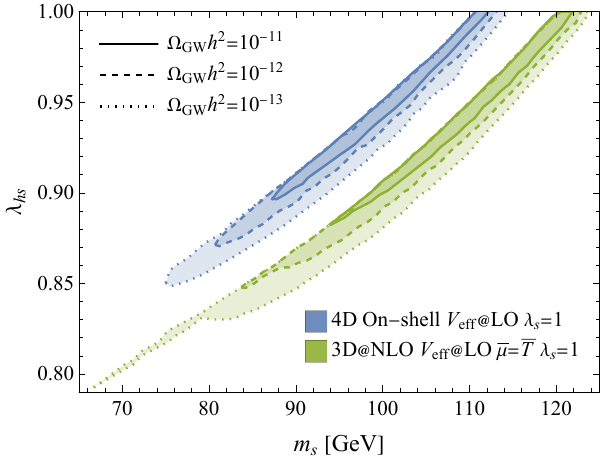}%
\\
\includegraphics[width=.49\textwidth]{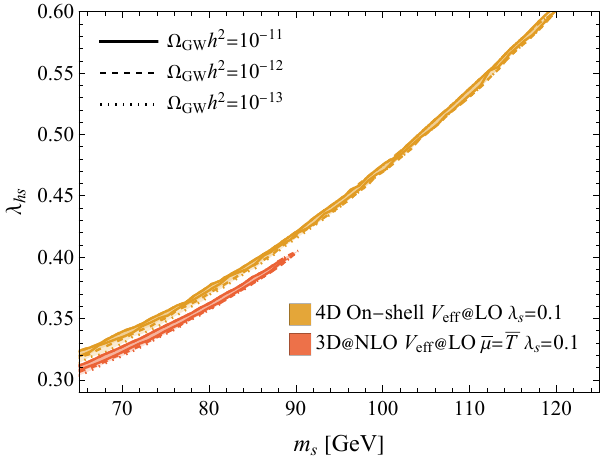}%
\caption{%
  Comparison of the parameter space upon varying
  the quartic singlet scalar self-coupling, $\lambda_s \in \{0.1,1.0\}$, for
  the on-shell~scheme and
  the 3D~EFT approach at 4D RG scale $\LamD = \bT$
  using (3D@NLO $\Veff$@LO) in the EFT.
  }
\label{fig:lambdas:comp}
\end{figure}

\subsection{One-loop corrected vacuum renormalization}
\label{sec:vacren:3d}

In the 3D~EFT, we use one-loop corrected vacuum values for the above couplings,
as described in appendix~A of~\cite{Niemi:2021qvp}
and~\cite{Kajantie:1995dw}.
The xSM model parameters are fixed
by the following physical scheme:
\begin{itemize}
  \item[\bf In:]
    Parameters $\Ms,\lambda_{s},\lambda_{hs}$, and
    pole masses at numerical values eq.~\eqref{eq:pole:mass}.
  \item[(i)]
    minimize the scalar potential at tree level (cf.\ eq.~\eqref{eq:MSbar:RGcondition:V'}),
    with $\langle S \rangle = 0$
    to determine the Higgs VEV,
  \item[(ii)]
    solve the renormalized parameters at \MSbar-scale
    $\LamD = \MZ$ from the one-loop corrected relations~\cite{Kajantie:1995dw};
    see~\cite{Niemi:2021qvp} for the xSM,
    \cite{Croon:2020cgk} for the SM, and
    \cite{Kainulainen:2019kyp} for the Two Higgs Doublet model,
  \item[(iii)]
    run the parameters to the matching scale $\LamD = X \bT$  using
    one-loop $\beta$-functions \eqref{eq:beta:g1}--\eqref{eq:beta:ms}.
    Here, $X$ is a constant typically varied to quantify
    the importance of higher-order corrections.
  \item[\bf Out:]
    \MSbar-parameters as function of physical parameters and $\bT$.
\end{itemize}
By relating pole masses to physical two-point functions,
this scheme ensures that higher order corrections in
the renormalization conditions are included and
the momentum dependence of the pole masses is respected.

The effective potential is not a physical quantity and
is independent of momentum.
Momentum dependence is necessary to physically fix the pole mass.
Since we evaluate
the vacuum renormalization at $\MZ$,
appendix~\ref{sec:vacren:4d} (cf.\ \cite{Croon:2020cgk})
is a naive approximation of the prescription in appendix~\ref{sec:vacren:3d}
and reproduces
the renormalization condition
of the Higgs sector.
To also relate parameters beyond
$\mu_h$ and
$\lambda$
to physical observables,
the full one-loop corrected vacuum renormalization of appendix~\ref{sec:vacren:3d}
is required~\cite{Niemi:2021qvp}.

\section{Impact of the scalar self-coupling $\lambda_s$}
\label{sec:lambdas}

The main body discusses parameter space scans in
the $(m_{s},\lambda_{hs},\lambda_{s})$-hyperplane.
The singlet quartic self-coupling is fixed at $\lambda_{s} = 1$.
To demonstrate that these results are qualitatively $\lambda_{s}$-independent,
we display in fig.~\ref{fig:lambdas:comp}
the parameter spaces at fixed $\lambda_{s} = 0.1$ in
the on-shell~scheme and
the 3D~EFT.
For the 3D~EFT analysis,
we again employ
(3D@NLO $\Veff$@LO),
namely a
one-loop effective potential with
NLO dimensional reduction.

For both approximations, the detectable parameter space moves towards smaller values of $\lambda_{hs}$.
Due to the decreasing mass range that produces observable GW signals,
the overall relevant parameter space shrinks.
By contrasting both approaches,
a similar $\mathcal{O}(1\%)$ uncertainty in
the signal reconstruction is expected --
similar to the $\lambda_{s} = 1$ case.
Hence,
we conclude that our results are robust including variations of
the quartic self-coupling of the singlet scalar.

\begin{figure}[t]
\centering
\includegraphics[width=.49\textwidth]{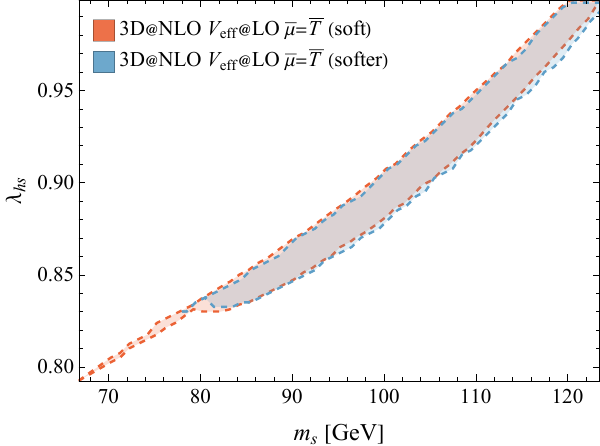}%
\caption{%
  Comparison of the parameter spaces with $\Omega_\rmii{GW}h^2 \geq 10^{-13}$
  predicted from the
  (3D@NLO $\Veff$@LO) potential
  using
  NLO dimensional reduction and
  LO effective potential either in
  the soft or
  the softer EFT.
  The soft EFT retains
  explicit terms of
  $J_3(m_{X_0}^{2})$
  for (adjoint) temporal scalars $X_0 \in \{A_0,B_0,C_0\}$.
  }
\label{fig:softer}
\end{figure}

\section{Impact of soft-scale effects on 3D~EFT}
\label{sec:soft:softer}
This section investigates the effects
of missing higher-order terms in the $v/T$ and $x/T$ expansions
at the softer scale by retaining
the full soft dynamics of the 3D~EFT at one-loop level.

To this end,
we use the soft potential at one-loop level from~\cite{Niemi:2021qvp}
now also in the background of the (adjoint) temporal gauge fields
$X_0 \in \{A_0,B_0,C_0\}$.
Due to an effective explicit center symmetry breaking during the EFT construction,
in the 3D~EFT
the only minimum that can be resolved is
$\langle X_0 \rangle \sim 0$
such that
$X_0$-backgrounds
vanish identically at the corresponding minimum~\cite{Kajantie:1998yc}.

By expanding the functions $J_3(m_{X_0}^{2})$
for (adjoint) temporal scalars $X_0$
in terms of their
mass eigenvalues,
we see that the ultrasoft matching relations contain
merely the first few terms in a $\frac{h_3 v^2}{\mD}$ expansion~\cite{Kajantie:1995dw}.
Here, $h_3$ is
the coupling between Lorentz and
temporal scalars
$X_0$~\cite{Kajantie:1995dw,Gould:2023ovu,Kierkla:2023von}.
To include also effects of large field values which is the case
in strong transitions as studied in this \article{} one needs to
monitor the effect of these soft corrections.
Here, we report that they are subdominant to the two-loop contributions in
resummation as seen
in fig.~\ref{fig:softer}.

\bibliographystyle{apsrev4-1}
\bibliography{ref}

\begin{thebibliography}{140}%
\makeatletter
\providecommand \@ifxundefined [1]{%
 \@ifx{#1\undefined}
}%
\providecommand \@ifnum [1]{%
 \ifnum #1\expandafter \@firstoftwo
 \else \expandafter \@secondoftwo
 \fi
}%
\providecommand \@ifx [1]{%
 \ifx #1\expandafter \@firstoftwo
 \else \expandafter \@secondoftwo
 \fi
}%
\providecommand \natexlab [1]{#1}%
\providecommand \enquote  [1]{``#1''}%
\providecommand \bibnamefont  [1]{#1}%
\providecommand \bibfnamefont [1]{#1}%
\providecommand \citenamefont [1]{#1}%
\providecommand \href@noop [0]{\@secondoftwo}%
\providecommand \href [0]{\begingroup \@sanitize@url \@href}%
\providecommand \@href[1]{\@@startlink{#1}\@@href}%
\providecommand \@@href[1]{\endgroup#1\@@endlink}%
\providecommand \@sanitize@url [0]{\catcode `\\12\catcode `\$12\catcode
  `\&12\catcode `\#12\catcode `\^12\catcode `\_12\catcode `\%12\relax}%
\providecommand \@@startlink[1]{}%
\providecommand \@@endlink[0]{}%
\providecommand \url  [0]{\begingroup\@sanitize@url \@url }%
\providecommand \@url [1]{\endgroup\@href {#1}{\urlprefix }}%
\providecommand \urlprefix  [0]{URL }%
\providecommand \Eprint [0]{\href }%
\providecommand \doibase [0]{http://dx.doi.org/}%
\providecommand \selectlanguage [0]{\@gobble}%
\providecommand \bibinfo  [0]{\@secondoftwo}%
\providecommand \bibfield  [0]{\@secondoftwo}%
\providecommand \translation [1]{[#1]}%
\providecommand \BibitemOpen [0]{}%
\providecommand \bibitemStop [0]{}%
\providecommand \bibitemNoStop [0]{.\EOS\space}%
\providecommand \EOS [0]{\spacefactor3000\relax}%
\providecommand \BibitemShut  [1]{\csname bibitem#1\endcsname}%
\let\auto@bib@innerbib\@empty
\bibitem [{\citenamefont {Agazie}\ \emph
  {et~al.}(2023{\natexlab{a}})\citenamefont {Agazie} \emph
  {et~al.}}]{NANOGrav:2023gor}%
  \BibitemOpen
  \bibfield  {author} {\bibinfo {author} {\bibfnamefont {G.}~\bibnamefont
  {Agazie}} \emph {et~al.} (\bibinfo {collaboration} {NANOGrav}),\ }\href
  {\doibase 10.3847/2041-8213/acdac6} {\bibfield  {journal} {\bibinfo
  {journal} {Astrophys. J. Lett.}\ }\textbf {\bibinfo {volume} {951}},\
  \bibinfo {pages} {L8} (\bibinfo {year} {2023}{\natexlab{a}})},\ \Eprint
  {http://arxiv.org/abs/2306.16213} {arXiv:2306.16213 [astro-ph.HE]}
  \BibitemShut {NoStop}%
\bibitem [{\citenamefont {Agazie}\ \emph
  {et~al.}(2023{\natexlab{b}})\citenamefont {Agazie} \emph
  {et~al.}}]{NANOGrav:2023hde}%
  \BibitemOpen
  \bibfield  {author} {\bibinfo {author} {\bibfnamefont {G.}~\bibnamefont
  {Agazie}} \emph {et~al.} (\bibinfo {collaboration} {NANOGrav}),\ }\href
  {\doibase 10.3847/2041-8213/acda9a} {\bibfield  {journal} {\bibinfo
  {journal} {Astrophys. J. Lett.}\ }\textbf {\bibinfo {volume} {951}},\
  \bibinfo {pages} {L9} (\bibinfo {year} {2023}{\natexlab{b}})},\ \Eprint
  {http://arxiv.org/abs/2306.16217} {arXiv:2306.16217 [astro-ph.HE]}
  \BibitemShut {NoStop}%
\bibitem [{\citenamefont {Antoniadis}\ \emph
  {et~al.}(2023{\natexlab{a}})\citenamefont {Antoniadis} \emph
  {et~al.}}]{EPTA:2023sfo}%
  \BibitemOpen
  \bibfield  {author} {\bibinfo {author} {\bibfnamefont {J.}~\bibnamefont
  {Antoniadis}} \emph {et~al.} (\bibinfo {collaboration} {EPTA}),\ }\href
  {\doibase 10.1051/0004-6361/202346841} {\bibfield  {journal} {\bibinfo
  {journal} {Astron. Astrophys.}\ }\textbf {\bibinfo {volume} {678}},\ \bibinfo
  {pages} {A48} (\bibinfo {year} {2023}{\natexlab{a}})},\ \Eprint
  {http://arxiv.org/abs/2306.16224} {arXiv:2306.16224 [astro-ph.HE]}
  \BibitemShut {NoStop}%
\bibitem [{\citenamefont {Antoniadis}\ \emph
  {et~al.}(2023{\natexlab{b}})\citenamefont {Antoniadis} \emph
  {et~al.}}]{EPTA:2023fyk}%
  \BibitemOpen
  \bibfield  {author} {\bibinfo {author} {\bibfnamefont {J.}~\bibnamefont
  {Antoniadis}} \emph {et~al.} (\bibinfo {collaboration} {EPTA, InPTA:}),\
  }\href {\doibase 10.1051/0004-6361/202346844} {\bibfield  {journal} {\bibinfo
   {journal} {Astron. Astrophys.}\ }\textbf {\bibinfo {volume} {678}},\
  \bibinfo {pages} {A50} (\bibinfo {year} {2023}{\natexlab{b}})},\ \Eprint
  {http://arxiv.org/abs/2306.16214} {arXiv:2306.16214 [astro-ph.HE]}
  \BibitemShut {NoStop}%
\bibitem [{\citenamefont {Zic}\ \emph {et~al.}(2023)\citenamefont {Zic} \emph
  {et~al.}}]{Zic:2023gta}%
  \BibitemOpen
  \bibfield  {author} {\bibinfo {author} {\bibfnamefont {A.}~\bibnamefont
  {Zic}} \emph {et~al.},\ }\href {\doibase 10.1017/pasa.2023.36} {\bibfield
  {journal} {\bibinfo  {journal} {Publ. Astron. Soc. Austral.}\ }\textbf
  {\bibinfo {volume} {40}},\ \bibinfo {pages} {e049} (\bibinfo {year}
  {2023})},\ \Eprint {http://arxiv.org/abs/2306.16230} {arXiv:2306.16230
  [astro-ph.HE]} \BibitemShut {NoStop}%
\bibitem [{\citenamefont {Reardon}\ \emph {et~al.}(2023)\citenamefont {Reardon}
  \emph {et~al.}}]{Reardon:2023gzh}%
  \BibitemOpen
  \bibfield  {author} {\bibinfo {author} {\bibfnamefont {D.~J.}\ \bibnamefont
  {Reardon}} \emph {et~al.},\ }\href {\doibase 10.3847/2041-8213/acdd02}
  {\bibfield  {journal} {\bibinfo  {journal} {Astrophys. J. Lett.}\ }\textbf
  {\bibinfo {volume} {951}},\ \bibinfo {pages} {L6} (\bibinfo {year} {2023})},\
  \Eprint {http://arxiv.org/abs/2306.16215} {arXiv:2306.16215 [astro-ph.HE]}
  \BibitemShut {NoStop}%
\bibitem [{\citenamefont {Xu}\ \emph {et~al.}(2023)\citenamefont {Xu} \emph
  {et~al.}}]{Xu:2023wog}%
  \BibitemOpen
  \bibfield  {author} {\bibinfo {author} {\bibfnamefont {H.}~\bibnamefont {Xu}}
  \emph {et~al.},\ }\href {\doibase 10.1088/1674-4527/acdfa5} {\bibfield
  {journal} {\bibinfo  {journal} {Res. Astron. Astrophys.}\ }\textbf {\bibinfo
  {volume} {23}},\ \bibinfo {pages} {075024} (\bibinfo {year} {2023})},\
  \Eprint {http://arxiv.org/abs/2306.16216} {arXiv:2306.16216 [astro-ph.HE]}
  \BibitemShut {NoStop}%
\bibitem [{\citenamefont {Afzal}\ \emph {et~al.}(2023)\citenamefont {Afzal}
  \emph {et~al.}}]{NANOGrav:2023hvm}%
  \BibitemOpen
  \bibfield  {author} {\bibinfo {author} {\bibfnamefont {A.}~\bibnamefont
  {Afzal}} \emph {et~al.} (\bibinfo {collaboration} {NANOGrav}),\ }\href
  {\doibase 10.3847/2041-8213/acdc91} {\bibfield  {journal} {\bibinfo
  {journal} {Astrophys. J. Lett.}\ }\textbf {\bibinfo {volume} {951}},\
  \bibinfo {pages} {L11} (\bibinfo {year} {2023})},\ \Eprint
  {http://arxiv.org/abs/2306.16219} {arXiv:2306.16219 [astro-ph.HE]}
  \BibitemShut {NoStop}%
\bibitem [{\citenamefont {Figueroa}\ \emph {et~al.}(2024)\citenamefont
  {Figueroa}, \citenamefont {Pieroni}, \citenamefont {Ricciardone},\ and\
  \citenamefont {Simakachorn}}]{Figueroa:2023zhu}%
  \BibitemOpen
  \bibfield  {author} {\bibinfo {author} {\bibfnamefont {D.~G.}\ \bibnamefont
  {Figueroa}}, \bibinfo {author} {\bibfnamefont {M.}~\bibnamefont {Pieroni}},
  \bibinfo {author} {\bibfnamefont {A.}~\bibnamefont {Ricciardone}}, \ and\
  \bibinfo {author} {\bibfnamefont {P.}~\bibnamefont {Simakachorn}},\ }\href
  {\doibase 10.1103/PhysRevLett.132.171002} {\bibfield  {journal} {\bibinfo
  {journal} {Phys. Rev. Lett.}\ }\textbf {\bibinfo {volume} {132}},\ \bibinfo
  {pages} {171002} (\bibinfo {year} {2024})},\ \Eprint
  {http://arxiv.org/abs/2307.02399} {arXiv:2307.02399 [astro-ph.CO]}
  \BibitemShut {NoStop}%
\bibitem [{\citenamefont {Ellis}\ \emph {et~al.}(2024)\citenamefont {Ellis},
  \citenamefont {Fairbairn}, \citenamefont {Franciolini}, \citenamefont
  {H\"utsi}, \citenamefont {Iovino}, \citenamefont {Lewicki}, \citenamefont
  {Raidal}, \citenamefont {Urrutia}, \citenamefont {Vaskonen},\ and\
  \citenamefont {Veerm\"ae}}]{Ellis:2023oxs}%
  \BibitemOpen
  \bibfield  {author} {\bibinfo {author} {\bibfnamefont {J.}~\bibnamefont
  {Ellis}}, \bibinfo {author} {\bibfnamefont {M.}~\bibnamefont {Fairbairn}},
  \bibinfo {author} {\bibfnamefont {G.}~\bibnamefont {Franciolini}}, \bibinfo
  {author} {\bibfnamefont {G.}~\bibnamefont {H\"utsi}}, \bibinfo {author}
  {\bibfnamefont {A.}~\bibnamefont {Iovino}}, \bibinfo {author} {\bibfnamefont
  {M.}~\bibnamefont {Lewicki}}, \bibinfo {author} {\bibfnamefont
  {M.}~\bibnamefont {Raidal}}, \bibinfo {author} {\bibfnamefont
  {J.}~\bibnamefont {Urrutia}}, \bibinfo {author} {\bibfnamefont
  {V.}~\bibnamefont {Vaskonen}}, \ and\ \bibinfo {author} {\bibfnamefont
  {H.}~\bibnamefont {Veerm\"ae}},\ }\href {\doibase
  10.1103/PhysRevD.109.023522} {\bibfield  {journal} {\bibinfo  {journal}
  {Phys. Rev. D}\ }\textbf {\bibinfo {volume} {109}},\ \bibinfo {pages}
  {023522} (\bibinfo {year} {2024})},\ \Eprint
  {http://arxiv.org/abs/2308.08546} {arXiv:2308.08546 [astro-ph.CO]}
  \BibitemShut {NoStop}%
\bibitem [{\citenamefont {Caprini}\ \emph {et~al.}(2016)\citenamefont {Caprini}
  \emph {et~al.}}]{Caprini:2015zlo}%
  \BibitemOpen
  \bibfield  {author} {\bibinfo {author} {\bibfnamefont {C.}~\bibnamefont
  {Caprini}} \emph {et~al.},\ }\href {\doibase 10.1088/1475-7516/2016/04/001}
  {\bibfield  {journal} {\bibinfo  {journal} {JCAP}\ }\textbf {\bibinfo
  {volume} {04}},\ \bibinfo {pages} {001} (\bibinfo {year} {2016})},\ \Eprint
  {http://arxiv.org/abs/1512.06239} {arXiv:1512.06239 [astro-ph.CO]}
  \BibitemShut {NoStop}%
\bibitem [{\citenamefont {Caprini}\ \emph {et~al.}(2020)\citenamefont {Caprini}
  \emph {et~al.}}]{Caprini:2019egz}%
  \BibitemOpen
  \bibfield  {author} {\bibinfo {author} {\bibfnamefont {C.}~\bibnamefont
  {Caprini}} \emph {et~al.},\ }\href {\doibase 10.1088/1475-7516/2020/03/024}
  {\bibfield  {journal} {\bibinfo  {journal} {JCAP}\ }\textbf {\bibinfo
  {volume} {03}},\ \bibinfo {pages} {024} (\bibinfo {year} {2020})},\ \Eprint
  {http://arxiv.org/abs/1910.13125} {arXiv:1910.13125 [astro-ph.CO]}
  \BibitemShut {NoStop}%
\bibitem [{\citenamefont {Auclair}\ \emph {et~al.}(2023)\citenamefont {Auclair}
  \emph {et~al.}}]{LISACosmologyWorkingGroup:2022jok}%
  \BibitemOpen
  \bibfield  {author} {\bibinfo {author} {\bibfnamefont {P.}~\bibnamefont
  {Auclair}} \emph {et~al.} (\bibinfo {collaboration} {LISA Cosmology Working
  Group}),\ }\href {\doibase 10.1007/s41114-023-00045-2} {\bibfield  {journal}
  {\bibinfo  {journal} {Living Rev. Rel.}\ }\textbf {\bibinfo {volume} {26}},\
  \bibinfo {pages} {5} (\bibinfo {year} {2023})},\ \Eprint
  {http://arxiv.org/abs/2204.05434} {arXiv:2204.05434 [astro-ph.CO]}
  \BibitemShut {NoStop}%
\bibitem [{\citenamefont {Farakos}\ \emph {et~al.}(1995)\citenamefont
  {Farakos}, \citenamefont {Kajantie}, \citenamefont {Rummukainen},\ and\
  \citenamefont {Shaposhnikov}}]{Farakos:1994xh}%
  \BibitemOpen
  \bibfield  {author} {\bibinfo {author} {\bibfnamefont {K.}~\bibnamefont
  {Farakos}}, \bibinfo {author} {\bibfnamefont {K.}~\bibnamefont {Kajantie}},
  \bibinfo {author} {\bibfnamefont {K.}~\bibnamefont {Rummukainen}}, \ and\
  \bibinfo {author} {\bibfnamefont {M.~E.}\ \bibnamefont {Shaposhnikov}},\
  }\href {\doibase 10.1016/0550-3213(95)80129-4} {\bibfield  {journal}
  {\bibinfo  {journal} {Nucl. Phys. B}\ }\textbf {\bibinfo {volume} {442}},\
  \bibinfo {pages} {317} (\bibinfo {year} {1995})},\ \Eprint
  {http://arxiv.org/abs/hep-lat/9412091} {arXiv:hep-lat/9412091} \BibitemShut
  {NoStop}%
\bibitem [{\citenamefont {Kajantie}\ \emph
  {et~al.}(1996{\natexlab{a}})\citenamefont {Kajantie}, \citenamefont {Laine},
  \citenamefont {Rummukainen},\ and\ \citenamefont
  {Shaposhnikov}}]{Kajantie:1995kf}%
  \BibitemOpen
  \bibfield  {author} {\bibinfo {author} {\bibfnamefont {K.}~\bibnamefont
  {Kajantie}}, \bibinfo {author} {\bibfnamefont {M.}~\bibnamefont {Laine}},
  \bibinfo {author} {\bibfnamefont {K.}~\bibnamefont {Rummukainen}}, \ and\
  \bibinfo {author} {\bibfnamefont {M.~E.}\ \bibnamefont {Shaposhnikov}},\
  }\href {\doibase 10.1016/0550-3213(96)00052-1} {\bibfield  {journal}
  {\bibinfo  {journal} {Nucl. Phys. B}\ }\textbf {\bibinfo {volume} {466}},\
  \bibinfo {pages} {189} (\bibinfo {year} {1996}{\natexlab{a}})},\ \Eprint
  {http://arxiv.org/abs/hep-lat/9510020} {arXiv:hep-lat/9510020} \BibitemShut
  {NoStop}%
\bibitem [{\citenamefont {Moore}\ and\ \citenamefont
  {Rummukainen}(2001)}]{Moore:2000jw}%
  \BibitemOpen
  \bibfield  {author} {\bibinfo {author} {\bibfnamefont {G.~D.}\ \bibnamefont
  {Moore}}\ and\ \bibinfo {author} {\bibfnamefont {K.}~\bibnamefont
  {Rummukainen}},\ }\href {\doibase 10.1103/PhysRevD.63.045002} {\bibfield
  {journal} {\bibinfo  {journal} {Phys.\ Rev.}\ }\textbf {\bibinfo {volume}
  {D63}},\ \bibinfo {pages} {045002} (\bibinfo {year} {2001})},\ \Eprint
  {http://arxiv.org/abs/hep-ph/0009132} {arXiv:hep-ph/0009132 [hep-ph]}
  \BibitemShut {NoStop}%
\bibitem [{\citenamefont {Gould}\ \emph {et~al.}(2022)\citenamefont {Gould},
  \citenamefont {G\"uyer},\ and\ \citenamefont {Rummukainen}}]{Gould:2022ran}%
  \BibitemOpen
  \bibfield  {author} {\bibinfo {author} {\bibfnamefont {O.}~\bibnamefont
  {Gould}}, \bibinfo {author} {\bibfnamefont {S.}~\bibnamefont {G\"uyer}}, \
  and\ \bibinfo {author} {\bibfnamefont {K.}~\bibnamefont {Rummukainen}},\
  }\href {\doibase 10.1103/PhysRevD.106.114507} {\bibfield  {journal} {\bibinfo
   {journal} {Phys. Rev. D}\ }\textbf {\bibinfo {volume} {106}},\ \bibinfo
  {pages} {114507} (\bibinfo {year} {2022})},\ \Eprint
  {http://arxiv.org/abs/2205.07238} {arXiv:2205.07238 [hep-lat]} \BibitemShut
  {NoStop}%
\bibitem [{\citenamefont {Ginsparg}(1980)}]{Ginsparg:1980ef}%
  \BibitemOpen
  \bibfield  {author} {\bibinfo {author} {\bibfnamefont {P.~H.}\ \bibnamefont
  {Ginsparg}},\ }\href {\doibase 10.1016/0550-3213(80)90418-6} {\bibfield
  {journal} {\bibinfo  {journal} {Nucl. Phys.}\ }\textbf {\bibinfo {volume}
  {B170}},\ \bibinfo {pages} {388} (\bibinfo {year} {1980})}\BibitemShut
  {NoStop}%
\bibitem [{\citenamefont {Appelquist}\ and\ \citenamefont
  {Pisarski}(1981)}]{Appelquist:1981vg}%
  \BibitemOpen
  \bibfield  {author} {\bibinfo {author} {\bibfnamefont {T.}~\bibnamefont
  {Appelquist}}\ and\ \bibinfo {author} {\bibfnamefont {R.~D.}\ \bibnamefont
  {Pisarski}},\ }\href {\doibase 10.1103/PhysRevD.23.2305} {\bibfield
  {journal} {\bibinfo  {journal} {Phys. Rev.}\ }\textbf {\bibinfo {volume}
  {D23}},\ \bibinfo {pages} {2305} (\bibinfo {year} {1981})}\BibitemShut
  {NoStop}%
\bibitem [{\citenamefont {Gould}\ and\ \citenamefont
  {Tenkanen}(2024)}]{Gould:2023ovu}%
  \BibitemOpen
  \bibfield  {author} {\bibinfo {author} {\bibfnamefont {O.}~\bibnamefont
  {Gould}}\ and\ \bibinfo {author} {\bibfnamefont {T.~V.~I.}\ \bibnamefont
  {Tenkanen}},\ }\href {\doibase 10.1007/JHEP01(2024)048} {\bibfield  {journal}
  {\bibinfo  {journal} {JHEP}\ }\textbf {\bibinfo {volume} {01}},\ \bibinfo
  {pages} {048} (\bibinfo {year} {2024})},\ \Eprint
  {http://arxiv.org/abs/2309.01672} {arXiv:2309.01672 [hep-ph]} \BibitemShut
  {NoStop}%
\bibitem [{\citenamefont {Delaunay}\ \emph {et~al.}(2008)\citenamefont
  {Delaunay}, \citenamefont {Grojean},\ and\ \citenamefont
  {Wells}}]{Delaunay:2007wb}%
  \BibitemOpen
  \bibfield  {author} {\bibinfo {author} {\bibfnamefont {C.}~\bibnamefont
  {Delaunay}}, \bibinfo {author} {\bibfnamefont {C.}~\bibnamefont {Grojean}}, \
  and\ \bibinfo {author} {\bibfnamefont {J.~D.}\ \bibnamefont {Wells}},\ }\href
  {\doibase 10.1088/1126-6708/2008/04/029} {\bibfield  {journal} {\bibinfo
  {journal} {JHEP}\ }\textbf {\bibinfo {volume} {04}},\ \bibinfo {pages} {029}
  (\bibinfo {year} {2008})},\ \Eprint {http://arxiv.org/abs/0711.2511}
  {arXiv:0711.2511 [hep-ph]} \BibitemShut {NoStop}%
\bibitem [{\citenamefont {Patel}\ and\ \citenamefont
  {Ramsey-Musolf}(2011)}]{Patel:2011th}%
  \BibitemOpen
  \bibfield  {author} {\bibinfo {author} {\bibfnamefont {H.~H.}\ \bibnamefont
  {Patel}}\ and\ \bibinfo {author} {\bibfnamefont {M.~J.}\ \bibnamefont
  {Ramsey-Musolf}},\ }\href {\doibase 10.1007/JHEP07(2011)029} {\bibfield
  {journal} {\bibinfo  {journal} {JHEP}\ }\textbf {\bibinfo {volume} {07}},\
  \bibinfo {pages} {029} (\bibinfo {year} {2011})},\ \Eprint
  {http://arxiv.org/abs/1101.4665} {arXiv:1101.4665 [hep-ph]} \BibitemShut
  {NoStop}%
\bibitem [{\citenamefont {Athron}\ \emph {et~al.}(2024)\citenamefont {Athron},
  \citenamefont {Bal\'azs}, \citenamefont {Fowlie}, \citenamefont {Morris},\
  and\ \citenamefont {Wu}}]{Athron:2023xlk}%
  \BibitemOpen
  \bibfield  {author} {\bibinfo {author} {\bibfnamefont {P.}~\bibnamefont
  {Athron}}, \bibinfo {author} {\bibfnamefont {C.}~\bibnamefont {Bal\'azs}},
  \bibinfo {author} {\bibfnamefont {A.}~\bibnamefont {Fowlie}}, \bibinfo
  {author} {\bibfnamefont {L.}~\bibnamefont {Morris}}, \ and\ \bibinfo {author}
  {\bibfnamefont {L.}~\bibnamefont {Wu}},\ }\href {\doibase
  10.1016/j.ppnp.2023.104094} {\bibfield  {journal} {\bibinfo  {journal} {Prog.
  Part. Nucl. Phys.}\ }\textbf {\bibinfo {volume} {135}},\ \bibinfo {pages}
  {104094} (\bibinfo {year} {2024})},\ \Eprint
  {http://arxiv.org/abs/2305.02357} {arXiv:2305.02357 [hep-ph]} \BibitemShut
  {NoStop}%
\bibitem [{\citenamefont {Caprini}\ \emph {et~al.}(2024)\citenamefont
  {Caprini}, \citenamefont {Jinno}, \citenamefont {Lewicki}, \citenamefont
  {Madge}, \citenamefont {Merchand}, \citenamefont {Nardini}, \citenamefont
  {Pieroni}, \citenamefont {Roper~Pol},\ and\ \citenamefont
  {Vaskonen}}]{Caprini:2024hue}%
  \BibitemOpen
  \bibfield  {author} {\bibinfo {author} {\bibfnamefont {C.}~\bibnamefont
  {Caprini}}, \bibinfo {author} {\bibfnamefont {R.}~\bibnamefont {Jinno}},
  \bibinfo {author} {\bibfnamefont {M.}~\bibnamefont {Lewicki}}, \bibinfo
  {author} {\bibfnamefont {E.}~\bibnamefont {Madge}}, \bibinfo {author}
  {\bibfnamefont {M.}~\bibnamefont {Merchand}}, \bibinfo {author}
  {\bibfnamefont {G.}~\bibnamefont {Nardini}}, \bibinfo {author} {\bibfnamefont
  {M.}~\bibnamefont {Pieroni}}, \bibinfo {author} {\bibfnamefont
  {A.}~\bibnamefont {Roper~Pol}}, \ and\ \bibinfo {author} {\bibfnamefont
  {V.}~\bibnamefont {Vaskonen}} (\bibinfo {collaboration} {LISA Cosmology
  Working Group}),\ }\href {\doibase 10.1088/1475-7516/2024/10/020} {\bibfield
  {journal} {\bibinfo  {journal} {JCAP}\ }\textbf {\bibinfo {volume} {10}},\
  \bibinfo {pages} {020} (\bibinfo {year} {2024})},\ \Eprint
  {http://arxiv.org/abs/2403.03723} {arXiv:2403.03723 [astro-ph.CO]}
  \BibitemShut {NoStop}%
\bibitem [{\citenamefont {Grojean}\ and\ \citenamefont
  {Servant}(2007)}]{Grojean:2006bp}%
  \BibitemOpen
  \bibfield  {author} {\bibinfo {author} {\bibfnamefont {C.}~\bibnamefont
  {Grojean}}\ and\ \bibinfo {author} {\bibfnamefont {G.}~\bibnamefont
  {Servant}},\ }\href {\doibase 10.1103/PhysRevD.75.043507} {\bibfield
  {journal} {\bibinfo  {journal} {Phys. Rev. D}\ }\textbf {\bibinfo {volume}
  {75}},\ \bibinfo {pages} {043507} (\bibinfo {year} {2007})},\ \Eprint
  {http://arxiv.org/abs/hep-ph/0607107} {arXiv:hep-ph/0607107} \BibitemShut
  {NoStop}%
\bibitem [{\citenamefont {Kang}\ \emph {et~al.}(2018)\citenamefont {Kang},
  \citenamefont {Ko},\ and\ \citenamefont {Matsui}}]{Kang:2017mkl}%
  \BibitemOpen
  \bibfield  {author} {\bibinfo {author} {\bibfnamefont {Z.}~\bibnamefont
  {Kang}}, \bibinfo {author} {\bibfnamefont {P.}~\bibnamefont {Ko}}, \ and\
  \bibinfo {author} {\bibfnamefont {T.}~\bibnamefont {Matsui}},\ }\href
  {\doibase 10.1007/JHEP02(2018)115} {\bibfield  {journal} {\bibinfo  {journal}
  {JHEP}\ }\textbf {\bibinfo {volume} {02}},\ \bibinfo {pages} {115} (\bibinfo
  {year} {2018})},\ \Eprint {http://arxiv.org/abs/1706.09721} {arXiv:1706.09721
  [hep-ph]} \BibitemShut {NoStop}%
\bibitem [{\citenamefont {Hashino}\ \emph {et~al.}(2017)\citenamefont
  {Hashino}, \citenamefont {Kakizaki}, \citenamefont {Kanemura}, \citenamefont
  {Ko},\ and\ \citenamefont {Matsui}}]{Hashino:2016xoj}%
  \BibitemOpen
  \bibfield  {author} {\bibinfo {author} {\bibfnamefont {K.}~\bibnamefont
  {Hashino}}, \bibinfo {author} {\bibfnamefont {M.}~\bibnamefont {Kakizaki}},
  \bibinfo {author} {\bibfnamefont {S.}~\bibnamefont {Kanemura}}, \bibinfo
  {author} {\bibfnamefont {P.}~\bibnamefont {Ko}}, \ and\ \bibinfo {author}
  {\bibfnamefont {T.}~\bibnamefont {Matsui}},\ }\href {\doibase
  10.1016/j.physletb.2016.12.052} {\bibfield  {journal} {\bibinfo  {journal}
  {Phys. Lett. B}\ }\textbf {\bibinfo {volume} {766}},\ \bibinfo {pages} {49}
  (\bibinfo {year} {2017})},\ \Eprint {http://arxiv.org/abs/1609.00297}
  {arXiv:1609.00297 [hep-ph]} \BibitemShut {NoStop}%
\bibitem [{\citenamefont {Bian}\ \emph {et~al.}(2020)\citenamefont {Bian},
  \citenamefont {Guo}, \citenamefont {Wu},\ and\ \citenamefont
  {Zhou}}]{Bian:2019bsn}%
  \BibitemOpen
  \bibfield  {author} {\bibinfo {author} {\bibfnamefont {L.}~\bibnamefont
  {Bian}}, \bibinfo {author} {\bibfnamefont {H.-K.}\ \bibnamefont {Guo}},
  \bibinfo {author} {\bibfnamefont {Y.}~\bibnamefont {Wu}}, \ and\ \bibinfo
  {author} {\bibfnamefont {R.}~\bibnamefont {Zhou}},\ }\href {\doibase
  10.1103/PhysRevD.101.035011} {\bibfield  {journal} {\bibinfo  {journal}
  {Phys. Rev. D}\ }\textbf {\bibinfo {volume} {101}},\ \bibinfo {pages}
  {035011} (\bibinfo {year} {2020})},\ \Eprint
  {http://arxiv.org/abs/1906.11664} {arXiv:1906.11664 [hep-ph]} \BibitemShut
  {NoStop}%
\bibitem [{\citenamefont {Friedrich}\ \emph {et~al.}(2022)\citenamefont
  {Friedrich}, \citenamefont {Ramsey-Musolf}, \citenamefont {Tenkanen},\ and\
  \citenamefont {Tran}}]{Friedrich:2022cak}%
  \BibitemOpen
  \bibfield  {author} {\bibinfo {author} {\bibfnamefont {L.~S.}\ \bibnamefont
  {Friedrich}}, \bibinfo {author} {\bibfnamefont {M.~J.}\ \bibnamefont
  {Ramsey-Musolf}}, \bibinfo {author} {\bibfnamefont {T.~V.~I.}\ \bibnamefont
  {Tenkanen}}, \ and\ \bibinfo {author} {\bibfnamefont {V.~Q.}\ \bibnamefont
  {Tran}},\ }\href@noop {} {\  (\bibinfo {year} {2022})},\ \Eprint
  {http://arxiv.org/abs/2203.05889} {arXiv:2203.05889 [hep-ph]} \BibitemShut
  {NoStop}%
\bibitem [{\citenamefont {Croon}\ \emph {et~al.}(2021)\citenamefont {Croon},
  \citenamefont {Gould}, \citenamefont {Schicho}, \citenamefont {Tenkanen},\
  and\ \citenamefont {White}}]{Croon:2020cgk}%
  \BibitemOpen
  \bibfield  {author} {\bibinfo {author} {\bibfnamefont {D.}~\bibnamefont
  {Croon}}, \bibinfo {author} {\bibfnamefont {O.}~\bibnamefont {Gould}},
  \bibinfo {author} {\bibfnamefont {P.}~\bibnamefont {Schicho}}, \bibinfo
  {author} {\bibfnamefont {T.~V.~I.}\ \bibnamefont {Tenkanen}}, \ and\ \bibinfo
  {author} {\bibfnamefont {G.}~\bibnamefont {White}},\ }\href {\doibase
  10.1007/JHEP04(2021)055} {\bibfield  {journal} {\bibinfo  {journal} {JHEP}\
  }\textbf {\bibinfo {volume} {04}},\ \bibinfo {pages} {055} (\bibinfo {year}
  {2021})},\ \Eprint {http://arxiv.org/abs/2009.10080} {arXiv:2009.10080
  [hep-ph]} \BibitemShut {NoStop}%
\bibitem [{\citenamefont {Gould}\ and\ \citenamefont
  {Xie}(2023)}]{Gould:2023jbz}%
  \BibitemOpen
  \bibfield  {author} {\bibinfo {author} {\bibfnamefont {O.}~\bibnamefont
  {Gould}}\ and\ \bibinfo {author} {\bibfnamefont {C.}~\bibnamefont {Xie}},\
  }\href {\doibase 10.1007/JHEP12(2023)049} {\bibfield  {journal} {\bibinfo
  {journal} {JHEP}\ }\textbf {\bibinfo {volume} {12}},\ \bibinfo {pages} {049}
  (\bibinfo {year} {2023})},\ \Eprint {http://arxiv.org/abs/2310.02308}
  {arXiv:2310.02308 [hep-ph]} \BibitemShut {NoStop}%
\bibitem [{\citenamefont {Athron}\ \emph {et~al.}(2023)\citenamefont {Athron},
  \citenamefont {Balazs}, \citenamefont {Fowlie}, \citenamefont {Morris},
  \citenamefont {White},\ and\ \citenamefont {Zhang}}]{Athron:2022jyi}%
  \BibitemOpen
  \bibfield  {author} {\bibinfo {author} {\bibfnamefont {P.}~\bibnamefont
  {Athron}}, \bibinfo {author} {\bibfnamefont {C.}~\bibnamefont {Balazs}},
  \bibinfo {author} {\bibfnamefont {A.}~\bibnamefont {Fowlie}}, \bibinfo
  {author} {\bibfnamefont {L.}~\bibnamefont {Morris}}, \bibinfo {author}
  {\bibfnamefont {G.}~\bibnamefont {White}}, \ and\ \bibinfo {author}
  {\bibfnamefont {Y.}~\bibnamefont {Zhang}},\ }\href {\doibase
  10.1007/JHEP01(2023)050} {\bibfield  {journal} {\bibinfo  {journal} {JHEP}\
  }\textbf {\bibinfo {volume} {01}},\ \bibinfo {pages} {050} (\bibinfo {year}
  {2023})},\ \Eprint {http://arxiv.org/abs/2208.01319} {arXiv:2208.01319
  [hep-ph]} \BibitemShut {NoStop}%
\bibitem [{\citenamefont {Gould}\ \emph {et~al.}(2019)\citenamefont {Gould},
  \citenamefont {Kozaczuk}, \citenamefont {Niemi}, \citenamefont
  {Ramsey-Musolf}, \citenamefont {Tenkanen},\ and\ \citenamefont
  {Weir}}]{Gould:2019qek}%
  \BibitemOpen
  \bibfield  {author} {\bibinfo {author} {\bibfnamefont {O.}~\bibnamefont
  {Gould}}, \bibinfo {author} {\bibfnamefont {J.}~\bibnamefont {Kozaczuk}},
  \bibinfo {author} {\bibfnamefont {L.}~\bibnamefont {Niemi}}, \bibinfo
  {author} {\bibfnamefont {M.~J.}\ \bibnamefont {Ramsey-Musolf}}, \bibinfo
  {author} {\bibfnamefont {T.~V.~I.}\ \bibnamefont {Tenkanen}}, \ and\ \bibinfo
  {author} {\bibfnamefont {D.~J.}\ \bibnamefont {Weir}},\ }\href {\doibase
  10.1103/PhysRevD.100.115024} {\bibfield  {journal} {\bibinfo  {journal}
  {Phys. Rev. D}\ }\textbf {\bibinfo {volume} {100}},\ \bibinfo {pages}
  {115024} (\bibinfo {year} {2019})},\ \Eprint
  {http://arxiv.org/abs/1903.11604} {arXiv:1903.11604 [hep-ph]} \BibitemShut
  {NoStop}%
\bibitem [{\citenamefont {Ellis}\ \emph {et~al.}(2020)\citenamefont {Ellis},
  \citenamefont {Lewicki},\ and\ \citenamefont {No}}]{Ellis:2020awk}%
  \BibitemOpen
  \bibfield  {author} {\bibinfo {author} {\bibfnamefont {J.}~\bibnamefont
  {Ellis}}, \bibinfo {author} {\bibfnamefont {M.}~\bibnamefont {Lewicki}}, \
  and\ \bibinfo {author} {\bibfnamefont {J.~M.}\ \bibnamefont {No}},\ }\href
  {\doibase 10.1088/1475-7516/2020/07/050} {\bibfield  {journal} {\bibinfo
  {journal} {JCAP}\ }\textbf {\bibinfo {volume} {07}},\ \bibinfo {pages} {050}
  (\bibinfo {year} {2020})},\ \Eprint {http://arxiv.org/abs/2003.07360}
  {arXiv:2003.07360 [hep-ph]} \BibitemShut {NoStop}%
\bibitem [{\citenamefont {Schicho}\ \emph {et~al.}(2021)\citenamefont
  {Schicho}, \citenamefont {Tenkanen},\ and\ \citenamefont
  {\"Osterman}}]{Schicho:2021gca}%
  \BibitemOpen
  \bibfield  {author} {\bibinfo {author} {\bibfnamefont {P.~M.}\ \bibnamefont
  {Schicho}}, \bibinfo {author} {\bibfnamefont {T.~V.~I.}\ \bibnamefont
  {Tenkanen}}, \ and\ \bibinfo {author} {\bibfnamefont {J.}~\bibnamefont
  {\"Osterman}},\ }\href {\doibase 10.1007/JHEP06(2021)130} {\bibfield
  {journal} {\bibinfo  {journal} {JHEP}\ }\textbf {\bibinfo {volume} {06}},\
  \bibinfo {pages} {130} (\bibinfo {year} {2021})},\ \Eprint
  {http://arxiv.org/abs/2102.11145} {arXiv:2102.11145 [hep-ph]} \BibitemShut
  {NoStop}%
\bibitem [{\citenamefont {Niemi}\ \emph
  {et~al.}(2021{\natexlab{a}})\citenamefont {Niemi}, \citenamefont {Schicho},\
  and\ \citenamefont {Tenkanen}}]{Niemi:2021qvp}%
  \BibitemOpen
  \bibfield  {author} {\bibinfo {author} {\bibfnamefont {L.}~\bibnamefont
  {Niemi}}, \bibinfo {author} {\bibfnamefont {P.}~\bibnamefont {Schicho}}, \
  and\ \bibinfo {author} {\bibfnamefont {T.~V.~I.}\ \bibnamefont {Tenkanen}},\
  }\href {\doibase 10.1103/PhysRevD.103.115035} {\bibfield  {journal} {\bibinfo
   {journal} {Phys. Rev. D}\ }\textbf {\bibinfo {volume} {103}},\ \bibinfo
  {pages} {115035} (\bibinfo {year} {2021}{\natexlab{a}})},\ \bibinfo {note}
  {[Erratum: Phys.Rev.D 109, 039902 (2024)]},\ \Eprint
  {http://arxiv.org/abs/2103.07467} {arXiv:2103.07467 [hep-ph]} \BibitemShut
  {NoStop}%
\bibitem [{\citenamefont {Gould}\ and\ \citenamefont
  {Tenkanen}(2021)}]{Gould:2021oba}%
  \BibitemOpen
  \bibfield  {author} {\bibinfo {author} {\bibfnamefont {O.}~\bibnamefont
  {Gould}}\ and\ \bibinfo {author} {\bibfnamefont {T.~V.~I.}\ \bibnamefont
  {Tenkanen}},\ }\href {\doibase 10.1007/JHEP06(2021)069} {\bibfield  {journal}
  {\bibinfo  {journal} {JHEP}\ }\textbf {\bibinfo {volume} {06}},\ \bibinfo
  {pages} {069} (\bibinfo {year} {2021})},\ \Eprint
  {http://arxiv.org/abs/2104.04399} {arXiv:2104.04399 [hep-ph]} \BibitemShut
  {NoStop}%
\bibitem [{\citenamefont {Ashoorioon}\ and\ \citenamefont
  {Konstandin}(2009)}]{Ashoorioon:2009nf}%
  \BibitemOpen
  \bibfield  {author} {\bibinfo {author} {\bibfnamefont {A.}~\bibnamefont
  {Ashoorioon}}\ and\ \bibinfo {author} {\bibfnamefont {T.}~\bibnamefont
  {Konstandin}},\ }\href {\doibase 10.1088/1126-6708/2009/07/086} {\bibfield
  {journal} {\bibinfo  {journal} {JHEP}\ }\textbf {\bibinfo {volume} {07}},\
  \bibinfo {pages} {086} (\bibinfo {year} {2009})},\ \Eprint
  {http://arxiv.org/abs/0904.0353} {arXiv:0904.0353 [hep-ph]} \BibitemShut
  {NoStop}%
\bibitem [{\citenamefont {Espinosa}\ \emph {et~al.}(2012)\citenamefont
  {Espinosa}, \citenamefont {Konstandin},\ and\ \citenamefont
  {Riva}}]{Espinosa:2011ax}%
  \BibitemOpen
  \bibfield  {author} {\bibinfo {author} {\bibfnamefont {J.~R.}\ \bibnamefont
  {Espinosa}}, \bibinfo {author} {\bibfnamefont {T.}~\bibnamefont
  {Konstandin}}, \ and\ \bibinfo {author} {\bibfnamefont {F.}~\bibnamefont
  {Riva}},\ }\href {\doibase 10.1016/j.nuclphysb.2011.09.010} {\bibfield
  {journal} {\bibinfo  {journal} {Nucl. Phys. B}\ }\textbf {\bibinfo {volume}
  {854}},\ \bibinfo {pages} {592} (\bibinfo {year} {2012})},\ \Eprint
  {http://arxiv.org/abs/1107.5441} {arXiv:1107.5441 [hep-ph]} \BibitemShut
  {NoStop}%
\bibitem [{\citenamefont {Profumo}\ \emph {et~al.}(2015)\citenamefont
  {Profumo}, \citenamefont {Ramsey-Musolf}, \citenamefont {Wainwright},\ and\
  \citenamefont {Winslow}}]{Profumo:2014opa}%
  \BibitemOpen
  \bibfield  {author} {\bibinfo {author} {\bibfnamefont {S.}~\bibnamefont
  {Profumo}}, \bibinfo {author} {\bibfnamefont {M.~J.}\ \bibnamefont
  {Ramsey-Musolf}}, \bibinfo {author} {\bibfnamefont {C.~L.}\ \bibnamefont
  {Wainwright}}, \ and\ \bibinfo {author} {\bibfnamefont {P.}~\bibnamefont
  {Winslow}},\ }\href {\doibase 10.1103/PhysRevD.91.035018} {\bibfield
  {journal} {\bibinfo  {journal} {Phys. Rev. D}\ }\textbf {\bibinfo {volume}
  {91}},\ \bibinfo {pages} {035018} (\bibinfo {year} {2015})},\ \Eprint
  {http://arxiv.org/abs/1407.5342} {arXiv:1407.5342 [hep-ph]} \BibitemShut
  {NoStop}%
\bibitem [{\citenamefont {Brauner}\ \emph {et~al.}(2017)\citenamefont
  {Brauner}, \citenamefont {Tenkanen}, \citenamefont {Tranberg}, \citenamefont
  {Vuorinen},\ and\ \citenamefont {Weir}}]{Brauner:2016fla}%
  \BibitemOpen
  \bibfield  {author} {\bibinfo {author} {\bibfnamefont {T.}~\bibnamefont
  {Brauner}}, \bibinfo {author} {\bibfnamefont {T.~V.~I.}\ \bibnamefont
  {Tenkanen}}, \bibinfo {author} {\bibfnamefont {A.}~\bibnamefont {Tranberg}},
  \bibinfo {author} {\bibfnamefont {A.}~\bibnamefont {Vuorinen}}, \ and\
  \bibinfo {author} {\bibfnamefont {D.~J.}\ \bibnamefont {Weir}},\ }\href
  {\doibase 10.1007/JHEP03(2017)007} {\bibfield  {journal} {\bibinfo  {journal}
  {JHEP}\ }\textbf {\bibinfo {volume} {03}},\ \bibinfo {pages} {007} (\bibinfo
  {year} {2017})},\ \Eprint {http://arxiv.org/abs/1609.06230} {arXiv:1609.06230
  [hep-ph]} \BibitemShut {NoStop}%
\bibitem [{\citenamefont {Sagunski}\ \emph {et~al.}(2023)\citenamefont
  {Sagunski}, \citenamefont {Schicho},\ and\ \citenamefont
  {Schmitt}}]{Sagunski:2023ynd}%
  \BibitemOpen
  \bibfield  {author} {\bibinfo {author} {\bibfnamefont {L.}~\bibnamefont
  {Sagunski}}, \bibinfo {author} {\bibfnamefont {P.}~\bibnamefont {Schicho}}, \
  and\ \bibinfo {author} {\bibfnamefont {D.}~\bibnamefont {Schmitt}},\ }\href
  {\doibase 10.1103/PhysRevD.107.123512} {\bibfield  {journal} {\bibinfo
  {journal} {Phys. Rev. D}\ }\textbf {\bibinfo {volume} {107}},\ \bibinfo
  {pages} {123512} (\bibinfo {year} {2023})},\ \Eprint
  {http://arxiv.org/abs/2303.02450} {arXiv:2303.02450 [hep-ph]} \BibitemShut
  {NoStop}%
\bibitem [{\citenamefont {Martin}\ and\ \citenamefont
  {Patel}(2018)}]{Martin:2018emo}%
  \BibitemOpen
  \bibfield  {author} {\bibinfo {author} {\bibfnamefont {S.~P.}\ \bibnamefont
  {Martin}}\ and\ \bibinfo {author} {\bibfnamefont {H.~H.}\ \bibnamefont
  {Patel}},\ }\href {\doibase 10.1103/PhysRevD.98.076008} {\bibfield  {journal}
  {\bibinfo  {journal} {Phys. Rev. D}\ }\textbf {\bibinfo {volume} {98}},\
  \bibinfo {pages} {076008} (\bibinfo {year} {2018})},\ \Eprint
  {http://arxiv.org/abs/1808.07615} {arXiv:1808.07615 [hep-ph]} \BibitemShut
  {NoStop}%
\bibitem [{\citenamefont {Linde}(1980)}]{Linde:1980ts}%
  \BibitemOpen
  \bibfield  {author} {\bibinfo {author} {\bibfnamefont {A.~D.}\ \bibnamefont
  {Linde}},\ }\href {\doibase 10.1016/0370-2693(80)90769-8} {\bibfield
  {journal} {\bibinfo  {journal} {Phys. Lett. B}\ }\textbf {\bibinfo {volume}
  {96}},\ \bibinfo {pages} {289} (\bibinfo {year} {1980})}\BibitemShut
  {NoStop}%
\bibitem [{\citenamefont {L\"ofgren}(2023)}]{Lofgren:2023sep}%
  \BibitemOpen
  \bibfield  {author} {\bibinfo {author} {\bibfnamefont {J.}~\bibnamefont
  {L\"ofgren}},\ }\href {\doibase 10.1088/1361-6471/ad074b} {\bibfield
  {journal} {\bibinfo  {journal} {J. Phys. G}\ }\textbf {\bibinfo {volume}
  {50}},\ \bibinfo {pages} {125008} (\bibinfo {year} {2023})},\ \Eprint
  {http://arxiv.org/abs/2301.05197} {arXiv:2301.05197 [hep-ph]} \BibitemShut
  {NoStop}%
\bibitem [{\citenamefont {Ekstedt}\ \emph
  {et~al.}(2023{\natexlab{a}})\citenamefont {Ekstedt}, \citenamefont
  {Schicho},\ and\ \citenamefont {Tenkanen}}]{Ekstedt:2022bff}%
  \BibitemOpen
  \bibfield  {author} {\bibinfo {author} {\bibfnamefont {A.}~\bibnamefont
  {Ekstedt}}, \bibinfo {author} {\bibfnamefont {P.}~\bibnamefont {Schicho}}, \
  and\ \bibinfo {author} {\bibfnamefont {T.~V.~I.}\ \bibnamefont {Tenkanen}},\
  }\href {\doibase 10.1016/j.cpc.2023.108725} {\bibfield  {journal} {\bibinfo
  {journal} {Comput. Phys. Commun.}\ }\textbf {\bibinfo {volume} {288}},\
  \bibinfo {pages} {108725} (\bibinfo {year} {2023}{\natexlab{a}})},\ \Eprint
  {http://arxiv.org/abs/2205.08815} {arXiv:2205.08815 [hep-ph]} \BibitemShut
  {NoStop}%
\bibitem [{\citenamefont {Coleman}\ and\ \citenamefont
  {Weinberg}(1973)}]{Coleman:1973jx}%
  \BibitemOpen
  \bibfield  {author} {\bibinfo {author} {\bibfnamefont {S.~R.}\ \bibnamefont
  {Coleman}}\ and\ \bibinfo {author} {\bibfnamefont {E.~J.}\ \bibnamefont
  {Weinberg}},\ }\href {\doibase 10.1103/PhysRevD.7.1888} {\bibfield  {journal}
  {\bibinfo  {journal} {Phys. Rev. D}\ }\textbf {\bibinfo {volume} {7}},\
  \bibinfo {pages} {1888} (\bibinfo {year} {1973})}\BibitemShut {NoStop}%
\bibitem [{\citenamefont {Jackiw}(1974)}]{Jackiw:1974cv}%
  \BibitemOpen
  \bibfield  {author} {\bibinfo {author} {\bibfnamefont {R.}~\bibnamefont
  {Jackiw}},\ }\href {\doibase 10.1103/PhysRevD.9.1686} {\bibfield  {journal}
  {\bibinfo  {journal} {Phys. Rev. D}\ }\textbf {\bibinfo {volume} {9}},\
  \bibinfo {pages} {1686} (\bibinfo {year} {1974})}\BibitemShut {NoStop}%
\bibitem [{\citenamefont {Anderson}\ and\ \citenamefont
  {Hall}(1992)}]{Anderson:1991zb}%
  \BibitemOpen
  \bibfield  {author} {\bibinfo {author} {\bibfnamefont {G.~W.}\ \bibnamefont
  {Anderson}}\ and\ \bibinfo {author} {\bibfnamefont {L.~J.}\ \bibnamefont
  {Hall}},\ }\href {\doibase 10.1103/PhysRevD.45.2685} {\bibfield  {journal}
  {\bibinfo  {journal} {Phys. Rev. D}\ }\textbf {\bibinfo {volume} {45}},\
  \bibinfo {pages} {2685} (\bibinfo {year} {1992})}\BibitemShut {NoStop}%
\bibitem [{\citenamefont {Curtin}\ \emph {et~al.}(2014)\citenamefont {Curtin},
  \citenamefont {Meade},\ and\ \citenamefont {Yu}}]{Curtin:2014jma}%
  \BibitemOpen
  \bibfield  {author} {\bibinfo {author} {\bibfnamefont {D.}~\bibnamefont
  {Curtin}}, \bibinfo {author} {\bibfnamefont {P.}~\bibnamefont {Meade}}, \
  and\ \bibinfo {author} {\bibfnamefont {C.-T.}\ \bibnamefont {Yu}},\ }\href
  {\doibase 10.1007/JHEP11(2014)127} {\bibfield  {journal} {\bibinfo  {journal}
  {JHEP}\ }\textbf {\bibinfo {volume} {11}},\ \bibinfo {pages} {127} (\bibinfo
  {year} {2014})},\ \Eprint {http://arxiv.org/abs/1409.0005} {arXiv:1409.0005
  [hep-ph]} \BibitemShut {NoStop}%
\bibitem [{\citenamefont {Quiros}(1999)}]{Quiros:1999jp}%
  \BibitemOpen
  \bibfield  {author} {\bibinfo {author} {\bibfnamefont {M.}~\bibnamefont
  {Quiros}},\ }in\ \href@noop {} {\emph {\bibinfo {booktitle} {{ICTP Summer
  School in High-Energy Physics and Cosmology}}}}\ (\bibinfo {year} {1999})\
  pp.\ \bibinfo {pages} {187--259},\ \Eprint
  {http://arxiv.org/abs/hep-ph/9901312} {arXiv:hep-ph/9901312} \BibitemShut
  {NoStop}%
\bibitem [{\citenamefont {Parwani}(1992)}]{Parwani:1991gq}%
  \BibitemOpen
  \bibfield  {author} {\bibinfo {author} {\bibfnamefont {R.~R.}\ \bibnamefont
  {Parwani}},\ }\href {\doibase 10.1103/PhysRevD.45.4695} {\bibfield  {journal}
  {\bibinfo  {journal} {Phys. Rev. D}\ }\textbf {\bibinfo {volume} {45}},\
  \bibinfo {pages} {4695} (\bibinfo {year} {1992})},\ \bibinfo {note}
  {[Erratum: Phys.Rev.D 48, 5965 (1993)]},\ \Eprint
  {http://arxiv.org/abs/hep-ph/9204216} {arXiv:hep-ph/9204216} \BibitemShut
  {NoStop}%
\bibitem [{\citenamefont {Curtin}\ \emph {et~al.}(2018)\citenamefont {Curtin},
  \citenamefont {Meade},\ and\ \citenamefont {Ramani}}]{Curtin:2016urg}%
  \BibitemOpen
  \bibfield  {author} {\bibinfo {author} {\bibfnamefont {D.}~\bibnamefont
  {Curtin}}, \bibinfo {author} {\bibfnamefont {P.}~\bibnamefont {Meade}}, \
  and\ \bibinfo {author} {\bibfnamefont {H.}~\bibnamefont {Ramani}},\ }\href
  {\doibase 10.1140/epjc/s10052-018-6268-0} {\bibfield  {journal} {\bibinfo
  {journal} {Eur. Phys. J. C}\ }\textbf {\bibinfo {volume} {78}},\ \bibinfo
  {pages} {787} (\bibinfo {year} {2018})},\ \Eprint
  {http://arxiv.org/abs/1612.00466} {arXiv:1612.00466 [hep-ph]} \BibitemShut
  {NoStop}%
\bibitem [{\citenamefont {Kajantie}\ \emph
  {et~al.}(1996{\natexlab{b}})\citenamefont {Kajantie}, \citenamefont {Laine},
  \citenamefont {Rummukainen},\ and\ \citenamefont
  {Shaposhnikov}}]{Kajantie:1995dw}%
  \BibitemOpen
  \bibfield  {author} {\bibinfo {author} {\bibfnamefont {K.}~\bibnamefont
  {Kajantie}}, \bibinfo {author} {\bibfnamefont {M.}~\bibnamefont {Laine}},
  \bibinfo {author} {\bibfnamefont {K.}~\bibnamefont {Rummukainen}}, \ and\
  \bibinfo {author} {\bibfnamefont {M.~E.}\ \bibnamefont {Shaposhnikov}},\
  }\href {\doibase 10.1016/0550-3213(95)00549-8} {\bibfield  {journal}
  {\bibinfo  {journal} {Nucl. Phys. B}\ }\textbf {\bibinfo {volume} {458}},\
  \bibinfo {pages} {90} (\bibinfo {year} {1996}{\natexlab{b}})},\ \Eprint
  {http://arxiv.org/abs/hep-ph/9508379} {arXiv:hep-ph/9508379} \BibitemShut
  {NoStop}%
\bibitem [{\citenamefont {Ruijl}\ \emph {et~al.}(2017)\citenamefont {Ruijl},
  \citenamefont {Ueda},\ and\ \citenamefont {Vermaseren}}]{Ruijl:2017dtg}%
  \BibitemOpen
  \bibfield  {author} {\bibinfo {author} {\bibfnamefont {B.}~\bibnamefont
  {Ruijl}}, \bibinfo {author} {\bibfnamefont {T.}~\bibnamefont {Ueda}}, \ and\
  \bibinfo {author} {\bibfnamefont {J.}~\bibnamefont {Vermaseren}},\
  }\href@noop {} {\  (\bibinfo {year} {2017})},\ \Eprint
  {http://arxiv.org/abs/1707.06453} {arXiv:1707.06453 [hep-ph]} \BibitemShut
  {NoStop}%
\bibitem [{\citenamefont {Fonseca}(2021)}]{Fonseca:2020vke}%
  \BibitemOpen
  \bibfield  {author} {\bibinfo {author} {\bibfnamefont {R.~M.}\ \bibnamefont
  {Fonseca}},\ }\href {\doibase 10.1016/j.cpc.2021.108085} {\bibfield
  {journal} {\bibinfo  {journal} {Comput. Phys. Commun.}\ }\textbf {\bibinfo
  {volume} {267}},\ \bibinfo {pages} {108085} (\bibinfo {year} {2021})},\
  \Eprint {http://arxiv.org/abs/2011.01764} {arXiv:2011.01764 [hep-th]}
  \BibitemShut {NoStop}%
\bibitem [{\citenamefont {Weinberg}\ and\ \citenamefont
  {Wu}(1987)}]{Weinberg:1987vp}%
  \BibitemOpen
  \bibfield  {author} {\bibinfo {author} {\bibfnamefont {E.~J.}\ \bibnamefont
  {Weinberg}}\ and\ \bibinfo {author} {\bibfnamefont {A.-q.}\ \bibnamefont
  {Wu}},\ }\href {\doibase 10.1103/PhysRevD.36.2474} {\bibfield  {journal}
  {\bibinfo  {journal} {Phys. Rev. D}\ }\textbf {\bibinfo {volume} {36}},\
  \bibinfo {pages} {2474} (\bibinfo {year} {1987})}\BibitemShut {NoStop}%
\bibitem [{\citenamefont {Laine}(1995)}]{Laine:1994zq}%
  \BibitemOpen
  \bibfield  {author} {\bibinfo {author} {\bibfnamefont {M.}~\bibnamefont
  {Laine}},\ }\href {\doibase 10.1103/PhysRevD.51.4525} {\bibfield  {journal}
  {\bibinfo  {journal} {Phys. Rev. D}\ }\textbf {\bibinfo {volume} {51}},\
  \bibinfo {pages} {4525} (\bibinfo {year} {1995})},\ \Eprint
  {http://arxiv.org/abs/hep-ph/9411252} {arXiv:hep-ph/9411252} \BibitemShut
  {NoStop}%
\bibitem [{\citenamefont {Wainwright}\ \emph {et~al.}(2012)\citenamefont
  {Wainwright}, \citenamefont {Profumo},\ and\ \citenamefont
  {Ramsey-Musolf}}]{Wainwright:2012zn}%
  \BibitemOpen
  \bibfield  {author} {\bibinfo {author} {\bibfnamefont {C.~L.}\ \bibnamefont
  {Wainwright}}, \bibinfo {author} {\bibfnamefont {S.}~\bibnamefont {Profumo}},
  \ and\ \bibinfo {author} {\bibfnamefont {M.~J.}\ \bibnamefont
  {Ramsey-Musolf}},\ }\href {\doibase 10.1103/PhysRevD.86.083537} {\bibfield
  {journal} {\bibinfo  {journal} {Phys. Rev. D}\ }\textbf {\bibinfo {volume}
  {86}},\ \bibinfo {pages} {083537} (\bibinfo {year} {2012})},\ \Eprint
  {http://arxiv.org/abs/1204.5464} {arXiv:1204.5464 [hep-ph]} \BibitemShut
  {NoStop}%
\bibitem [{\citenamefont {Fukuda}\ and\ \citenamefont
  {Kugo}(1976)}]{Fukuda:1975di}%
  \BibitemOpen
  \bibfield  {author} {\bibinfo {author} {\bibfnamefont {R.}~\bibnamefont
  {Fukuda}}\ and\ \bibinfo {author} {\bibfnamefont {T.}~\bibnamefont {Kugo}},\
  }\href {\doibase 10.1103/PhysRevD.13.3469} {\bibfield  {journal} {\bibinfo
  {journal} {Phys. Rev. D}\ }\textbf {\bibinfo {volume} {13}},\ \bibinfo
  {pages} {3469} (\bibinfo {year} {1976})}\BibitemShut {NoStop}%
\bibitem [{\citenamefont {Ekstedt}\ \emph {et~al.}(2022)\citenamefont
  {Ekstedt}, \citenamefont {Gould},\ and\ \citenamefont
  {L\"ofgren}}]{Ekstedt:2022zro}%
  \BibitemOpen
  \bibfield  {author} {\bibinfo {author} {\bibfnamefont {A.}~\bibnamefont
  {Ekstedt}}, \bibinfo {author} {\bibfnamefont {O.}~\bibnamefont {Gould}}, \
  and\ \bibinfo {author} {\bibfnamefont {J.}~\bibnamefont {L\"ofgren}},\ }\href
  {\doibase 10.1103/PhysRevD.106.036012} {\bibfield  {journal} {\bibinfo
  {journal} {Phys. Rev. D}\ }\textbf {\bibinfo {volume} {106}},\ \bibinfo
  {pages} {036012} (\bibinfo {year} {2022})},\ \Eprint
  {http://arxiv.org/abs/2205.07241} {arXiv:2205.07241 [hep-ph]} \BibitemShut
  {NoStop}%
\bibitem [{\citenamefont {Niemi}\ \emph
  {et~al.}(2021{\natexlab{b}})\citenamefont {Niemi}, \citenamefont
  {Ramsey-Musolf}, \citenamefont {Tenkanen},\ and\ \citenamefont
  {Weir}}]{Niemi:2020hto}%
  \BibitemOpen
  \bibfield  {author} {\bibinfo {author} {\bibfnamefont {L.}~\bibnamefont
  {Niemi}}, \bibinfo {author} {\bibfnamefont {M.~J.}\ \bibnamefont
  {Ramsey-Musolf}}, \bibinfo {author} {\bibfnamefont {T.~V.~I.}\ \bibnamefont
  {Tenkanen}}, \ and\ \bibinfo {author} {\bibfnamefont {D.~J.}\ \bibnamefont
  {Weir}},\ }\href {\doibase 10.1103/PhysRevLett.126.171802} {\bibfield
  {journal} {\bibinfo  {journal} {Phys. Rev. Lett.}\ }\textbf {\bibinfo
  {volume} {126}},\ \bibinfo {pages} {171802} (\bibinfo {year}
  {2021}{\natexlab{b}})},\ \Eprint {http://arxiv.org/abs/2005.11332}
  {arXiv:2005.11332 [hep-ph]} \BibitemShut {NoStop}%
\bibitem [{\citenamefont {Niemi}\ \emph {et~al.}(2019)\citenamefont {Niemi},
  \citenamefont {Patel}, \citenamefont {Ramsey-Musolf}, \citenamefont
  {Tenkanen},\ and\ \citenamefont {Weir}}]{Niemi:2018asa}%
  \BibitemOpen
  \bibfield  {author} {\bibinfo {author} {\bibfnamefont {L.}~\bibnamefont
  {Niemi}}, \bibinfo {author} {\bibfnamefont {H.~H.}\ \bibnamefont {Patel}},
  \bibinfo {author} {\bibfnamefont {M.~J.}\ \bibnamefont {Ramsey-Musolf}},
  \bibinfo {author} {\bibfnamefont {T.~V.~I.}\ \bibnamefont {Tenkanen}}, \ and\
  \bibinfo {author} {\bibfnamefont {D.~J.}\ \bibnamefont {Weir}},\ }\href
  {\doibase 10.1103/PhysRevD.100.035002} {\bibfield  {journal} {\bibinfo
  {journal} {Phys. Rev. D}\ }\textbf {\bibinfo {volume} {100}},\ \bibinfo
  {pages} {035002} (\bibinfo {year} {2019})},\ \Eprint
  {http://arxiv.org/abs/1802.10500} {arXiv:1802.10500 [hep-ph]} \BibitemShut
  {NoStop}%
\bibitem [{\citenamefont {Kierkla}\ \emph {et~al.}(2024)\citenamefont
  {Kierkla}, \citenamefont {Swiezewska}, \citenamefont {Tenkanen},\ and\
  \citenamefont {van~de Vis}}]{Kierkla:2023von}%
  \BibitemOpen
  \bibfield  {author} {\bibinfo {author} {\bibfnamefont {M.}~\bibnamefont
  {Kierkla}}, \bibinfo {author} {\bibfnamefont {B.}~\bibnamefont {Swiezewska}},
  \bibinfo {author} {\bibfnamefont {T.~V.~I.}\ \bibnamefont {Tenkanen}}, \ and\
  \bibinfo {author} {\bibfnamefont {J.}~\bibnamefont {van~de Vis}},\ }\href
  {\doibase 10.1007/JHEP02(2024)234} {\bibfield  {journal} {\bibinfo  {journal}
  {JHEP}\ }\textbf {\bibinfo {volume} {02}},\ \bibinfo {pages} {234} (\bibinfo
  {year} {2024})},\ \Eprint {http://arxiv.org/abs/2312.12413} {arXiv:2312.12413
  [hep-ph]} \BibitemShut {NoStop}%
\bibitem [{\citenamefont {Langer}(1969)}]{Langer:1969bc}%
  \BibitemOpen
  \bibfield  {author} {\bibinfo {author} {\bibfnamefont {J.~S.}\ \bibnamefont
  {Langer}},\ }\href {\doibase 10.1016/0003-4916(69)90153-5} {\bibfield
  {journal} {\bibinfo  {journal} {Annals Phys.}\ }\textbf {\bibinfo {volume}
  {54}},\ \bibinfo {pages} {258} (\bibinfo {year} {1969})}\BibitemShut
  {NoStop}%
\bibitem [{\citenamefont {Coleman}(1977)}]{Coleman:1977py}%
  \BibitemOpen
  \bibfield  {author} {\bibinfo {author} {\bibfnamefont {S.~R.}\ \bibnamefont
  {Coleman}},\ }\href {\doibase 10.1103/PhysRevD.16.1248} {\bibfield  {journal}
  {\bibinfo  {journal} {Phys. Rev. D}\ }\textbf {\bibinfo {volume} {15}},\
  \bibinfo {pages} {2929} (\bibinfo {year} {1977})},\ \bibinfo {note}
  {[Erratum: Phys.Rev.D 16, 1248 (1977)]}\BibitemShut {NoStop}%
\bibitem [{\citenamefont {Linde}(1981)}]{Linde:1980tt}%
  \BibitemOpen
  \bibfield  {author} {\bibinfo {author} {\bibfnamefont {A.~D.}\ \bibnamefont
  {Linde}},\ }\href {\doibase 10.1016/0370-2693(81)90281-1} {\bibfield
  {journal} {\bibinfo  {journal} {Phys. Lett. B}\ }\textbf {\bibinfo {volume}
  {100}},\ \bibinfo {pages} {37} (\bibinfo {year} {1981})}\BibitemShut
  {NoStop}%
\bibitem [{\citenamefont {Linde}(1983)}]{Linde:1981zj}%
  \BibitemOpen
  \bibfield  {author} {\bibinfo {author} {\bibfnamefont {A.~D.}\ \bibnamefont
  {Linde}},\ }\href {\doibase 10.1016/0550-3213(83)90072-X} {\bibfield
  {journal} {\bibinfo  {journal} {Nucl. Phys. B}\ }\textbf {\bibinfo {volume}
  {216}},\ \bibinfo {pages} {421} (\bibinfo {year} {1983})},\ \bibinfo {note}
  {[Erratum: Nucl.Phys.B 223, 544 (1983)]}\BibitemShut {NoStop}%
\bibitem [{\citenamefont {Affleck}(1981)}]{Affleck:1980ac}%
  \BibitemOpen
  \bibfield  {author} {\bibinfo {author} {\bibfnamefont {I.}~\bibnamefont
  {Affleck}},\ }\href {\doibase 10.1103/PhysRevLett.46.388} {\bibfield
  {journal} {\bibinfo  {journal} {Phys. Rev. Lett.}\ }\textbf {\bibinfo
  {volume} {46}},\ \bibinfo {pages} {388} (\bibinfo {year} {1981})}\BibitemShut
  {NoStop}%
\bibitem [{\citenamefont {Ekstedt}(2022{\natexlab{a}})}]{Ekstedt:2022tqk}%
  \BibitemOpen
  \bibfield  {author} {\bibinfo {author} {\bibfnamefont {A.}~\bibnamefont
  {Ekstedt}},\ }\href {\doibase 10.1007/JHEP08(2022)115} {\bibfield  {journal}
  {\bibinfo  {journal} {JHEP}\ }\textbf {\bibinfo {volume} {08}},\ \bibinfo
  {pages} {115} (\bibinfo {year} {2022}{\natexlab{a}})},\ \Eprint
  {http://arxiv.org/abs/2201.07331} {arXiv:2201.07331 [hep-ph]} \BibitemShut
  {NoStop}%
\bibitem [{\citenamefont {Ekstedt}(2022{\natexlab{b}})}]{Ekstedt:2021kyx}%
  \BibitemOpen
  \bibfield  {author} {\bibinfo {author} {\bibfnamefont {A.}~\bibnamefont
  {Ekstedt}},\ }\href {\doibase 10.1140/epjc/s10052-022-10130-5} {\bibfield
  {journal} {\bibinfo  {journal} {Eur. Phys. J. C}\ }\textbf {\bibinfo {volume}
  {82}},\ \bibinfo {pages} {173} (\bibinfo {year} {2022}{\natexlab{b}})},\
  \Eprint {http://arxiv.org/abs/2104.11804} {arXiv:2104.11804 [hep-ph]}
  \BibitemShut {NoStop}%
\bibitem [{\citenamefont {Gould}\ and\ \citenamefont
  {Hirvonen}(2021)}]{Gould:2021ccf}%
  \BibitemOpen
  \bibfield  {author} {\bibinfo {author} {\bibfnamefont {O.}~\bibnamefont
  {Gould}}\ and\ \bibinfo {author} {\bibfnamefont {J.}~\bibnamefont
  {Hirvonen}},\ }\href {\doibase 10.1103/PhysRevD.104.096015} {\bibfield
  {journal} {\bibinfo  {journal} {Phys. Rev. D}\ }\textbf {\bibinfo {volume}
  {104}},\ \bibinfo {pages} {096015} (\bibinfo {year} {2021})},\ \Eprint
  {http://arxiv.org/abs/2108.04377} {arXiv:2108.04377 [hep-ph]} \BibitemShut
  {NoStop}%
\bibitem [{\citenamefont {Ai}\ \emph {et~al.}(2024)\citenamefont {Ai},
  \citenamefont {Alexandre},\ and\ \citenamefont {Sarkar}}]{Ai:2023yce}%
  \BibitemOpen
  \bibfield  {author} {\bibinfo {author} {\bibfnamefont {W.-Y.}\ \bibnamefont
  {Ai}}, \bibinfo {author} {\bibfnamefont {J.}~\bibnamefont {Alexandre}}, \
  and\ \bibinfo {author} {\bibfnamefont {S.}~\bibnamefont {Sarkar}},\ }\href
  {\doibase 10.1103/PhysRevD.109.045010} {\bibfield  {journal} {\bibinfo
  {journal} {Phys. Rev. D}\ }\textbf {\bibinfo {volume} {109}},\ \bibinfo
  {pages} {045010} (\bibinfo {year} {2024})},\ \Eprint
  {http://arxiv.org/abs/2312.04482} {arXiv:2312.04482 [hep-ph]} \BibitemShut
  {NoStop}%
\bibitem [{\citenamefont {Ekstedt}\ \emph
  {et~al.}(2023{\natexlab{b}})\citenamefont {Ekstedt}, \citenamefont {Gould},\
  and\ \citenamefont {Hirvonen}}]{Ekstedt:2023sqc}%
  \BibitemOpen
  \bibfield  {author} {\bibinfo {author} {\bibfnamefont {A.}~\bibnamefont
  {Ekstedt}}, \bibinfo {author} {\bibfnamefont {O.}~\bibnamefont {Gould}}, \
  and\ \bibinfo {author} {\bibfnamefont {J.}~\bibnamefont {Hirvonen}},\ }\href
  {\doibase 10.1007/JHEP12(2023)056} {\bibfield  {journal} {\bibinfo  {journal}
  {JHEP}\ }\textbf {\bibinfo {volume} {12}},\ \bibinfo {pages} {056} (\bibinfo
  {year} {2023}{\natexlab{b}})},\ \Eprint {http://arxiv.org/abs/2308.15652}
  {arXiv:2308.15652 [hep-ph]} \BibitemShut {NoStop}%
\bibitem [{\citenamefont {L\"ofgren}\ \emph {et~al.}(2023)\citenamefont
  {L\"ofgren}, \citenamefont {Ramsey-Musolf}, \citenamefont {Schicho},\ and\
  \citenamefont {Tenkanen}}]{Lofgren:2021ogg}%
  \BibitemOpen
  \bibfield  {author} {\bibinfo {author} {\bibfnamefont {J.}~\bibnamefont
  {L\"ofgren}}, \bibinfo {author} {\bibfnamefont {M.~J.}\ \bibnamefont
  {Ramsey-Musolf}}, \bibinfo {author} {\bibfnamefont {P.}~\bibnamefont
  {Schicho}}, \ and\ \bibinfo {author} {\bibfnamefont {T.~V.~I.}\ \bibnamefont
  {Tenkanen}},\ }\href {\doibase 10.1103/PhysRevLett.130.251801} {\bibfield
  {journal} {\bibinfo  {journal} {Phys. Rev. Lett.}\ }\textbf {\bibinfo
  {volume} {130}},\ \bibinfo {pages} {251801} (\bibinfo {year} {2023})},\
  \Eprint {http://arxiv.org/abs/2112.05472} {arXiv:2112.05472 [hep-ph]}
  \BibitemShut {NoStop}%
\bibitem [{\citenamefont {Hirvonen}\ \emph {et~al.}(2022)\citenamefont
  {Hirvonen}, \citenamefont {L\"ofgren}, \citenamefont {Ramsey-Musolf},
  \citenamefont {Schicho},\ and\ \citenamefont {Tenkanen}}]{Hirvonen:2021zej}%
  \BibitemOpen
  \bibfield  {author} {\bibinfo {author} {\bibfnamefont {J.}~\bibnamefont
  {Hirvonen}}, \bibinfo {author} {\bibfnamefont {J.}~\bibnamefont {L\"ofgren}},
  \bibinfo {author} {\bibfnamefont {M.~J.}\ \bibnamefont {Ramsey-Musolf}},
  \bibinfo {author} {\bibfnamefont {P.}~\bibnamefont {Schicho}}, \ and\
  \bibinfo {author} {\bibfnamefont {T.~V.~I.}\ \bibnamefont {Tenkanen}},\
  }\href {\doibase 10.1007/JHEP07(2022)135} {\bibfield  {journal} {\bibinfo
  {journal} {JHEP}\ }\textbf {\bibinfo {volume} {07}},\ \bibinfo {pages} {135}
  (\bibinfo {year} {2022})},\ \Eprint {http://arxiv.org/abs/2112.08912}
  {arXiv:2112.08912 [hep-ph]} \BibitemShut {NoStop}%
\bibitem [{\citenamefont {Saikawa}\ and\ \citenamefont
  {Shirai}(2018)}]{Saikawa:2018rcs}%
  \BibitemOpen
  \bibfield  {author} {\bibinfo {author} {\bibfnamefont {K.}~\bibnamefont
  {Saikawa}}\ and\ \bibinfo {author} {\bibfnamefont {S.}~\bibnamefont
  {Shirai}},\ }\href {\doibase 10.1088/1475-7516/2018/05/035} {\bibfield
  {journal} {\bibinfo  {journal} {JCAP}\ }\textbf {\bibinfo {volume} {05}},\
  \bibinfo {pages} {035} (\bibinfo {year} {2018})},\ \Eprint
  {http://arxiv.org/abs/1803.01038} {arXiv:1803.01038 [hep-ph]} \BibitemShut
  {NoStop}%
\bibitem [{\citenamefont {Ellis}\ \emph
  {et~al.}(2019{\natexlab{a}})\citenamefont {Ellis}, \citenamefont {Lewicki},\
  and\ \citenamefont {No}}]{Ellis:2018mja}%
  \BibitemOpen
  \bibfield  {author} {\bibinfo {author} {\bibfnamefont {J.}~\bibnamefont
  {Ellis}}, \bibinfo {author} {\bibfnamefont {M.}~\bibnamefont {Lewicki}}, \
  and\ \bibinfo {author} {\bibfnamefont {J.~M.}\ \bibnamefont {No}},\ }\href
  {\doibase 10.1088/1475-7516/2019/04/003} {\bibfield  {journal} {\bibinfo
  {journal} {JCAP}\ }\textbf {\bibinfo {volume} {04}},\ \bibinfo {pages} {003}
  (\bibinfo {year} {2019}{\natexlab{a}})},\ \Eprint
  {http://arxiv.org/abs/1809.08242} {arXiv:1809.08242 [hep-ph]} \BibitemShut
  {NoStop}%
\bibitem [{\citenamefont {Hindmarsh}\ \emph {et~al.}(2017)\citenamefont
  {Hindmarsh}, \citenamefont {Huber}, \citenamefont {Rummukainen},\ and\
  \citenamefont {Weir}}]{Hindmarsh:2017gnf}%
  \BibitemOpen
  \bibfield  {author} {\bibinfo {author} {\bibfnamefont {M.}~\bibnamefont
  {Hindmarsh}}, \bibinfo {author} {\bibfnamefont {S.~J.}\ \bibnamefont
  {Huber}}, \bibinfo {author} {\bibfnamefont {K.}~\bibnamefont {Rummukainen}},
  \ and\ \bibinfo {author} {\bibfnamefont {D.~J.}\ \bibnamefont {Weir}},\
  }\href {\doibase 10.1103/PhysRevD.96.103520} {\bibfield  {journal} {\bibinfo
  {journal} {Phys. Rev. D}\ }\textbf {\bibinfo {volume} {96}},\ \bibinfo
  {pages} {103520} (\bibinfo {year} {2017})},\ \bibinfo {note} {[Erratum:
  Phys.Rev.D 101, 089902 (2020)]},\ \Eprint {http://arxiv.org/abs/1704.05871}
  {arXiv:1704.05871 [astro-ph.CO]} \BibitemShut {NoStop}%
\bibitem [{\citenamefont {Giese}\ \emph {et~al.}(2020)\citenamefont {Giese},
  \citenamefont {Konstandin},\ and\ \citenamefont {van~de
  Vis}}]{Giese:2020rtr}%
  \BibitemOpen
  \bibfield  {author} {\bibinfo {author} {\bibfnamefont {F.}~\bibnamefont
  {Giese}}, \bibinfo {author} {\bibfnamefont {T.}~\bibnamefont {Konstandin}}, \
  and\ \bibinfo {author} {\bibfnamefont {J.}~\bibnamefont {van~de Vis}},\
  }\href {\doibase 10.1088/1475-7516/2020/07/057} {\bibfield  {journal}
  {\bibinfo  {journal} {JCAP}\ }\textbf {\bibinfo {volume} {07}},\ \bibinfo
  {pages} {057} (\bibinfo {year} {2020})},\ \Eprint
  {http://arxiv.org/abs/2004.06995} {arXiv:2004.06995 [astro-ph.CO]}
  \BibitemShut {NoStop}%
\bibitem [{\citenamefont {Tenkanen}\ and\ \citenamefont {van~de
  Vis}(2022)}]{Tenkanen:2022tly}%
  \BibitemOpen
  \bibfield  {author} {\bibinfo {author} {\bibfnamefont {T.~V.~I.}\
  \bibnamefont {Tenkanen}}\ and\ \bibinfo {author} {\bibfnamefont
  {J.}~\bibnamefont {van~de Vis}},\ }\href {\doibase 10.1007/JHEP08(2022)302}
  {\bibfield  {journal} {\bibinfo  {journal} {JHEP}\ }\textbf {\bibinfo
  {volume} {08}},\ \bibinfo {pages} {302} (\bibinfo {year} {2022})},\ \Eprint
  {http://arxiv.org/abs/2206.01130} {arXiv:2206.01130 [hep-ph]} \BibitemShut
  {NoStop}%
\bibitem [{\citenamefont {Konstandin}\ \emph {et~al.}(2014)\citenamefont
  {Konstandin}, \citenamefont {Nardini},\ and\ \citenamefont
  {Rues}}]{Konstandin:2014zta}%
  \BibitemOpen
  \bibfield  {author} {\bibinfo {author} {\bibfnamefont {T.}~\bibnamefont
  {Konstandin}}, \bibinfo {author} {\bibfnamefont {G.}~\bibnamefont {Nardini}},
  \ and\ \bibinfo {author} {\bibfnamefont {I.}~\bibnamefont {Rues}},\ }\href
  {\doibase 10.1088/1475-7516/2014/09/028} {\bibfield  {journal} {\bibinfo
  {journal} {JCAP}\ }\textbf {\bibinfo {volume} {09}},\ \bibinfo {pages} {028}
  (\bibinfo {year} {2014})},\ \Eprint {http://arxiv.org/abs/1407.3132}
  {arXiv:1407.3132 [hep-ph]} \BibitemShut {NoStop}%
\bibitem [{\citenamefont {Kozaczuk}(2015)}]{Kozaczuk:2015owa}%
  \BibitemOpen
  \bibfield  {author} {\bibinfo {author} {\bibfnamefont {J.}~\bibnamefont
  {Kozaczuk}},\ }\href {\doibase 10.1007/JHEP10(2015)135} {\bibfield  {journal}
  {\bibinfo  {journal} {JHEP}\ }\textbf {\bibinfo {volume} {10}},\ \bibinfo
  {pages} {135} (\bibinfo {year} {2015})},\ \Eprint
  {http://arxiv.org/abs/1506.04741} {arXiv:1506.04741 [hep-ph]} \BibitemShut
  {NoStop}%
\bibitem [{\citenamefont {Friedlander}\ \emph {et~al.}(2021)\citenamefont
  {Friedlander}, \citenamefont {Banta}, \citenamefont {Cline},\ and\
  \citenamefont {Tucker-Smith}}]{Friedlander:2020tnq}%
  \BibitemOpen
  \bibfield  {author} {\bibinfo {author} {\bibfnamefont {A.}~\bibnamefont
  {Friedlander}}, \bibinfo {author} {\bibfnamefont {I.}~\bibnamefont {Banta}},
  \bibinfo {author} {\bibfnamefont {J.~M.}\ \bibnamefont {Cline}}, \ and\
  \bibinfo {author} {\bibfnamefont {D.}~\bibnamefont {Tucker-Smith}},\ }\href
  {\doibase 10.1103/PhysRevD.103.055020} {\bibfield  {journal} {\bibinfo
  {journal} {Phys. Rev. D}\ }\textbf {\bibinfo {volume} {103}},\ \bibinfo
  {pages} {055020} (\bibinfo {year} {2021})},\ \Eprint
  {http://arxiv.org/abs/2009.14295} {arXiv:2009.14295 [hep-ph]} \BibitemShut
  {NoStop}%
\bibitem [{\citenamefont {Moore}\ and\ \citenamefont
  {Prokopec}(1995{\natexlab{a}})}]{Moore:1995ua}%
  \BibitemOpen
  \bibfield  {author} {\bibinfo {author} {\bibfnamefont {G.~D.}\ \bibnamefont
  {Moore}}\ and\ \bibinfo {author} {\bibfnamefont {T.}~\bibnamefont
  {Prokopec}},\ }\href {\doibase 10.1103/PhysRevLett.75.777} {\bibfield
  {journal} {\bibinfo  {journal} {Phys. Rev. Lett.}\ }\textbf {\bibinfo
  {volume} {75}},\ \bibinfo {pages} {777} (\bibinfo {year}
  {1995}{\natexlab{a}})},\ \Eprint {http://arxiv.org/abs/hep-ph/9503296}
  {arXiv:hep-ph/9503296} \BibitemShut {NoStop}%
\bibitem [{\citenamefont {Moore}\ and\ \citenamefont
  {Prokopec}(1995{\natexlab{b}})}]{Moore:1995si}%
  \BibitemOpen
  \bibfield  {author} {\bibinfo {author} {\bibfnamefont {G.~D.}\ \bibnamefont
  {Moore}}\ and\ \bibinfo {author} {\bibfnamefont {T.}~\bibnamefont
  {Prokopec}},\ }\href {\doibase 10.1103/PhysRevD.52.7182} {\bibfield
  {journal} {\bibinfo  {journal} {Phys. Rev. D}\ }\textbf {\bibinfo {volume}
  {52}},\ \bibinfo {pages} {7182} (\bibinfo {year} {1995}{\natexlab{b}})},\
  \Eprint {http://arxiv.org/abs/hep-ph/9506475} {arXiv:hep-ph/9506475}
  \BibitemShut {NoStop}%
\bibitem [{\citenamefont {De~Curtis}\ \emph {et~al.}(2022)\citenamefont
  {De~Curtis}, \citenamefont {Rose}, \citenamefont {Guiggiani}, \citenamefont
  {Muyor},\ and\ \citenamefont {Panico}}]{DeCurtis:2022hlx}%
  \BibitemOpen
  \bibfield  {author} {\bibinfo {author} {\bibfnamefont {S.}~\bibnamefont
  {De~Curtis}}, \bibinfo {author} {\bibfnamefont {L.~D.}\ \bibnamefont {Rose}},
  \bibinfo {author} {\bibfnamefont {A.}~\bibnamefont {Guiggiani}}, \bibinfo
  {author} {\bibfnamefont {A.~G.}\ \bibnamefont {Muyor}}, \ and\ \bibinfo
  {author} {\bibfnamefont {G.}~\bibnamefont {Panico}},\ }\href {\doibase
  10.1007/JHEP03(2022)163} {\bibfield  {journal} {\bibinfo  {journal} {JHEP}\
  }\textbf {\bibinfo {volume} {03}},\ \bibinfo {pages} {163} (\bibinfo {year}
  {2022})},\ \Eprint {http://arxiv.org/abs/2201.08220} {arXiv:2201.08220
  [hep-ph]} \BibitemShut {NoStop}%
\bibitem [{\citenamefont {Dorsch}\ \emph {et~al.}(2022)\citenamefont {Dorsch},
  \citenamefont {Huber},\ and\ \citenamefont {Konstandin}}]{Dorsch:2021nje}%
  \BibitemOpen
  \bibfield  {author} {\bibinfo {author} {\bibfnamefont {G.~C.}\ \bibnamefont
  {Dorsch}}, \bibinfo {author} {\bibfnamefont {S.~J.}\ \bibnamefont {Huber}}, \
  and\ \bibinfo {author} {\bibfnamefont {T.}~\bibnamefont {Konstandin}},\
  }\href {\doibase 10.1088/1475-7516/2022/04/010} {\bibfield  {journal}
  {\bibinfo  {journal} {JCAP}\ }\textbf {\bibinfo {volume} {04}},\ \bibinfo
  {pages} {010} (\bibinfo {year} {2022})},\ \Eprint
  {http://arxiv.org/abs/2112.12548} {arXiv:2112.12548 [hep-ph]} \BibitemShut
  {NoStop}%
\bibitem [{\citenamefont {Bodeker}\ and\ \citenamefont
  {Moore}(2017)}]{Bodeker:2017cim}%
  \BibitemOpen
  \bibfield  {author} {\bibinfo {author} {\bibfnamefont {D.}~\bibnamefont
  {Bodeker}}\ and\ \bibinfo {author} {\bibfnamefont {G.~D.}\ \bibnamefont
  {Moore}},\ }\href {\doibase 10.1088/1475-7516/2017/05/025} {\bibfield
  {journal} {\bibinfo  {journal} {JCAP}\ }\textbf {\bibinfo {volume} {05}},\
  \bibinfo {pages} {025} (\bibinfo {year} {2017})},\ \Eprint
  {http://arxiv.org/abs/1703.08215} {arXiv:1703.08215 [hep-ph]} \BibitemShut
  {NoStop}%
\bibitem [{\citenamefont {Liu}\ \emph {et~al.}(1992)\citenamefont {Liu},
  \citenamefont {McLerran},\ and\ \citenamefont {Turok}}]{Liu:1992tn}%
  \BibitemOpen
  \bibfield  {author} {\bibinfo {author} {\bibfnamefont {B.-H.}\ \bibnamefont
  {Liu}}, \bibinfo {author} {\bibfnamefont {L.~D.}\ \bibnamefont {McLerran}}, \
  and\ \bibinfo {author} {\bibfnamefont {N.}~\bibnamefont {Turok}},\ }\href
  {\doibase 10.1103/PhysRevD.46.2668} {\bibfield  {journal} {\bibinfo
  {journal} {Phys. Rev. D}\ }\textbf {\bibinfo {volume} {46}},\ \bibinfo
  {pages} {2668} (\bibinfo {year} {1992})}\BibitemShut {NoStop}%
\bibitem [{\citenamefont {Cline}\ \emph {et~al.}(2021)\citenamefont {Cline},
  \citenamefont {Friedlander}, \citenamefont {He}, \citenamefont {Kainulainen},
  \citenamefont {Laurent},\ and\ \citenamefont {Tucker-Smith}}]{Cline:2021iff}%
  \BibitemOpen
  \bibfield  {author} {\bibinfo {author} {\bibfnamefont {J.~M.}\ \bibnamefont
  {Cline}}, \bibinfo {author} {\bibfnamefont {A.}~\bibnamefont {Friedlander}},
  \bibinfo {author} {\bibfnamefont {D.-M.}\ \bibnamefont {He}}, \bibinfo
  {author} {\bibfnamefont {K.}~\bibnamefont {Kainulainen}}, \bibinfo {author}
  {\bibfnamefont {B.}~\bibnamefont {Laurent}}, \ and\ \bibinfo {author}
  {\bibfnamefont {D.}~\bibnamefont {Tucker-Smith}},\ }\href {\doibase
  10.1103/PhysRevD.103.123529} {\bibfield  {journal} {\bibinfo  {journal}
  {Phys. Rev. D}\ }\textbf {\bibinfo {volume} {103}},\ \bibinfo {pages}
  {123529} (\bibinfo {year} {2021})},\ \Eprint
  {http://arxiv.org/abs/2102.12490} {arXiv:2102.12490 [hep-ph]} \BibitemShut
  {NoStop}%
\bibitem [{\citenamefont {Dine}\ \emph {et~al.}(1992)\citenamefont {Dine},
  \citenamefont {Leigh}, \citenamefont {Huet}, \citenamefont {Linde},\ and\
  \citenamefont {Linde}}]{Dine:1992wr}%
  \BibitemOpen
  \bibfield  {author} {\bibinfo {author} {\bibfnamefont {M.}~\bibnamefont
  {Dine}}, \bibinfo {author} {\bibfnamefont {R.~G.}\ \bibnamefont {Leigh}},
  \bibinfo {author} {\bibfnamefont {P.~Y.}\ \bibnamefont {Huet}}, \bibinfo
  {author} {\bibfnamefont {A.~D.}\ \bibnamefont {Linde}}, \ and\ \bibinfo
  {author} {\bibfnamefont {D.~A.}\ \bibnamefont {Linde}},\ }\href {\doibase
  10.1103/PhysRevD.46.550} {\bibfield  {journal} {\bibinfo  {journal} {Phys.
  Rev. D}\ }\textbf {\bibinfo {volume} {46}},\ \bibinfo {pages} {550} (\bibinfo
  {year} {1992})},\ \Eprint {http://arxiv.org/abs/hep-ph/9203203}
  {arXiv:hep-ph/9203203} \BibitemShut {NoStop}%
\bibitem [{\citenamefont {Laurent}\ and\ \citenamefont
  {Cline}(2020)}]{Laurent:2020gpg}%
  \BibitemOpen
  \bibfield  {author} {\bibinfo {author} {\bibfnamefont {B.}~\bibnamefont
  {Laurent}}\ and\ \bibinfo {author} {\bibfnamefont {J.~M.}\ \bibnamefont
  {Cline}},\ }\href {\doibase 10.1103/PhysRevD.102.063516} {\bibfield
  {journal} {\bibinfo  {journal} {Phys. Rev. D}\ }\textbf {\bibinfo {volume}
  {102}},\ \bibinfo {pages} {063516} (\bibinfo {year} {2020})},\ \Eprint
  {http://arxiv.org/abs/2007.10935} {arXiv:2007.10935 [hep-ph]} \BibitemShut
  {NoStop}%
\bibitem [{\citenamefont {Laurent}\ and\ \citenamefont
  {Cline}(2022)}]{Laurent:2022jrs}%
  \BibitemOpen
  \bibfield  {author} {\bibinfo {author} {\bibfnamefont {B.}~\bibnamefont
  {Laurent}}\ and\ \bibinfo {author} {\bibfnamefont {J.~M.}\ \bibnamefont
  {Cline}},\ }\href {\doibase 10.1103/PhysRevD.106.023501} {\bibfield
  {journal} {\bibinfo  {journal} {Phys. Rev. D}\ }\textbf {\bibinfo {volume}
  {106}},\ \bibinfo {pages} {023501} (\bibinfo {year} {2022})},\ \Eprint
  {http://arxiv.org/abs/2204.13120} {arXiv:2204.13120 [hep-ph]} \BibitemShut
  {NoStop}%
\bibitem [{\citenamefont {Ai}\ \emph {et~al.}(2023)\citenamefont {Ai},
  \citenamefont {Laurent},\ and\ \citenamefont {van~de Vis}}]{Ai:2023see}%
  \BibitemOpen
  \bibfield  {author} {\bibinfo {author} {\bibfnamefont {W.-Y.}\ \bibnamefont
  {Ai}}, \bibinfo {author} {\bibfnamefont {B.}~\bibnamefont {Laurent}}, \ and\
  \bibinfo {author} {\bibfnamefont {J.}~\bibnamefont {van~de Vis}},\ }\href
  {\doibase 10.1088/1475-7516/2023/07/002} {\bibfield  {journal} {\bibinfo
  {journal} {JCAP}\ }\textbf {\bibinfo {volume} {07}},\ \bibinfo {pages} {002}
  (\bibinfo {year} {2023})},\ \Eprint {http://arxiv.org/abs/2303.10171}
  {arXiv:2303.10171 [astro-ph.CO]} \BibitemShut {NoStop}%
\bibitem [{\citenamefont {Krajewski}\ \emph {et~al.}(2024)\citenamefont
  {Krajewski}, \citenamefont {Lewicki},\ and\ \citenamefont
  {Zych}}]{Krajewski:2024gma}%
  \BibitemOpen
  \bibfield  {author} {\bibinfo {author} {\bibfnamefont {T.}~\bibnamefont
  {Krajewski}}, \bibinfo {author} {\bibfnamefont {M.}~\bibnamefont {Lewicki}},
  \ and\ \bibinfo {author} {\bibfnamefont {M.}~\bibnamefont {Zych}},\ }\href
  {\doibase 10.1007/JHEP05(2024)011} {\bibfield  {journal} {\bibinfo  {journal}
  {JHEP}\ }\textbf {\bibinfo {volume} {05}},\ \bibinfo {pages} {011} (\bibinfo
  {year} {2024})},\ \Eprint {http://arxiv.org/abs/2402.15408} {arXiv:2402.15408
  [astro-ph.CO]} \BibitemShut {NoStop}%
\bibitem [{\citenamefont {Lewicki}\ \emph {et~al.}(2022)\citenamefont
  {Lewicki}, \citenamefont {Merchand},\ and\ \citenamefont
  {Zych}}]{Lewicki:2021pgr}%
  \BibitemOpen
  \bibfield  {author} {\bibinfo {author} {\bibfnamefont {M.}~\bibnamefont
  {Lewicki}}, \bibinfo {author} {\bibfnamefont {M.}~\bibnamefont {Merchand}}, \
  and\ \bibinfo {author} {\bibfnamefont {M.}~\bibnamefont {Zych}},\ }\href
  {\doibase 10.1007/JHEP02(2022)017} {\bibfield  {journal} {\bibinfo  {journal}
  {JHEP}\ }\textbf {\bibinfo {volume} {02}},\ \bibinfo {pages} {017} (\bibinfo
  {year} {2022})},\ \Eprint {http://arxiv.org/abs/2111.02393} {arXiv:2111.02393
  [astro-ph.CO]} \BibitemShut {NoStop}%
\bibitem [{\citenamefont {Wainwright}(2012)}]{Wainwright:2011kj}%
  \BibitemOpen
  \bibfield  {author} {\bibinfo {author} {\bibfnamefont {C.~L.}\ \bibnamefont
  {Wainwright}},\ }\href {\doibase 10.1016/j.cpc.2012.04.004} {\bibfield
  {journal} {\bibinfo  {journal} {Comput. Phys. Commun.}\ }\textbf {\bibinfo
  {volume} {183}},\ \bibinfo {pages} {2006} (\bibinfo {year} {2012})},\ \Eprint
  {http://arxiv.org/abs/1109.4189} {arXiv:1109.4189 [hep-ph]} \BibitemShut
  {NoStop}%
\bibitem [{\citenamefont {Schicho}\ and\ \citenamefont
  {Schmitt}(2024)}]{DRansitions}%
  \BibitemOpen
  \bibfield  {author} {\bibinfo {author} {\bibfnamefont {P.}~\bibnamefont
  {Schicho}}\ and\ \bibinfo {author} {\bibfnamefont {D.}~\bibnamefont
  {Schmitt}},\ }\href@noop {} {}\bibinfo {howpublished}
  {\url{https://github.com/DMGW-Goethe/DRansitions}} (\bibinfo {year}
  {2024})\BibitemShut {NoStop}%
\bibitem [{\citenamefont {Ellis}\ \emph
  {et~al.}(2019{\natexlab{b}})\citenamefont {Ellis}, \citenamefont {Lewicki},
  \citenamefont {No},\ and\ \citenamefont {Vaskonen}}]{Ellis:2019oqb}%
  \BibitemOpen
  \bibfield  {author} {\bibinfo {author} {\bibfnamefont {J.}~\bibnamefont
  {Ellis}}, \bibinfo {author} {\bibfnamefont {M.}~\bibnamefont {Lewicki}},
  \bibinfo {author} {\bibfnamefont {J.~M.}\ \bibnamefont {No}}, \ and\ \bibinfo
  {author} {\bibfnamefont {V.}~\bibnamefont {Vaskonen}},\ }\href {\doibase
  10.1088/1475-7516/2019/06/024} {\bibfield  {journal} {\bibinfo  {journal}
  {JCAP}\ }\textbf {\bibinfo {volume} {06}},\ \bibinfo {pages} {024} (\bibinfo
  {year} {2019}{\natexlab{b}})},\ \Eprint {http://arxiv.org/abs/1903.09642}
  {arXiv:1903.09642 [hep-ph]} \BibitemShut {NoStop}%
\bibitem [{\citenamefont {Lewicki}\ and\ \citenamefont
  {Vaskonen}(2020{\natexlab{a}})}]{Lewicki:2019gmv}%
  \BibitemOpen
  \bibfield  {author} {\bibinfo {author} {\bibfnamefont {M.}~\bibnamefont
  {Lewicki}}\ and\ \bibinfo {author} {\bibfnamefont {V.}~\bibnamefont
  {Vaskonen}},\ }\href {\doibase 10.1016/j.dark.2020.100672} {\bibfield
  {journal} {\bibinfo  {journal} {Phys. Dark Univ.}\ }\textbf {\bibinfo
  {volume} {30}},\ \bibinfo {pages} {100672} (\bibinfo {year}
  {2020}{\natexlab{a}})},\ \Eprint {http://arxiv.org/abs/1912.00997}
  {arXiv:1912.00997 [astro-ph.CO]} \BibitemShut {NoStop}%
\bibitem [{\citenamefont {Lewicki}\ and\ \citenamefont
  {Vaskonen}(2020{\natexlab{b}})}]{Lewicki:2020jiv}%
  \BibitemOpen
  \bibfield  {author} {\bibinfo {author} {\bibfnamefont {M.}~\bibnamefont
  {Lewicki}}\ and\ \bibinfo {author} {\bibfnamefont {V.}~\bibnamefont
  {Vaskonen}},\ }\href {\doibase 10.1140/epjc/s10052-020-08589-1} {\bibfield
  {journal} {\bibinfo  {journal} {Eur. Phys. J. C}\ }\textbf {\bibinfo {volume}
  {80}},\ \bibinfo {pages} {1003} (\bibinfo {year} {2020}{\natexlab{b}})},\
  \Eprint {http://arxiv.org/abs/2007.04967} {arXiv:2007.04967 [astro-ph.CO]}
  \BibitemShut {NoStop}%
\bibitem [{\citenamefont {Lewicki}\ and\ \citenamefont
  {Vaskonen}(2021)}]{Lewicki:2020azd}%
  \BibitemOpen
  \bibfield  {author} {\bibinfo {author} {\bibfnamefont {M.}~\bibnamefont
  {Lewicki}}\ and\ \bibinfo {author} {\bibfnamefont {V.}~\bibnamefont
  {Vaskonen}},\ }\href {\doibase 10.1140/epjc/s10052-021-09232-3} {\bibfield
  {journal} {\bibinfo  {journal} {Eur. Phys. J. C}\ }\textbf {\bibinfo {volume}
  {81}},\ \bibinfo {pages} {437} (\bibinfo {year} {2021})},\ \Eprint
  {http://arxiv.org/abs/2012.07826} {arXiv:2012.07826 [astro-ph.CO]}
  \BibitemShut {NoStop}%
\bibitem [{\citenamefont {Lewicki}\ and\ \citenamefont
  {Vaskonen}(2023{\natexlab{a}})}]{Lewicki:2022pdb}%
  \BibitemOpen
  \bibfield  {author} {\bibinfo {author} {\bibfnamefont {M.}~\bibnamefont
  {Lewicki}}\ and\ \bibinfo {author} {\bibfnamefont {V.}~\bibnamefont
  {Vaskonen}},\ }\href {\doibase 10.1140/epjc/s10052-023-11241-3} {\bibfield
  {journal} {\bibinfo  {journal} {Eur. Phys. J. C}\ }\textbf {\bibinfo {volume}
  {83}},\ \bibinfo {pages} {109} (\bibinfo {year} {2023}{\natexlab{a}})},\
  \Eprint {http://arxiv.org/abs/2208.11697} {arXiv:2208.11697 [astro-ph.CO]}
  \BibitemShut {NoStop}%
\bibitem [{\citenamefont {Roper~Pol}\ \emph {et~al.}(2020)\citenamefont
  {Roper~Pol}, \citenamefont {Mandal}, \citenamefont {Brandenburg},
  \citenamefont {Kahniashvili},\ and\ \citenamefont
  {Kosowsky}}]{RoperPol:2019wvy}%
  \BibitemOpen
  \bibfield  {author} {\bibinfo {author} {\bibfnamefont {A.}~\bibnamefont
  {Roper~Pol}}, \bibinfo {author} {\bibfnamefont {S.}~\bibnamefont {Mandal}},
  \bibinfo {author} {\bibfnamefont {A.}~\bibnamefont {Brandenburg}}, \bibinfo
  {author} {\bibfnamefont {T.}~\bibnamefont {Kahniashvili}}, \ and\ \bibinfo
  {author} {\bibfnamefont {A.}~\bibnamefont {Kosowsky}},\ }\href {\doibase
  10.1103/PhysRevD.102.083512} {\bibfield  {journal} {\bibinfo  {journal}
  {Phys. Rev. D}\ }\textbf {\bibinfo {volume} {102}},\ \bibinfo {pages}
  {083512} (\bibinfo {year} {2020})},\ \Eprint
  {http://arxiv.org/abs/1903.08585} {arXiv:1903.08585 [astro-ph.CO]}
  \BibitemShut {NoStop}%
\bibitem [{\citenamefont {Kahniashvili}\ \emph {et~al.}(2021)\citenamefont
  {Kahniashvili}, \citenamefont {Brandenburg}, \citenamefont {Gogoberidze},
  \citenamefont {Mandal},\ and\ \citenamefont
  {Roper~Pol}}]{Kahniashvili:2020jgm}%
  \BibitemOpen
  \bibfield  {author} {\bibinfo {author} {\bibfnamefont {T.}~\bibnamefont
  {Kahniashvili}}, \bibinfo {author} {\bibfnamefont {A.}~\bibnamefont
  {Brandenburg}}, \bibinfo {author} {\bibfnamefont {G.}~\bibnamefont
  {Gogoberidze}}, \bibinfo {author} {\bibfnamefont {S.}~\bibnamefont {Mandal}},
  \ and\ \bibinfo {author} {\bibfnamefont {A.}~\bibnamefont {Roper~Pol}},\
  }\href {\doibase 10.1103/PhysRevResearch.3.013193} {\bibfield  {journal}
  {\bibinfo  {journal} {Phys. Rev. Res.}\ }\textbf {\bibinfo {volume} {3}},\
  \bibinfo {pages} {013193} (\bibinfo {year} {2021})},\ \Eprint
  {http://arxiv.org/abs/2011.05556} {arXiv:2011.05556 [astro-ph.CO]}
  \BibitemShut {NoStop}%
\bibitem [{\citenamefont {Roper~Pol}\ \emph {et~al.}(2022)\citenamefont
  {Roper~Pol}, \citenamefont {Mandal}, \citenamefont {Brandenburg},\ and\
  \citenamefont {Kahniashvili}}]{RoperPol:2021xnd}%
  \BibitemOpen
  \bibfield  {author} {\bibinfo {author} {\bibfnamefont {A.}~\bibnamefont
  {Roper~Pol}}, \bibinfo {author} {\bibfnamefont {S.}~\bibnamefont {Mandal}},
  \bibinfo {author} {\bibfnamefont {A.}~\bibnamefont {Brandenburg}}, \ and\
  \bibinfo {author} {\bibfnamefont {T.}~\bibnamefont {Kahniashvili}},\ }\href
  {\doibase 10.1088/1475-7516/2022/04/019} {\bibfield  {journal} {\bibinfo
  {journal} {JCAP}\ }\textbf {\bibinfo {volume} {04}},\ \bibinfo {pages} {019}
  (\bibinfo {year} {2022})},\ \Eprint {http://arxiv.org/abs/2107.05356}
  {arXiv:2107.05356 [gr-qc]} \BibitemShut {NoStop}%
\bibitem [{\citenamefont {Hindmarsh}(2018)}]{Hindmarsh:2016lnk}%
  \BibitemOpen
  \bibfield  {author} {\bibinfo {author} {\bibfnamefont {M.}~\bibnamefont
  {Hindmarsh}},\ }\href {\doibase 10.1103/PhysRevLett.120.071301} {\bibfield
  {journal} {\bibinfo  {journal} {Phys. Rev. Lett.}\ }\textbf {\bibinfo
  {volume} {120}},\ \bibinfo {pages} {071301} (\bibinfo {year} {2018})},\
  \Eprint {http://arxiv.org/abs/1608.04735} {arXiv:1608.04735 [astro-ph.CO]}
  \BibitemShut {NoStop}%
\bibitem [{\citenamefont {Hindmarsh}\ and\ \citenamefont
  {Hijazi}(2019)}]{Hindmarsh:2019phv}%
  \BibitemOpen
  \bibfield  {author} {\bibinfo {author} {\bibfnamefont {M.}~\bibnamefont
  {Hindmarsh}}\ and\ \bibinfo {author} {\bibfnamefont {M.}~\bibnamefont
  {Hijazi}},\ }\href {\doibase 10.1088/1475-7516/2019/12/062} {\bibfield
  {journal} {\bibinfo  {journal} {JCAP}\ }\textbf {\bibinfo {volume} {12}},\
  \bibinfo {pages} {062} (\bibinfo {year} {2019})},\ \Eprint
  {http://arxiv.org/abs/1909.10040} {arXiv:1909.10040 [astro-ph.CO]}
  \BibitemShut {NoStop}%
\bibitem [{\citenamefont {Jinno}\ \emph {et~al.}(2021)\citenamefont {Jinno},
  \citenamefont {Konstandin},\ and\ \citenamefont {Rubira}}]{Jinno:2020eqg}%
  \BibitemOpen
  \bibfield  {author} {\bibinfo {author} {\bibfnamefont {R.}~\bibnamefont
  {Jinno}}, \bibinfo {author} {\bibfnamefont {T.}~\bibnamefont {Konstandin}}, \
  and\ \bibinfo {author} {\bibfnamefont {H.}~\bibnamefont {Rubira}},\ }\href
  {\doibase 10.1088/1475-7516/2021/04/014} {\bibfield  {journal} {\bibinfo
  {journal} {JCAP}\ }\textbf {\bibinfo {volume} {04}},\ \bibinfo {pages} {014}
  (\bibinfo {year} {2021})},\ \Eprint {http://arxiv.org/abs/2010.00971}
  {arXiv:2010.00971 [astro-ph.CO]} \BibitemShut {NoStop}%
\bibitem [{\citenamefont {Gowling}\ and\ \citenamefont
  {Hindmarsh}(2021)}]{Gowling:2021gcy}%
  \BibitemOpen
  \bibfield  {author} {\bibinfo {author} {\bibfnamefont {C.}~\bibnamefont
  {Gowling}}\ and\ \bibinfo {author} {\bibfnamefont {M.}~\bibnamefont
  {Hindmarsh}},\ }\href {\doibase 10.1088/1475-7516/2021/10/039} {\bibfield
  {journal} {\bibinfo  {journal} {JCAP}\ }\textbf {\bibinfo {volume} {10}},\
  \bibinfo {pages} {039} (\bibinfo {year} {2021})},\ \Eprint
  {http://arxiv.org/abs/2106.05984} {arXiv:2106.05984 [astro-ph.CO]}
  \BibitemShut {NoStop}%
\bibitem [{\citenamefont {Roper~Pol}\ \emph {et~al.}(2024)\citenamefont
  {Roper~Pol}, \citenamefont {Procacci},\ and\ \citenamefont
  {Caprini}}]{RoperPol:2023dzg}%
  \BibitemOpen
  \bibfield  {author} {\bibinfo {author} {\bibfnamefont {A.}~\bibnamefont
  {Roper~Pol}}, \bibinfo {author} {\bibfnamefont {S.}~\bibnamefont {Procacci}},
  \ and\ \bibinfo {author} {\bibfnamefont {C.}~\bibnamefont {Caprini}},\ }\href
  {\doibase 10.1103/PhysRevD.109.063531} {\bibfield  {journal} {\bibinfo
  {journal} {Phys. Rev. D}\ }\textbf {\bibinfo {volume} {109}},\ \bibinfo
  {pages} {063531} (\bibinfo {year} {2024})},\ \Eprint
  {http://arxiv.org/abs/2308.12943} {arXiv:2308.12943 [gr-qc]} \BibitemShut
  {NoStop}%
\bibitem [{\citenamefont {Sharma}\ \emph {et~al.}(2023)\citenamefont {Sharma},
  \citenamefont {Dahl}, \citenamefont {Brandenburg},\ and\ \citenamefont
  {Hindmarsh}}]{Sharma:2023mao}%
  \BibitemOpen
  \bibfield  {author} {\bibinfo {author} {\bibfnamefont {R.}~\bibnamefont
  {Sharma}}, \bibinfo {author} {\bibfnamefont {J.}~\bibnamefont {Dahl}},
  \bibinfo {author} {\bibfnamefont {A.}~\bibnamefont {Brandenburg}}, \ and\
  \bibinfo {author} {\bibfnamefont {M.}~\bibnamefont {Hindmarsh}},\ }\href
  {\doibase 10.1088/1475-7516/2023/12/042} {\bibfield  {journal} {\bibinfo
  {journal} {JCAP}\ }\textbf {\bibinfo {volume} {12}},\ \bibinfo {pages} {042}
  (\bibinfo {year} {2023})},\ \Eprint {http://arxiv.org/abs/2308.12916}
  {arXiv:2308.12916 [gr-qc]} \BibitemShut {NoStop}%
\bibitem [{\citenamefont {Ellis}\ \emph {et~al.}(2023)\citenamefont {Ellis},
  \citenamefont {Lewicki}, \citenamefont {Merchand}, \citenamefont {No},\ and\
  \citenamefont {Zych}}]{Ellis:2022lft}%
  \BibitemOpen
  \bibfield  {author} {\bibinfo {author} {\bibfnamefont {J.}~\bibnamefont
  {Ellis}}, \bibinfo {author} {\bibfnamefont {M.}~\bibnamefont {Lewicki}},
  \bibinfo {author} {\bibfnamefont {M.}~\bibnamefont {Merchand}}, \bibinfo
  {author} {\bibfnamefont {J.~M.}\ \bibnamefont {No}}, \ and\ \bibinfo {author}
  {\bibfnamefont {M.}~\bibnamefont {Zych}},\ }\href {\doibase
  10.1007/JHEP01(2023)093} {\bibfield  {journal} {\bibinfo  {journal} {JHEP}\
  }\textbf {\bibinfo {volume} {01}},\ \bibinfo {pages} {093} (\bibinfo {year}
  {2023})},\ \Eprint {http://arxiv.org/abs/2210.16305} {arXiv:2210.16305
  [hep-ph]} \BibitemShut {NoStop}%
\bibitem [{\citenamefont {Hindmarsh}\ \emph {et~al.}(2014)\citenamefont
  {Hindmarsh}, \citenamefont {Huber}, \citenamefont {Rummukainen},\ and\
  \citenamefont {Weir}}]{Hindmarsh:2013xza}%
  \BibitemOpen
  \bibfield  {author} {\bibinfo {author} {\bibfnamefont {M.}~\bibnamefont
  {Hindmarsh}}, \bibinfo {author} {\bibfnamefont {S.~J.}\ \bibnamefont
  {Huber}}, \bibinfo {author} {\bibfnamefont {K.}~\bibnamefont {Rummukainen}},
  \ and\ \bibinfo {author} {\bibfnamefont {D.~J.}\ \bibnamefont {Weir}},\
  }\href {\doibase 10.1103/PhysRevLett.112.041301} {\bibfield  {journal}
  {\bibinfo  {journal} {Phys. Rev. Lett.}\ }\textbf {\bibinfo {volume} {112}},\
  \bibinfo {pages} {041301} (\bibinfo {year} {2014})},\ \Eprint
  {http://arxiv.org/abs/1304.2433} {arXiv:1304.2433 [hep-ph]} \BibitemShut
  {NoStop}%
\bibitem [{\citenamefont {Hindmarsh}\ \emph {et~al.}(2015)\citenamefont
  {Hindmarsh}, \citenamefont {Huber}, \citenamefont {Rummukainen},\ and\
  \citenamefont {Weir}}]{Hindmarsh:2015qta}%
  \BibitemOpen
  \bibfield  {author} {\bibinfo {author} {\bibfnamefont {M.}~\bibnamefont
  {Hindmarsh}}, \bibinfo {author} {\bibfnamefont {S.~J.}\ \bibnamefont
  {Huber}}, \bibinfo {author} {\bibfnamefont {K.}~\bibnamefont {Rummukainen}},
  \ and\ \bibinfo {author} {\bibfnamefont {D.~J.}\ \bibnamefont {Weir}},\
  }\href {\doibase 10.1103/PhysRevD.92.123009} {\bibfield  {journal} {\bibinfo
  {journal} {Phys. Rev. D}\ }\textbf {\bibinfo {volume} {92}},\ \bibinfo
  {pages} {123009} (\bibinfo {year} {2015})},\ \Eprint
  {http://arxiv.org/abs/1504.03291} {arXiv:1504.03291 [astro-ph.CO]}
  \BibitemShut {NoStop}%
\bibitem [{\citenamefont {Guo}\ \emph {et~al.}(2021)\citenamefont {Guo},
  \citenamefont {Sinha}, \citenamefont {Vagie},\ and\ \citenamefont
  {White}}]{Guo:2020grp}%
  \BibitemOpen
  \bibfield  {author} {\bibinfo {author} {\bibfnamefont {H.-K.}\ \bibnamefont
  {Guo}}, \bibinfo {author} {\bibfnamefont {K.}~\bibnamefont {Sinha}}, \bibinfo
  {author} {\bibfnamefont {D.}~\bibnamefont {Vagie}}, \ and\ \bibinfo {author}
  {\bibfnamefont {G.}~\bibnamefont {White}},\ }\href {\doibase
  10.1088/1475-7516/2021/01/001} {\bibfield  {journal} {\bibinfo  {journal}
  {JCAP}\ }\textbf {\bibinfo {volume} {01}},\ \bibinfo {pages} {001} (\bibinfo
  {year} {2021})},\ \Eprint {http://arxiv.org/abs/2007.08537} {arXiv:2007.08537
  [hep-ph]} \BibitemShut {NoStop}%
\bibitem [{\citenamefont {Beniwal}\ \emph {et~al.}(2017)\citenamefont
  {Beniwal}, \citenamefont {Lewicki}, \citenamefont {Wells}, \citenamefont
  {White},\ and\ \citenamefont {Williams}}]{Beniwal:2017eik}%
  \BibitemOpen
  \bibfield  {author} {\bibinfo {author} {\bibfnamefont {A.}~\bibnamefont
  {Beniwal}}, \bibinfo {author} {\bibfnamefont {M.}~\bibnamefont {Lewicki}},
  \bibinfo {author} {\bibfnamefont {J.~D.}\ \bibnamefont {Wells}}, \bibinfo
  {author} {\bibfnamefont {M.}~\bibnamefont {White}}, \ and\ \bibinfo {author}
  {\bibfnamefont {A.~G.}\ \bibnamefont {Williams}},\ }\href {\doibase
  10.1007/JHEP08(2017)108} {\bibfield  {journal} {\bibinfo  {journal} {JHEP}\
  }\textbf {\bibinfo {volume} {08}},\ \bibinfo {pages} {108} (\bibinfo {year}
  {2017})},\ \Eprint {http://arxiv.org/abs/1702.06124} {arXiv:1702.06124
  [hep-ph]} \BibitemShut {NoStop}%
\bibitem [{\citenamefont {Blasi}\ and\ \citenamefont
  {Mariotti}(2022)}]{Blasi:2022woz}%
  \BibitemOpen
  \bibfield  {author} {\bibinfo {author} {\bibfnamefont {S.}~\bibnamefont
  {Blasi}}\ and\ \bibinfo {author} {\bibfnamefont {A.}~\bibnamefont
  {Mariotti}},\ }\href {\doibase 10.1103/PhysRevLett.129.261303} {\bibfield
  {journal} {\bibinfo  {journal} {Phys. Rev. Lett.}\ }\textbf {\bibinfo
  {volume} {129}},\ \bibinfo {pages} {261303} (\bibinfo {year} {2022})},\
  \Eprint {http://arxiv.org/abs/2203.16450} {arXiv:2203.16450 [hep-ph]}
  \BibitemShut {NoStop}%
\bibitem [{\citenamefont {Blasi}\ \emph {et~al.}(2023)\citenamefont {Blasi},
  \citenamefont {Jinno}, \citenamefont {Konstandin}, \citenamefont {Rubira},\
  and\ \citenamefont {Stomberg}}]{Blasi:2023rqi}%
  \BibitemOpen
  \bibfield  {author} {\bibinfo {author} {\bibfnamefont {S.}~\bibnamefont
  {Blasi}}, \bibinfo {author} {\bibfnamefont {R.}~\bibnamefont {Jinno}},
  \bibinfo {author} {\bibfnamefont {T.}~\bibnamefont {Konstandin}}, \bibinfo
  {author} {\bibfnamefont {H.}~\bibnamefont {Rubira}}, \ and\ \bibinfo {author}
  {\bibfnamefont {I.}~\bibnamefont {Stomberg}},\ }\href {\doibase
  10.1088/1475-7516/2023/10/051} {\bibfield  {journal} {\bibinfo  {journal}
  {JCAP}\ }\textbf {\bibinfo {volume} {10}},\ \bibinfo {pages} {051} (\bibinfo
  {year} {2023})},\ \Eprint {http://arxiv.org/abs/2302.06952} {arXiv:2302.06952
  [astro-ph.CO]} \BibitemShut {NoStop}%
\bibitem [{\citenamefont {Agrawal}\ \emph {et~al.}(2024)\citenamefont
  {Agrawal}, \citenamefont {Blasi}, \citenamefont {Mariotti},\ and\
  \citenamefont {Nee}}]{Agrawal:2023cgp}%
  \BibitemOpen
  \bibfield  {author} {\bibinfo {author} {\bibfnamefont {P.}~\bibnamefont
  {Agrawal}}, \bibinfo {author} {\bibfnamefont {S.}~\bibnamefont {Blasi}},
  \bibinfo {author} {\bibfnamefont {A.}~\bibnamefont {Mariotti}}, \ and\
  \bibinfo {author} {\bibfnamefont {M.}~\bibnamefont {Nee}},\ }\href {\doibase
  10.1007/JHEP06(2024)089} {\bibfield  {journal} {\bibinfo  {journal} {JHEP}\
  }\textbf {\bibinfo {volume} {06}},\ \bibinfo {pages} {089} (\bibinfo {year}
  {2024})},\ \Eprint {http://arxiv.org/abs/2312.06749} {arXiv:2312.06749
  [hep-ph]} \BibitemShut {NoStop}%
\bibitem [{\citenamefont {Li}\ \emph {et~al.}(2025)\citenamefont {Li},
  \citenamefont {Jia},\ and\ \citenamefont {Bian}}]{Li:2023yzq}%
  \BibitemOpen
  \bibfield  {author} {\bibinfo {author} {\bibfnamefont {Y.}~\bibnamefont
  {Li}}, \bibinfo {author} {\bibfnamefont {Y.}~\bibnamefont {Jia}}, \ and\
  \bibinfo {author} {\bibfnamefont {L.}~\bibnamefont {Bian}},\ }\href {\doibase
  10.1088/1475-7516/2025/02/038} {\bibfield  {journal} {\bibinfo  {journal}
  {JCAP}\ }\textbf {\bibinfo {volume} {02}},\ \bibinfo {pages} {038} (\bibinfo
  {year} {2025})},\ \Eprint {http://arxiv.org/abs/2304.05220} {arXiv:2304.05220
  [hep-ph]} \BibitemShut {NoStop}%
\bibitem [{\citenamefont {Beniwal}\ \emph {et~al.}(2019)\citenamefont
  {Beniwal}, \citenamefont {Lewicki}, \citenamefont {White},\ and\
  \citenamefont {Williams}}]{Beniwal:2018hyi}%
  \BibitemOpen
  \bibfield  {author} {\bibinfo {author} {\bibfnamefont {A.}~\bibnamefont
  {Beniwal}}, \bibinfo {author} {\bibfnamefont {M.}~\bibnamefont {Lewicki}},
  \bibinfo {author} {\bibfnamefont {M.}~\bibnamefont {White}}, \ and\ \bibinfo
  {author} {\bibfnamefont {A.~G.}\ \bibnamefont {Williams}},\ }\href {\doibase
  10.1007/JHEP02(2019)183} {\bibfield  {journal} {\bibinfo  {journal} {JHEP}\
  }\textbf {\bibinfo {volume} {02}},\ \bibinfo {pages} {183} (\bibinfo {year}
  {2019})},\ \Eprint {http://arxiv.org/abs/1810.02380} {arXiv:1810.02380
  [hep-ph]} \BibitemShut {NoStop}%
\bibitem [{\citenamefont {Bian}\ and\ \citenamefont
  {Tang}(2018)}]{Bian:2018mkl}%
  \BibitemOpen
  \bibfield  {author} {\bibinfo {author} {\bibfnamefont {L.}~\bibnamefont
  {Bian}}\ and\ \bibinfo {author} {\bibfnamefont {Y.-L.}\ \bibnamefont
  {Tang}},\ }\href {\doibase 10.1007/JHEP12(2018)006} {\bibfield  {journal}
  {\bibinfo  {journal} {JHEP}\ }\textbf {\bibinfo {volume} {12}},\ \bibinfo
  {pages} {006} (\bibinfo {year} {2018})},\ \Eprint
  {http://arxiv.org/abs/1810.03172} {arXiv:1810.03172 [hep-ph]} \BibitemShut
  {NoStop}%
\bibitem [{\citenamefont {Azatov}\ \emph {et~al.}(2022)\citenamefont {Azatov},
  \citenamefont {Barni}, \citenamefont {Chakraborty}, \citenamefont
  {Vanvlasselaer},\ and\ \citenamefont {Yin}}]{Azatov:2022tii}%
  \BibitemOpen
  \bibfield  {author} {\bibinfo {author} {\bibfnamefont {A.}~\bibnamefont
  {Azatov}}, \bibinfo {author} {\bibfnamefont {G.}~\bibnamefont {Barni}},
  \bibinfo {author} {\bibfnamefont {S.}~\bibnamefont {Chakraborty}}, \bibinfo
  {author} {\bibfnamefont {M.}~\bibnamefont {Vanvlasselaer}}, \ and\ \bibinfo
  {author} {\bibfnamefont {W.}~\bibnamefont {Yin}},\ }\href {\doibase
  10.1007/JHEP10(2022)017} {\bibfield  {journal} {\bibinfo  {journal} {JHEP}\
  }\textbf {\bibinfo {volume} {10}},\ \bibinfo {pages} {017} (\bibinfo {year}
  {2022})},\ \Eprint {http://arxiv.org/abs/2207.02230} {arXiv:2207.02230
  [hep-ph]} \BibitemShut {NoStop}%
\bibitem [{\citenamefont {Schicho}\ \emph {et~al.}(2022)\citenamefont
  {Schicho}, \citenamefont {Tenkanen},\ and\ \citenamefont
  {White}}]{Schicho:2022wty}%
  \BibitemOpen
  \bibfield  {author} {\bibinfo {author} {\bibfnamefont {P.}~\bibnamefont
  {Schicho}}, \bibinfo {author} {\bibfnamefont {T.~V.~I.}\ \bibnamefont
  {Tenkanen}}, \ and\ \bibinfo {author} {\bibfnamefont {G.}~\bibnamefont
  {White}},\ }\href {\doibase 10.1007/JHEP11(2022)047} {\bibfield  {journal}
  {\bibinfo  {journal} {JHEP}\ }\textbf {\bibinfo {volume} {11}},\ \bibinfo
  {pages} {047} (\bibinfo {year} {2022})},\ \Eprint
  {http://arxiv.org/abs/2203.04284} {arXiv:2203.04284 [hep-ph]} \BibitemShut
  {NoStop}%
\bibitem [{\citenamefont {Smith}\ \emph {et~al.}(2019)\citenamefont {Smith},
  \citenamefont {Smith}, \citenamefont {Caldwell},\ and\ \citenamefont
  {Caldwell}}]{Smith:2019wny}%
  \BibitemOpen
  \bibfield  {author} {\bibinfo {author} {\bibfnamefont {T.~L.}\ \bibnamefont
  {Smith}}, \bibinfo {author} {\bibfnamefont {T.~L.}\ \bibnamefont {Smith}},
  \bibinfo {author} {\bibfnamefont {R.~R.}\ \bibnamefont {Caldwell}}, \ and\
  \bibinfo {author} {\bibfnamefont {R.}~\bibnamefont {Caldwell}},\ }\href
  {\doibase 10.1103/PhysRevD.100.104055} {\bibfield  {journal} {\bibinfo
  {journal} {Phys. Rev. D}\ }\textbf {\bibinfo {volume} {100}},\ \bibinfo
  {pages} {104055} (\bibinfo {year} {2019})},\ \bibinfo {note} {[Erratum:
  Phys.Rev.D 105, 029902 (2022)]},\ \Eprint {http://arxiv.org/abs/1908.00546}
  {arXiv:1908.00546 [astro-ph.CO]} \BibitemShut {NoStop}%
\bibitem [{\citenamefont {Pieroni}\ and\ \citenamefont
  {Barausse}(2020)}]{Pieroni:2020rob}%
  \BibitemOpen
  \bibfield  {author} {\bibinfo {author} {\bibfnamefont {M.}~\bibnamefont
  {Pieroni}}\ and\ \bibinfo {author} {\bibfnamefont {E.}~\bibnamefont
  {Barausse}},\ }\href {\doibase 10.1088/1475-7516/2020/07/021} {\bibfield
  {journal} {\bibinfo  {journal} {JCAP}\ }\textbf {\bibinfo {volume} {07}},\
  \bibinfo {pages} {021} (\bibinfo {year} {2020})},\ \bibinfo {note} {[Erratum:
  JCAP 09, E01 (2020)]},\ \Eprint {http://arxiv.org/abs/2004.01135}
  {arXiv:2004.01135 [astro-ph.CO]} \BibitemShut {NoStop}%
\bibitem [{\citenamefont {Flauger}\ \emph {et~al.}(2021)\citenamefont
  {Flauger}, \citenamefont {Karnesis}, \citenamefont {Nardini}, \citenamefont
  {Pieroni}, \citenamefont {Ricciardone},\ and\ \citenamefont
  {Torrado}}]{Flauger:2020qyi}%
  \BibitemOpen
  \bibfield  {author} {\bibinfo {author} {\bibfnamefont {R.}~\bibnamefont
  {Flauger}}, \bibinfo {author} {\bibfnamefont {N.}~\bibnamefont {Karnesis}},
  \bibinfo {author} {\bibfnamefont {G.}~\bibnamefont {Nardini}}, \bibinfo
  {author} {\bibfnamefont {M.}~\bibnamefont {Pieroni}}, \bibinfo {author}
  {\bibfnamefont {A.}~\bibnamefont {Ricciardone}}, \ and\ \bibinfo {author}
  {\bibfnamefont {J.}~\bibnamefont {Torrado}},\ }\href {\doibase
  10.1088/1475-7516/2021/01/059} {\bibfield  {journal} {\bibinfo  {journal}
  {JCAP}\ }\textbf {\bibinfo {volume} {01}},\ \bibinfo {pages} {059} (\bibinfo
  {year} {2021})},\ \Eprint {http://arxiv.org/abs/2009.11845} {arXiv:2009.11845
  [astro-ph.CO]} \BibitemShut {NoStop}%
\bibitem [{\citenamefont {Fisher}(1922)}]{Fisher:1922saa}%
  \BibitemOpen
  \bibfield  {author} {\bibinfo {author} {\bibfnamefont {R.~A.}\ \bibnamefont
  {Fisher}},\ }\href {\doibase 10.1098/rsta.1922.0009} {\bibfield  {journal}
  {\bibinfo  {journal} {Phil. Trans. Roy. Soc. Lond. A}\ }\textbf {\bibinfo
  {volume} {222}},\ \bibinfo {pages} {309} (\bibinfo {year}
  {1922})}\BibitemShut {NoStop}%
\bibitem [{\citenamefont {Cornish}\ and\ \citenamefont
  {Robson}(2017)}]{Cornish:2017vip}%
  \BibitemOpen
  \bibfield  {author} {\bibinfo {author} {\bibfnamefont {N.}~\bibnamefont
  {Cornish}}\ and\ \bibinfo {author} {\bibfnamefont {T.}~\bibnamefont
  {Robson}},\ }\href {\doibase 10.1088/1742-6596/840/1/012024} {\bibfield
  {journal} {\bibinfo  {journal} {J. Phys. Conf. Ser.}\ }\textbf {\bibinfo
  {volume} {840}},\ \bibinfo {pages} {012024} (\bibinfo {year} {2017})},\
  \Eprint {http://arxiv.org/abs/1703.09858} {arXiv:1703.09858 [astro-ph.IM]}
  \BibitemShut {NoStop}%
\bibitem [{\citenamefont {Robson}\ \emph {et~al.}(2019)\citenamefont {Robson},
  \citenamefont {Cornish},\ and\ \citenamefont {Liu}}]{Robson:2018ifk}%
  \BibitemOpen
  \bibfield  {author} {\bibinfo {author} {\bibfnamefont {T.}~\bibnamefont
  {Robson}}, \bibinfo {author} {\bibfnamefont {N.~J.}\ \bibnamefont {Cornish}},
  \ and\ \bibinfo {author} {\bibfnamefont {C.}~\bibnamefont {Liu}},\ }\href
  {\doibase 10.1088/1361-6382/ab1101} {\bibfield  {journal} {\bibinfo
  {journal} {Class. Quant. Grav.}\ }\textbf {\bibinfo {volume} {36}},\ \bibinfo
  {pages} {105011} (\bibinfo {year} {2019})},\ \Eprint
  {http://arxiv.org/abs/1803.01944} {arXiv:1803.01944 [astro-ph.HE]}
  \BibitemShut {NoStop}%
\bibitem [{\citenamefont {Lewicki}\ and\ \citenamefont
  {Vaskonen}(2023{\natexlab{b}})}]{Lewicki:2021kmu}%
  \BibitemOpen
  \bibfield  {author} {\bibinfo {author} {\bibfnamefont {M.}~\bibnamefont
  {Lewicki}}\ and\ \bibinfo {author} {\bibfnamefont {V.}~\bibnamefont
  {Vaskonen}},\ }\href {\doibase 10.1140/epjc/s10052-023-11323-2} {\bibfield
  {journal} {\bibinfo  {journal} {Eur. Phys. J. C}\ }\textbf {\bibinfo {volume}
  {83}},\ \bibinfo {pages} {168} (\bibinfo {year} {2023}{\natexlab{b}})},\
  \Eprint {http://arxiv.org/abs/2111.05847} {arXiv:2111.05847 [astro-ph.CO]}
  \BibitemShut {NoStop}%
\bibitem [{\citenamefont {Hartwig}\ \emph {et~al.}(2023)\citenamefont
  {Hartwig}, \citenamefont {Lilley}, \citenamefont {Muratore},\ and\
  \citenamefont {Pieroni}}]{Hartwig:2023pft}%
  \BibitemOpen
  \bibfield  {author} {\bibinfo {author} {\bibfnamefont {O.}~\bibnamefont
  {Hartwig}}, \bibinfo {author} {\bibfnamefont {M.}~\bibnamefont {Lilley}},
  \bibinfo {author} {\bibfnamefont {M.}~\bibnamefont {Muratore}}, \ and\
  \bibinfo {author} {\bibfnamefont {M.}~\bibnamefont {Pieroni}},\ }\href
  {\doibase 10.1103/PhysRevD.107.123531} {\bibfield  {journal} {\bibinfo
  {journal} {Phys. Rev. D}\ }\textbf {\bibinfo {volume} {107}},\ \bibinfo
  {pages} {123531} (\bibinfo {year} {2023})},\ \Eprint
  {http://arxiv.org/abs/2303.15929} {arXiv:2303.15929 [gr-qc]} \BibitemShut
  {NoStop}%
\bibitem [{\citenamefont {Ekstedt}\ \emph {et~al.}(2024)\citenamefont
  {Ekstedt}, \citenamefont {Schicho},\ and\ \citenamefont
  {Tenkanen}}]{Ekstedt:2024etx}%
  \BibitemOpen
  \bibfield  {author} {\bibinfo {author} {\bibfnamefont {A.}~\bibnamefont
  {Ekstedt}}, \bibinfo {author} {\bibfnamefont {P.}~\bibnamefont {Schicho}}, \
  and\ \bibinfo {author} {\bibfnamefont {T.~V.~I.}\ \bibnamefont {Tenkanen}},\
  }\href {\doibase 10.1103/PhysRevD.110.096006} {\bibfield  {journal} {\bibinfo
   {journal} {Phys. Rev. D}\ }\textbf {\bibinfo {volume} {110}},\ \bibinfo
  {pages} {096006} (\bibinfo {year} {2024})},\ \Eprint
  {http://arxiv.org/abs/2405.18349} {arXiv:2405.18349 [hep-ph]} \BibitemShut
  {NoStop}%
\bibitem [{\citenamefont {Chiang}\ \emph {et~al.}(2019)\citenamefont {Chiang},
  \citenamefont {Li},\ and\ \citenamefont {Senaha}}]{Chiang:2018gsn}%
  \BibitemOpen
  \bibfield  {author} {\bibinfo {author} {\bibfnamefont {C.-W.}\ \bibnamefont
  {Chiang}}, \bibinfo {author} {\bibfnamefont {Y.-T.}\ \bibnamefont {Li}}, \
  and\ \bibinfo {author} {\bibfnamefont {E.}~\bibnamefont {Senaha}},\ }\href
  {\doibase 10.1016/j.physletb.2018.12.017} {\bibfield  {journal} {\bibinfo
  {journal} {Phys. Lett. B}\ }\textbf {\bibinfo {volume} {789}},\ \bibinfo
  {pages} {154} (\bibinfo {year} {2019})},\ \Eprint
  {http://arxiv.org/abs/1808.01098} {arXiv:1808.01098 [hep-ph]} \BibitemShut
  {NoStop}%
\bibitem [{\citenamefont {Gonderinger}\ \emph {et~al.}(2010)\citenamefont
  {Gonderinger}, \citenamefont {Li}, \citenamefont {Patel},\ and\ \citenamefont
  {Ramsey-Musolf}}]{Gonderinger:2009jp}%
  \BibitemOpen
  \bibfield  {author} {\bibinfo {author} {\bibfnamefont {M.}~\bibnamefont
  {Gonderinger}}, \bibinfo {author} {\bibfnamefont {Y.}~\bibnamefont {Li}},
  \bibinfo {author} {\bibfnamefont {H.}~\bibnamefont {Patel}}, \ and\ \bibinfo
  {author} {\bibfnamefont {M.~J.}\ \bibnamefont {Ramsey-Musolf}},\ }\href
  {\doibase 10.1007/JHEP01(2010)053} {\bibfield  {journal} {\bibinfo  {journal}
  {JHEP}\ }\textbf {\bibinfo {volume} {01}},\ \bibinfo {pages} {053} (\bibinfo
  {year} {2010})},\ \Eprint {http://arxiv.org/abs/0910.3167} {arXiv:0910.3167
  [hep-ph]} \BibitemShut {NoStop}%
\bibitem [{\citenamefont {Workman}\ and\ \citenamefont
  {Others}(2022)}]{Workman:2022ynf}%
  \BibitemOpen
  \bibfield  {author} {\bibinfo {author} {\bibfnamefont {R.~L.}\ \bibnamefont
  {Workman}}\ and\ \bibinfo {author} {\bibnamefont {Others}} (\bibinfo
  {collaboration} {Particle Data Group}),\ }\href {\doibase
  10.1093/ptep/ptac097} {\bibfield  {journal} {\bibinfo  {journal} {PTEP}\
  }\textbf {\bibinfo {volume} {2022}},\ \bibinfo {pages} {083C01} (\bibinfo
  {year} {2022})}\BibitemShut {NoStop}%
\bibitem [{\citenamefont {Kainulainen}\ \emph {et~al.}(2019)\citenamefont
  {Kainulainen}, \citenamefont {Keus}, \citenamefont {Niemi}, \citenamefont
  {Rummukainen}, \citenamefont {Tenkanen},\ and\ \citenamefont
  {Vaskonen}}]{Kainulainen:2019kyp}%
  \BibitemOpen
  \bibfield  {author} {\bibinfo {author} {\bibfnamefont {K.}~\bibnamefont
  {Kainulainen}}, \bibinfo {author} {\bibfnamefont {V.}~\bibnamefont {Keus}},
  \bibinfo {author} {\bibfnamefont {L.}~\bibnamefont {Niemi}}, \bibinfo
  {author} {\bibfnamefont {K.}~\bibnamefont {Rummukainen}}, \bibinfo {author}
  {\bibfnamefont {T.~V.~I.}\ \bibnamefont {Tenkanen}}, \ and\ \bibinfo {author}
  {\bibfnamefont {V.}~\bibnamefont {Vaskonen}},\ }\href {\doibase
  10.1007/JHEP06(2019)075} {\bibfield  {journal} {\bibinfo  {journal} {JHEP}\
  }\textbf {\bibinfo {volume} {06}},\ \bibinfo {pages} {075} (\bibinfo {year}
  {2019})},\ \Eprint {http://arxiv.org/abs/1904.01329} {arXiv:1904.01329
  [hep-ph]} \BibitemShut {NoStop}%
\bibitem [{\citenamefont {Kajantie}\ \emph {et~al.}(1998)\citenamefont
  {Kajantie}, \citenamefont {Laine}, \citenamefont {Rajantie}, \citenamefont
  {Rummukainen},\ and\ \citenamefont {Tsypin}}]{Kajantie:1998yc}%
  \BibitemOpen
  \bibfield  {author} {\bibinfo {author} {\bibfnamefont {K.}~\bibnamefont
  {Kajantie}}, \bibinfo {author} {\bibfnamefont {M.}~\bibnamefont {Laine}},
  \bibinfo {author} {\bibfnamefont {A.}~\bibnamefont {Rajantie}}, \bibinfo
  {author} {\bibfnamefont {K.}~\bibnamefont {Rummukainen}}, \ and\ \bibinfo
  {author} {\bibfnamefont {M.}~\bibnamefont {Tsypin}},\ }\href {\doibase
  10.1088/1126-6708/1998/11/011} {\bibfield  {journal} {\bibinfo  {journal}
  {JHEP}\ }\textbf {\bibinfo {volume} {11}},\ \bibinfo {pages} {011} (\bibinfo
  {year} {1998})},\ \Eprint {http://arxiv.org/abs/hep-lat/9811004}
  {arXiv:hep-lat/9811004} \BibitemShut {NoStop}%
\end{thebibliography}%
\end{document}